\documentclass[smallextended]{svjour3}
\usepackage[table]{xcolor}
\usepackage[utf8]{inputenc}
\usepackage{mathpazo}
\usepackage{amsmath}
\usepackage{url}
\usepackage[round]{natbib}
\usepackage[margin=1in]{geometry}
\usepackage{amssymb}
\usepackage{standalone}
\usepackage{graphicx}
\usepackage{epstopdf}
\usepackage{hyperref}
\usepackage{subfigure}

\usepackage[normalem]{ulem}

\usepackage{bm}

\usepackage{booktabs}

\usepackage{multicol} 
\usepackage[math]{cellspace}
\AtBeginDocument{
    \heavyrulewidth=.08em
    \lightrulewidth=.05em
    \cmidrulewidth=.03em
    \belowrulesep=.65ex
    \belowbottomsep=0pt
    \aboverulesep=.4ex
    \abovetopsep=0pt
    \cmidrulesep=\doublerulesep
    \cmidrulekern=.5em
    \defaultaddspace=.5em
}

\definecolor{MyBlue}{rgb}  {0.1,0.1,0.9}
\definecolor{MyRed}{rgb}   {0.9,0.1,0.1}
\definecolor{MyGreen}{rgb} {0.05,0.55,0.1}
\definecolor{burntorange}{rgb}{0.8, 0.33, 0.0}
\definecolor{NeilMagenta}{rgb}{0.8, 0.1, 0.8}
\definecolor{table1}{rgb}{0.9, 0.9, 0.98}
\definecolor{table2}{rgb}{0.8, 0.8, 1.0}

\newcommand{\temp}[1]{{ #1}}

\newcommand{\fd}[2]{\frac{\mathrm{d} #1 }{\mathrm{d} #2}}

\newcommand \beq{\begin{eqnarray}}
\newcommand \eeq{\end{eqnarray}}
\newcommand \beqno{\begin{eqnarray*}}
\newcommand \eeqno{\end{eqnarray*}}
\newcommand \bit{\begin{itemize}}
\newcommand \eit{\end{itemize}}

\begin{document}
\title{Pattern Formation as a Resilience Mechanism in Cancer Immunotherapy}
\titlerunning{Pattern Formation as a Resilience Mechanism in Cancer Immunotherapy}
\author{ Molly Brennan \and Andrew L. Krause\footnote{Corresponding author \email{andrew.krause@durham.ac.uk}} \and Edgardo Villar-Sep\'{u}lveda \and Christopher B. Prior}
\authorrunning{M. Brennan \and A. L. Krause \and E. Villar-Sep\'{u}veda \and C. B. Prior}

\institute{M. Brennan 
    \at Department of Mathematics, University College London, 25 Gordon Street, London WC1H 0AY, United Kingdom\\ \\
    {A. L. Krause, C. B. Prior}
    \at Mathematical Sciences Department, Durham University, Upper Mountjoy Campus, Stockton Rd, Durham DH1 3LE, United Kingdom\\ \\
    E. Villar-Sep\'{u}lveda 
\at {School of Engineering Mathematics and Technology}, {University of Bristol}, {{Ada Lovelace Building, Tankard's Cl, University Walk}, {Bristol}, {BS8 1TW}, United Kingdom }}

\date{Received: date / Accepted: date}

\maketitle

\begin{abstract}
Mathematical and computational modelling in oncology has played an increasingly important role in not only understanding the impact of various approaches to treatment on tumour growth, but in optimizing dosing regimens and aiding the development of treatment strategies. However, as with all modelling, only an approximation is made in the description of the biological and physical system. Here we show that tissue-scale spatial structure can have a profound impact on the resilience of tumours to immunotherapy using a classical model incorporating IL-2 compounds and effector cells as treatment parameters. Using linear stability analysis, numerical continuation, and direct simulations, we show that diffusing cancer cell populations can undergo pattern-forming (Turing) instabilities, leading to spatially-structured states that persist far into treatment regimes where the corresponding spatially homogeneous systems would uniformly predict a cancer-free state. These spatially-patterned states persist in a wide range of parameters, as well as under time-dependent treatment regimes. Incorporating treatment via domain boundaries can increase this resistance to treatment in the interior of the domain, further highlighting the importance of spatial modelling when designing treatment protocols informed by mathematical models. Counter-intuitively, this mechanism shows that increased effector cell mobility can increase the resilience of tumours to treatment. We conclude by discussing practical and theoretical considerations for understanding this kind of spatial resilience in other models of cancer treatment, in particular those incorporating more realistic spatial transport.

This paper belongs to the special collection: Problems, Progress and Perspectives in Mathematical and Computational Biology.
\end{abstract}
 

\maketitle

\section{Introduction}


Mathematical modelling has played an increasingly important role in both our basic understanding of cancer dynamics \citep{byrne2006modelling, altrock2015mathematics}, as well as in informing better treatment protocols \citep{enderling2006mathematical, zahid2023proliferation, browning2024predicting, pasetto2024calibrating}. Spatial modelling of solid tumours has a history dating at least to the 1960s, with diffusion of oxygen implicated as a crucial part of the growth process \citep{burton_rate_1966, greenspan1972models, araujo_history_2004}. Many frameworks now exist for studying the spatial distribution, growth, and invasion of solid tumours \citep{osborne2010hybrid, sfakianakis2020mathematical}, as well as the subsequent restructuring of tissue microvasculature \citep{anderson1998continuous}. These models have implicated important quantitative and qualitative impacts of spatial considerations, as compared to well-mixed (spatially-averaged) counterparts, especially in cases such as radiotherapy, where it is difficult to average out spatial effects due to inherent heterogeneity \citep{bull2022hallmarks}. In this paper, we consider a model of immunotherapeutic treatment of a solid tumour to uncover another important effect of spatial considerations: the role of subcritical Turing instabilities in providing resilience to treatment. \temp{Turing instabilities, whereby spatially inhomogeneous structures emerge from a spatially homogeneous state after a small perturbation, can therefore provide a novel route to treatment resistance, underscoring the importance of spatial frameworks.} This has important impacts when modelling realistic time-and-space-dependent treatment conditions.

There are a variety of immunotherapeutic treatments for cancer, which have been pursued as an approach to developing novel treatment strategies \citep{mellman2011cancer}. Mathematical modelling of immunotherapies has revealed a range of complex interactions between cancer and immune cell dynamics with differing treatment protocols. {We first describe spatially well-mixed models, in the form of ordinary differential equations (ODEs), of cancer immunotherapy.} Building on substantial previous modelling work, \cite{kuznetsov1994nonlinear} explored a reduced model capturing immune (effector) cells, tumour cells, and a tumour-immune complex, finding key links between the phase space dynamics and observed behaviours of tumour dormancy and tumour escape in cancer progression. They also discuss how perturbing conditions can be linked to immunostimulated tumour growth. \cite{kirschner1998modeling} extended this approach with an additional concentration representing Interleukin-2 (IL-2), an immune cell-stimulating protein often used in immunotherapy, and studied the occurrence of limit cycles which they interpreted as representing periodic tumour recurrence. These models were further extended with immune response time delays by \cite{banerjee_immunotherapy_2008}, who found that delay could crucially change the stability of the cancer-immune coexistence state, depending on the amount of IL-2 present. \cite{tsygvintsev2012mathematical} considered an alternative extension incorporating time-dependent treatment, and explored how to determine optimal treatment regimes for the elimination of the tumour. \temp{Optimal scheduling of dosing protocols is a nontrivial task, especially given the complexities of delayed responses of immunotherapy observed clinically for some kinds of treatment \citep{ji2024utility}.}

While these {non-spatial} systems provide useful insight into the key parameters and interactions that affect tumour growth in the presence of an immune response or immunotherapy, they do not account for spatial effects, which can alter the behaviour of the system leading to different conclusions, for example on necessary treatment parameters for cancer clearance. \cite{marongiu_cancer_2012} present an extensive overview of experimental evidence proposing altered tissue architecture and other spatially dependent effects play key roles in early cancer progression, \temp{especially emphasising the importance of morphogenetic mechanisms of tissue patterning and organization in tumour development which occurs differentially across space}. \cite{suddin2021reaction} extend the well-mixed system in \citep{kirschner1998modeling} to a partial differential equation (PDE) system by considering the effects of diffusion in one spatial dimension. Through numerical simulations they show that the spatiotemporal behaviour of the system strongly depends on the random motility coefficient of the tumour cell population achieving both one-dimensional stable ``spots" and temporal oscillations of an almost spatially homogeneous tumour cell population under the range of motility coefficients considered. \cite{matzavinos2004mathematical} present a spatiotemporal tumour-immune model representing early-stage solid tumour growth and the effect of immune response on cancer dormancy. Through carrying out numerical simulations they witnessed non-standard travelling wave-type behaviour with fronts of ``solitary-like waves" invading the tissue, leading to spatially heterogeneous tumour cell density. Through a numerical bifurcation analysis they attribute this behaviour to the presence of a stable limit cycle arising through a Hopf bifurcation in immune response parameters. \cite{dai_optimal_2019} consider the invasion of a solid tumour in the context of an optimal control problem for chemotherapy and immunotherapy, showing that optimal treatment can break the typical ring and aggregate patterns that form in the system limiting the spread of the tumour and metastasis. \cite{ko_stationary_2011} investigate the behaviour of a similar tumour-immune reaction-diffusion system as in \cite{suddin2021reaction}, finding that large diffusivity of effector cells allows for the existence of a spatially heterogeneous steady state in their system, and that in such regimes there are no stable spatially homogeneous positive (i.e.~cancer coexistence) states. Emergence of heterogeneous structures via Turing-type instabilities in cancers has been considered by several authors, such as \cite{chaplain1995reaction, chaplain2001spatio} who postulated periodic pattern formation as a possible mechanism for the columnar extensions in carcinomas as they invade surrounding tissue. 

Here we consider an alternative application of such periodic pattern-formation in cancer, drawing on ideas of resilience from spatial ecology and epidemiology. Understanding the resilience of ecosystems to external perturbations, and how many populations avoid extinction in the face of `tipping points' is a well-studied area of modern ecological theory; see the reviews \citep{folke2004regime, ives2007stability, scheffer2015generic, kefi2022scaling} on the ecological side, \citep{van2002reproduction} for a broad mathematical approach in the epidemiological context, and \citep{krakovska2024resilience} for several mathematical formalisms of resilience in general dynamical systems. In our context, the fundamental idea will be that subcritical Turing instabilities give rise to parameter regimes where the treatment parameters are such that the cancer-free state is linearly stable, but stable spatially-inhomogeneous solutions exist where the cancer cell population still resides. Such scenarios have been implicated in a range of models in spatial ecology \citep{rietkerk2004self, bonachela2015termite}, providing a mechanism of persistence of ecosystems even during severe regime shifts. {The role of such spatial instabilities as a resilience mechanism has not, as far as we know, been explored in the context of cancer despite the potential impacts it may have on tumour growth and treatment.}

\begin{figure}
    \centering
    \subfigure[]{\includegraphics[width=0.3\linewidth]{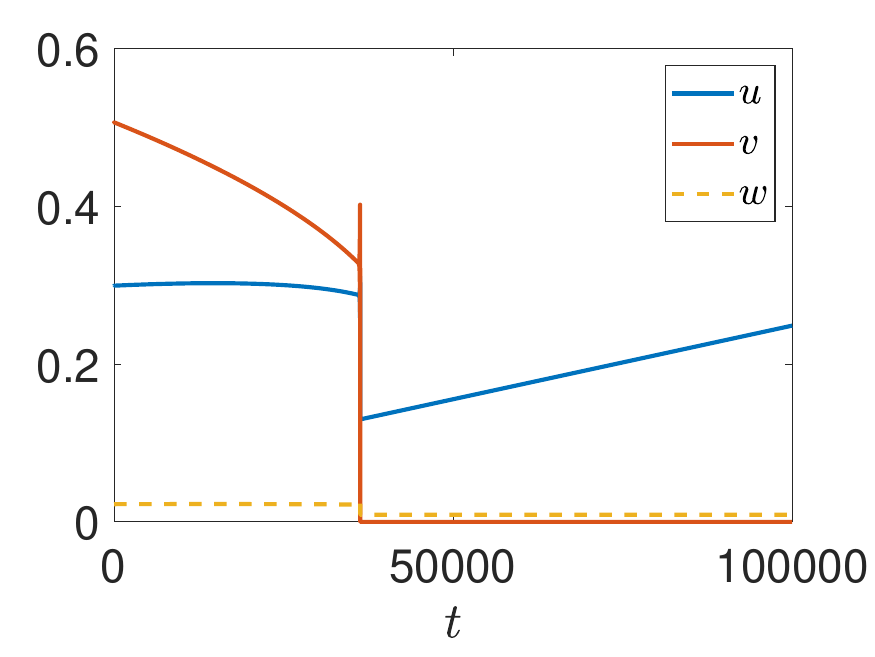}}
    \subfigure[]{\includegraphics[width =0.3\linewidth]{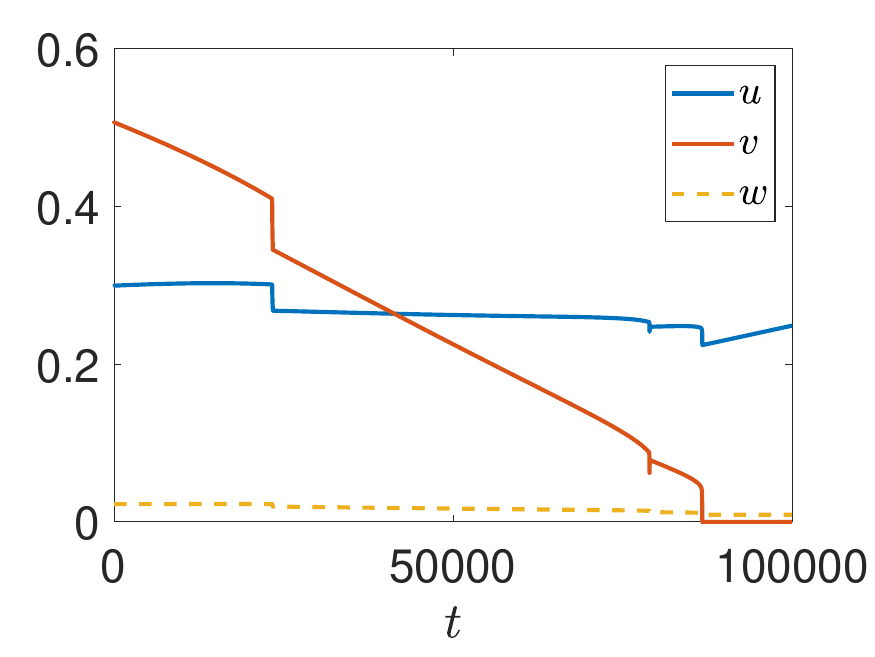}}
    \subfigure[]{\includegraphics[width = 0.3\linewidth]{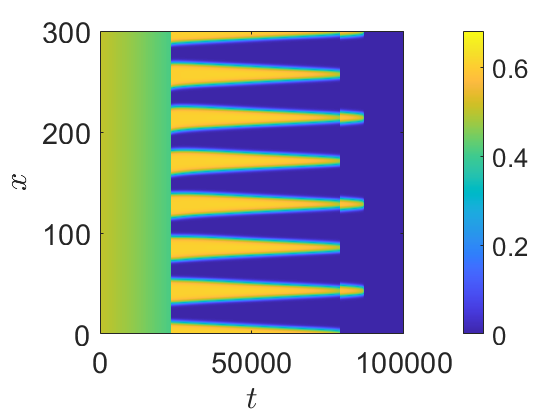}}

    \caption{(a): Simulations of the ODE model given by Equations \eqref{u_hom_eq}-\eqref{w_hom_eq}; the variables $u$ and $w$ correspond to immunotherapy-related quantities whereas $v$ represents the density of cancer cells. (b): Spatial averages from simulations of the PDE model given by Equations \eqref{u_nondim_eq}-\eqref{w_nondim_eq} with homogeneous Neumann boundary conditions. (c): Kymographs of $v$ from the spatial (PDE) simulations in (b). Parameters are given in Table \ref{parameter_table} with $\alpha=0.07$, $\delta_u=\delta_w=100$, where in (a)-(c) we have $\sigma_w=0.5$, and $\sigma_u = 0.01+0.03\frac{t}{T}$, with the final simulation time given by $T = 10^5$ in all cases.}
    \label{fig:overview_figure}
\end{figure}

{To give an idea of this mechanism, we show a pair of motivating simulations in Figure \ref{fig:overview_figure}}. This Figure compares ODE {(panel (a))} and PDE {(panels (b)-(c))} simulations of {an immunotherapy} model using identical parameters, where we increase one of the immunotherapy parameters in time via a slow, quasi-static ramp \temp{(modelling increasing doses of effector cells)}. The ODE model results in panel (a) show that the cancer cells ($v$) become extinct around time $t=3500$, whereas the PDE model results in panels (b)-(c) show a much longer persistence of the cancer cells, effectively demonstrating existence of cancer at nearly twice the treatment dose. \temp{Using the same parameters between spatial and non-spatial models is a crude method of comparison, but nevertheless this gives a picture of the impact of spatial considerations on the dynamics (see Section \ref{sec_discussion} for further discussion).} The mechanism of spatial resilience here is very similar to examples in spatial ecology, but has not, as far as we are aware, been emphasised in the cancer literature. Hence our goal will be to demonstrate this mechanism via an exemplar immunotherapy model, and discuss implications more broadly for the development of treatment protocols motivated by or which incorporate mechanistic modelling.

We will analyze a spatial variant of the model first proposed by \cite{kirschner1998modeling} by incorporating diffusion of cells and proteins as a simple spatial extension (essentially of the form proposed in \citep{suddin2021reaction}) in Section \ref{sec_modelling}. We will then use linear stability analysis of spatially homogeneous equilibria to find regions of coexistence and cancer-free stability, as well as regions where coexistence equilibria can undergo Turing instabilities, in Section \ref{sec_LSA}. We will extend this linear stability analysis to inhomogeneous equilibria via numerical continuation in Section \ref{sec_continuation}. We will then simulate this model in a variety of parameter regimes and geometries in Section \ref{sec_sim}, finding that patterned states not found in spatially homogeneous variants of the model can persist into regions where the cancer-free state is stable, implicating Turing instabilities as a mechanism for cancer resilience to treatment. We will focus on examples of time-dependent and boundary-driven treatment protocols which are in some sense `far' into parameter regions where the spatially homogeneous system would predict clearance of the cancer population. We discuss these results, including important caveats and generalizations, in Section \ref{sec_discussion}. Importantly, while we will study a rather simple spatial model incorporating only diffusion (neglecting nonlinear fluxes due to stromal/extracellular matrix interactions \citep{gatenby1996reaction, strobl2020mix, crossley2024modeling}, chemotaxis, or elastic effects \citep{walker2023minimal}), the mechanism of resilience we uncover is plausible in a range of forms of spatial interactions. Similarly we study a model of immunotherapy due to it being a paradigm of mathematical modelling in cancer, but anticipate that the general mechanisms observed are of wider interest well beyond this particular modelling framework and type of treatment protocol.

\section{Spatial Modelling of Tumour Immunotherapy via Effector Cells}\label{sec_modelling}

Our modelling framework will strike a balance between being complicated enough to be typical of more realistic reaction-transport models of cancer, but simple enough to analyze in detail to uncover fundamental mechanisms of resilience. This takes the form of the model presented in \cite{suddin2021reaction}, adapted from a purely temporal (ODE) formulation in \cite{kirschner1998modeling}. This model contains necessary tumour-immune components allowing us to consider the effects of immune response and immunotherapy, while being simple enough to admit some analytic analysis. 

We consider densities of tumour cells ($T$), effector cells ($E$) and IL-2 compounds ($I_L$). The latter two are part of the immune system and will be used within this model to represent both immune response, and immunotherapy via external input. The system is represented by the following set of equations:
\begin{align}
    \frac{\partial E}{\partial \overline{t}} =&d_E{\nabla}^2 E +cT -m_EE + \frac{p_EEI_L}{g_E + I_L} + s_E, \label{effectoreq}
    \\ 
    \frac{\partial T}{\partial \overline{t}} =&d_T{\nabla}^2 T + r_TT(1-bT) -\frac{p_TET}{g_T + T}, \label{tumoureq}
    \\ 
    \frac{\partial I_L}{\partial \overline{t}} =&d_{I_L}{\nabla}^2 I_L + \frac{p_{I_L}ET}{g_{I_L} + T} - m_{I_L}I_L + s_{I_L} \label{ILeq}\;, 
\end{align}
where all parameters are positive real constants. Equation \eqref{effectoreq} models the effector cell population incorporating tumour-induced growth following an immune response, with $c$ capturing the strength of this response. We model proliferation of the effector cell population stimulated by IL-2 concentration, and cell death at the rate $m_E$. Treatment is modelled by a constant external input of effector cells ($s_E$) which we later extend to investigate more realistic time-dependent treatments. The dynamics of the tumour cell population, given in Equation \eqref{tumoureq}, incorporates logistic growth and elimination of tumour cells upon interaction with the effector cells. The motivation for this form of interaction comes from both the effects of a solid mass of tumour cells, meaning the internal cells do not come into contact with the effector cells, and the limitations of the immune response \citep{kirschner1998modeling}. Finally, the dynamics of the IL-2 compounds in Equation \eqref{ILeq} include a source term stimulated by interaction between the tumour cells and the effector cells, degradation and treatment via external input of (typically synthetic) IL-2, which is often used both on its own and in conjunction with other treatment as a form of immunotherapy \citep{IL2}. We will specify the compact 1D and 2D geometries that we will consider later, with $\bm{x} \in \Omega$ denoting the spatial coordinate and domain. Except for a few illustrative cases, we will consider homogeneous Neumann boundary conditions for each species throughout for simplicity.

As an aside, we note that motility of effector and cancer cells is complicated by several factors. Cancer cells do not respond to normal cell signals, meaning they grow and divide uncontrollably \citep{cooper2007cell}, with rates of division varying across the population. Additionally, a simple diffusion model fails to account for the active movement of cells encountered in reality. For example, \cite{geiger2009metastasis} claim that tumour invasion occurs through movement as ``single cells or collectively in the form of clusters, or sheets", with the benefit of the latter being the ability of invading tumour cells with different advantageous adaptions working together to form new growths (as shown, for example, by \cite{strobl2020mix}). \cite{tracqui2009biophysical} discusses tumour movement via durotaxis, suggesting the cells move towards regions with a more rigid extracellular matrix. This is not unique to tumour cells and the general mechanistic role of the surroundings in influencing directional movement is explored by \cite{espina2022durotaxis}. Furthermore, the movement of tumour cells via chemotaxis, active transport along a concentration gradient, and haptotaxis, movement towards areas of higher adhesion properties, is investigated in \cite{chaplain2006mathematical} with the mathematical forms for such transport presented in a model that incorporates the extracellular matrix. The importance of such directional movement is highlighted as increased haptotaxis is seen to result in metastasis. Rather than consider these more complex forms of movement, we will use the crude diffusion model above to represent a simplified form of cell movement that we anticipate captures qualitatively similar features as more sophisticated models.

\subsection{Parameter Estimation}

We turn to previous studies on tumour-immune interactions for guidance on parameter values. We remark that while the precise details do matter for quantitative outcomes, our main focus here is to demonstrate qualitative phenomena possible within a realistic modelling exemplar that are likely to generalize to many other complex models in oncology.

We set $r_T=0.18 \text{\hspace{0.1cm}days}^{-1}$, as this value was stated in \cite{kuznetsov1994nonlinear} to produce growth graphs that best matched experimental data for tumour growth. We remark that for the parameter $b$ there is a large variation in the value (and interpretation) across the cited literature. For consistency with the scales in the simulations in \cite{suddin2021reaction}, we opt to use the value $b = 10^{-6}\text{\hspace{0.1cm} cm cells}^{-1}$. For the parameter $p_T$, representing the degradation rate of tumour cells due to effector cells, we note that \cite{tsygvintsev2012mathematical} took the range $0.01 \leq p_T \leq 100$ days$^{-1}$. This range was used to allow for varying the effector cells' ability to eliminate the tumour cells using gene therapy. Here we will use $p_T=0.1$days$^{-1}$, as in \citep{suddin2021reaction}, studying a value closer to the worst-case scenario in tumour-immune interactions. Aside from the diffusion parameters, which we explore separately below, we use all other parameters as estimated by \cite{kirschner1998modeling}, using experimental data, medical literature and sensitivity analysis. We note that a range of values are explored for the parameter $c$, as this represents the immune response that the tumour elicits which varies between patients, with higher values of $c$ implying a more active immune response to the tumour.

\subsubsection{Estimation of Diffusion Parameters}
Following the work by \cite{matzavinos2004mathematical} we obtain estimates for the range of diffusion parameters for each population, representing their random motility.

\subsubsection*{Effector cells}

The diffusion coefficient of the effector cells was estimated by using the Stokes-Einstein equation for the diffusion of spherical particles through a liquid with low Reynolds number:
\beq \label{einstein-stokes}
d_E = \Large  \frac{k_BT}{6 \pi R_1 \eta},
\eeq
where $T = 310K$ is the temperature in Kelvin, $R_1 = 4\mu$m is an average cell radius, $\eta$ = $\eta_{\text{water}} \approx$ $6.922\text{x}10^{-4}Kg m^{-1} s^{-1}$ is the viscosity of the medium, and $k_B$ is Boltzmann's constant. This leads to an estimated diffusion coefficient of $d_E = 7\text{x}10^{-5}\text{\hspace{0.1cm}cm}^2\text{\hspace{0.1cm}days}^{-1}$. This is a very simplistic representation of cell movement and in reality the effector cells are likely to actively move towards immunogenic regions or otherwise travel through less random means. We note that \cite{matzavinos2004mathematical} quotes a diffusion coefficient of up to $10^{-2}\text{\hspace{0.1cm}cm}^2\text{\hspace{0.1cm}days}^{-1}$, so we will consider a wide range of cell diffusion coefficients.

\subsubsection*{IL-2 compounds}

For the diffusion rate of IL-2 compounds, $d_3$, we take the range quoted in \cite{matzavinos2004mathematical} of $10^{-4} - 10^{-2}\text{\hspace{0.1cm}cm}^2\text{\hspace{0.1cm}days}^{-1}$ as these compounds are known to diffuse at a faster rate than cells, consistent with their smaller size in \eqref{einstein-stokes}. 

\subsubsection*{Tumour cells}

While we could use the same estimates of diffusion as the effector cells, due to the complexity of tumour movement and growth we opt instead to use the equation in \cite{prigogine1980stability}:
\begin{equation}
    d_T=D^2\lambda \label{diffusiontumour}
\end{equation}
which says the diffusion rate of the tumour cells is equal to its diameter, $D$, squared multiplied by the duplication rate of the tumour cells, $\lambda$. \cite{hao2018size} records tumour cell diameters on the order of $10\mu$m. According to \cite{cooper2007cell} a typical eukaryotic cell in humans divides every 24 hours. Using this as an estimate of the duplication rate, Equation \eqref{diffusiontumour} gives a motility estimate of $1.21\text{x}10^{-6}\text{\hspace{0.1cm}cm}^2\text{\hspace{0.1cm}days}^{-1}$. Varying $d_T$, i.e. varying the duplication rate and/or tumour cell radius in the derivation above, leads to large qualitative changes in the predicted behaviour highlighting the importance of tumour motility within immunotherapy models. We also remark that estimates of the treatment parameters, $S_E$ and $S_{I_L}$ will vary widely depending on dosing regimes. Determining such optimal dosing schedules, which typically involve balancing drug toxicity/side effects with achieving elimination or substantial reduction in solid tumour size, is a key goal of approaches in mathematical and computational oncology \citep{barbolosi2016computational}.

 \begin{table}
 \centering
 \setlength{\tabcolsep}{8pt} 
 \renewcommand{\arraystretch}{1.5}
 \rowcolors{2}{table1}{table2}
 \begin{tabular}{ | m{4cm} | m{2cm} | m{4cm} | m{3cm}|}
 \hline
 \textbf{Dimensional \newline parameters} & \textbf{Units} & \textbf{References} & \textbf{Nondimensional \newline parameters}\\
 \hline
 $7\times 10^{-5} \leq d_E \leq 10^{-2}$  &  cm$^{2}$days$^{-1}$& \cite{matzavinos2004mathematical}  & $0.747 \leq \delta_u \leq 8264$\\
 
  $1.21\times 10^{-6}\leq d_T \leq 9.375\times 10^{-5}$ & cm$^{2}$days$^{-1}$& \cite{matzavinos2004mathematical}, \newline \cite{prigogine1980stability} & $1.067 \leq \delta_w \leq 8264$\\
 
 $10^{-4}\leq d_{I_L} \leq 10^{-2}$ & cm$^{2}$days$^{-1}$& \cite{matzavinos2004mathematical} & $0 \leq \alpha \leq 0.154$\\

 $p_E=0.1245$ & days$^{-1}$& \cite{kirschner1998modeling} & $\mu_u = 0.167$\\

 $p_T=0.1$ & days$^{-1}$& \cite{suddin2021reaction}, \newline \cite{tsygvintsev2012mathematical} & $\rho_u =0.692$\\

 $p_{I_L} = 5$ & days$^{-1}$& \cite{kirschner1998modeling}& $\mu_w = 55.56$\\

 $g_E = 2\times 10^7$ &cells cm$^{-1}$ & \cite{kirschner1998modeling} & $\gamma_v = 0.1$\\

 $g_T = 1\times 10^5$ & cells cm$^{-1}$& \cite{kirschner1998modeling} & $\rho_w = 2.5$\\

 $g_{I_L} = 1\times 10^3$ & cells cm$^{-1}$& \cite{kirschner1998modeling} & $\gamma_w = 1\times10^{-3}$\\

 $m_E = 0.03$ & days$^{-1}$& \cite{kirschner1998modeling} & \\

 $m_{I_L} = 10$ & days$^{-1}$& \cite{kirschner1998modeling} & \\

 $r_T = 0.18$ & days$^{-1}$& \cite{kuznetsov1994nonlinear} & \\

 $b= 1.0\times 10^{-6}$ & cm cells$^{-1}$ & \cite{matzavinos2004mathematical}, \newline\cite{suddin2021reaction} & \\

 $ 0 \leq c \leq 0.05$ & days$^{-1}$& \cite{kirschner1998modeling} & \\
 \hline
 \end{tabular}
 \caption{Dimensional and nondimensional parameters in the model.}
 \label{parameter_table}
 \end{table}

\subsection{Nondimensionalisation}

\temp{In order to simplify the subsequent analysis, we perform a standard nondimensionalisation to reduce the overall number of parameters in the model.} We substitute scalings of each variable given by
\begin{equation}
    E = E_0u, \quad T = T_0v, \quad I_L = I_{L_0}w, \quad \overline{t} = t_s t, \quad \bm{x} = X\bm{\overline{x}}
\end{equation}
into \eqref{effectoreq}-\eqref{ILeq} to arrive at
\begin{align}
    \frac{\partial u}{\partial {t}} =&\frac{t_s d_E}{X^2}{\nabla}^2u  +\frac{t_s T_0c}{E_0}v -t_sm_E u + \frac{p_Et_s I_{L_0}uw}{g_E + I_{L_0}w} + \frac{t_s s_E}{E_0},
    \\ 
    \frac{\partial v}{\partial {t}} =& \frac{t_s d_T}{X^2}{\nabla}^2 v + r_Tt_s v(1-bT_0v) -\frac{p_TE_0t_s uv}{g_T + T_0v},
    \\ 
    \frac{\partial w}{\partial {t}} =&\frac{t_s d_{I_L}}{X^2}{\nabla}^2 w + \frac{p_{I_L}t_sE_0T_0uv}{I_{L_0}g_{I_L} + I_{L_0}T_0v} - t_sm_{I_L} w  + \frac{t_s s_{I_L}}{I_{L_0}}.
\end{align}
We can hence simplify the system slightly by setting,
\begin{equation}
    t_s = \frac{1}{r_T}, \quad E_0 = \frac{r_T}{bp_T}\quad  T_0 = \frac{1}{b}, \quad I_{L_0} = g_E, \quad X = \sqrt{\frac{d_T}{r_T}},
\end{equation}
finally leading to the system,
\begin{align}
    \frac{\partial u}{\partial {t}} =&\delta_u{\nabla}^2u  +\alpha v -\mu_u u + \frac{\rho_u uw}{1 + w} + \sigma_u, \label{u_nondim_eq}\\ 
    \frac{\partial v}{\partial {t}} =& {\nabla}^2 v +  v(1-v) -\frac{uv}{\gamma_v + v}, \label{v_nondim_eq}\\ 
    \frac{\partial w}{\partial {t}} =&\delta_w{\nabla}^2 w + \frac{\rho_w uv}{\gamma_w + v} - \mu_w w  +\sigma_w,\label{w_nondim_eq}
\end{align}
where the nondimensional parameters are given by,
\begin{equation}\begin{aligned}
   &\delta_u = \frac{d_E}{d_T}, \quad \alpha = \frac{p_T c}{r_T^2}\quad  \mu_u = \frac{m_E}{r_T}, \quad \rho_u = \frac{p_E}{r_T}, \quad \sigma_u = \frac{bp_T s_E}{r_T^2}, \quad \gamma_v = g_T b,\\
   &\delta_w = \frac{d_{I_L}}{d_T}, \quad \rho_w = \frac{p_{I_L}}{bp_Tg_E}, \quad \gamma_w = g_{I_L}b, \quad \mu_w = \frac{m_{I_L}}{r_T}, \quad \sigma_w = \frac{ s_{I_L}}{g_E r_T}.
\end{aligned}\end{equation}


 









We note that  $t_s$ is a characteristic time scale of tumour growth. This sets a bound on the minimum time for tumour eradication upon interaction with effector cells on a small scale, as in \cite{allison2004mathematical}. We note that our choices of $T_0$ and $t_s$ agree with \cite{suddin2021reaction}, but we have opted to use a different scaling of $E_0$, $I_{L_0}$, and $X$ in order to reduce the number of parameters in the nondimensional model. We note that the scale for $\bm{x}$ is then on the order of millimeters, and $t$ is approximately on the scale of weeks. This nondimensionalisation differs from most of the previous literature, but allows for a simpler interpretation of the five main bifurcation parameters, $\alpha$, $\sigma_u$, $\sigma_w$, $\delta_u$, and $\delta_w$ relative to the timescale of tumour growth and diffusion, which are both set to unity. Our scaling is somewhat of an aesthetic choice rather than one with a large impact on the quantitative and qualitative analysis that follows. \temp{We remark that we fix these parameters for clarity here, but one would expect possibly significant variation in clinical settings with human tissues (see the discussion by \cite{kuznetsov1994nonlinear} on global parameter sensitivity for example). Our focus will be on exhibiting a specific mechanism and understanding it in the context of a `realistic' immunotherapy model, rather than a detailed multiparameter sensitivity analysis. Even within the limited range of parameters we explore, we find rich dynamics which plausibly play roles outside of this particular modelling framework and parameter regime.}



\section{Linear Stability Analysis} \label{sec_LSA}

\subsection{Spatially Homogeneous Stability Analysis}

We first consider the spatially homogeneous ODE model:
\begin{equation}\label{u_hom_eq}
    \fd{u}{t} = \alpha v -\mu_u u + \frac{\rho_u uw}{1 + w} + \sigma_u,
\end{equation}
\begin{equation}\label{v_hom_eq}
    \fd{v}{t} = v(1-v) -\frac{uv}{\gamma_v + v},
\end{equation}
\begin{equation}\label{w_hom_eq}
    \fd{w}{t}=\frac{\rho_w uv}{\gamma_w + v} - \mu_w w  +\sigma_w.
\end{equation}
The Jacobian matrix of the system is given by,
\begin{equation}\label{homog_jacobian}
    \bm{J}(u,v,w) = \begin{pmatrix} J_{11}  & J_{12} & J_{13} \\ J_{21} & J_{22}  & J_{23} \\ J_{31} & J_{32} & J_{33}
    \end{pmatrix} = 
    \begin{pmatrix}
 \frac{\rho_{u}w}{w+1}-\mu_{u} & \alpha & \frac{\rho_{u}u}{w+1}-\frac{\rho_{u}uw}{{\left(w+1\right)}^2}\\ -\frac{v}{\gamma_{v}+v} & \frac{uv}{{\left(\gamma_{v}+v\right)}^2}-\frac{u}{\gamma_{v}+v}-2v+1 & 0\\ \frac{\rho_{w}v}{\gamma_{w}+v} & \frac{\rho_{w}u}{\gamma_{w}+v}-\frac{\rho_{w}uv}{{\left(\gamma_{w}+v\right)}^2} & -\mu_{w} 
\end{pmatrix}.
\end{equation}

There is a unique cancer free $(v_0=0)$ equilibrium for the spatially homogeneous model,  as in \citep{kirschner1998modeling, suddin2021reaction}, given by:
\begin{equation}
   (u_0, v_0, w_0) =  \left ( \frac{\sigma_u(\mu_w + \sigma_w)}{\mu_u (\mu_w + \sigma_w) -  \rho_u \sigma_w}, 0, \frac{\sigma_{w}}{\mu_w} \right ),
\end{equation}
which we see is feasible as long as $\mu_u (\mu_w + \sigma_w) > \rho_u \sigma_w$. The eigenvalues of the Jacobian matrix evaluated at this equilibrium are given by
\begin{equation}
    \lambda_1 = 1-\frac{\sigma_{u}(\mu_{w}+\sigma_{w})}{\gamma_{v}(\mu_{u}\mu_{w}+\mu_{u}\sigma_{w}-\rho_{u}\sigma_{w})}, \quad \lambda_2 =\frac{\rho_{u}\sigma_{w}-\mu_{u}(\mu_{w}+\sigma_{w})}{\mu_{w}+\sigma_{w}}, \quad  \lambda_3 = -\mu_w.
\end{equation}
To ensure $\lambda_2<0$ we need $\mu_{u}(\mu_{w}+\sigma_{w})>\rho_{u}\sigma_{w}$, which is also the feasibility criterion. This automatically ensures that the denominator in the second term of $\lambda_1$ is positive, so we only additionally require that denominator to be smaller than the numerator to ensure $\lambda_1<0$. In summary then, to ensure stability of the cancer-free state (and thus potential elimination of the cancer) we need the parameters to satisfy
\begin{equation}\label{stab_CF}
 \sigma_{u}(\mu_{w}+\sigma_{w})>\gamma_{v}(\mu_{u}\mu_{w}+\mu_{u}\sigma_{w}-\rho_{u}\sigma_{w})>0.
\end{equation}
Hence we see that the cancer free state is unstable with treatment solely through IL-2 input ($\sigma_w>0$ if $\sigma_u=0$), indicating that treatment via IL-2 compounds alone is not enough to ensure elimination of the tumour. This is consistent with the biology, as the IL-2 compounds do not directly affect the tumour cells themselves, but simply stimulate proliferation in the effector cells.

Looking for strictly positive steady states (`coexistence steady states'), we can assume $v_0 \neq 0$. 
Dividing the steady state form of \eqref{v_hom_eq} by $v_0$ and solving for $u_0$ we find,
\begin{equation}
 u_0 = (\gamma_v + v_0)(1-v_0). \label{uss}  
\end{equation}
Solving the steady state version of \eqref{w_hom_eq} for $w_0$ and substituting in the expression \eqref{uss}, we find,
\begin{equation}
w_0 = \frac{\sigma _{w}+\frac{\rho _{w}u_0v_0}{\gamma _{w}+v_0}}{\mu _{w}} = \frac{\sigma _{w}+\frac{\rho _{w}v_0\left(\gamma _{v}+v_0\right)\left(1-v_0\right)}{\gamma _{w}+v_0}}{\mu _{w}}.    \label{wss}
\end{equation}

We can substitute the expressions, \eqref{uss} and \eqref{wss}, for $w_0$ and $u_0$ into the steady state form of \eqref{u_hom_eq} to obtain a quintic equation for $v_0$ after carefully clearing denominators (see Appendix \ref{appendix_equilibria} for the expression of this polynomial). We can then consider only sets of parameters which admit roots $v_0 \in (0,1)$, as anything outside of this range will lead to an infeasible value of $u_0$. Without making further assumptions on the parameters, we then have to proceed numerically to search for solutions and check their stability via the eigenvalues of \eqref{homog_jacobian}. For parameters in the ranges given by Table \ref{parameter_table}, we only ever observe at most two stable homogeneous steady state solutions with $v_0 \in (0, 1)$.

We note that, while the coexistence equilibria are not simple to analyze directly, parameters exhibiting transcritical bifurcations between coexistence and cancer-free states occur exactly when the constant term in the quintic equation for such equilibria becomes zero (equation \eqref{quintic}); that is, transcritical bifurcations involving $v_0 = 0$ can only occur along the curve $a_0 = \gamma_w(\gamma_v\rho_u\sigma_w +(\mu_w + \sigma_w)(\sigma_u - \gamma_v \mu_u))= 0$. We note that, assuming $\gamma_w > 0$, these parameters are exactly when Equation \eqref{stab_CF} becomes an equality. Hence the only instabilities of the cancer-free state occur due to transcritical bifurcations with a coexistence equilibrium.

\begin{figure}
    \centering
    \subfigure[$\alpha=0.01$]{\includegraphics[width=0.8\linewidth]{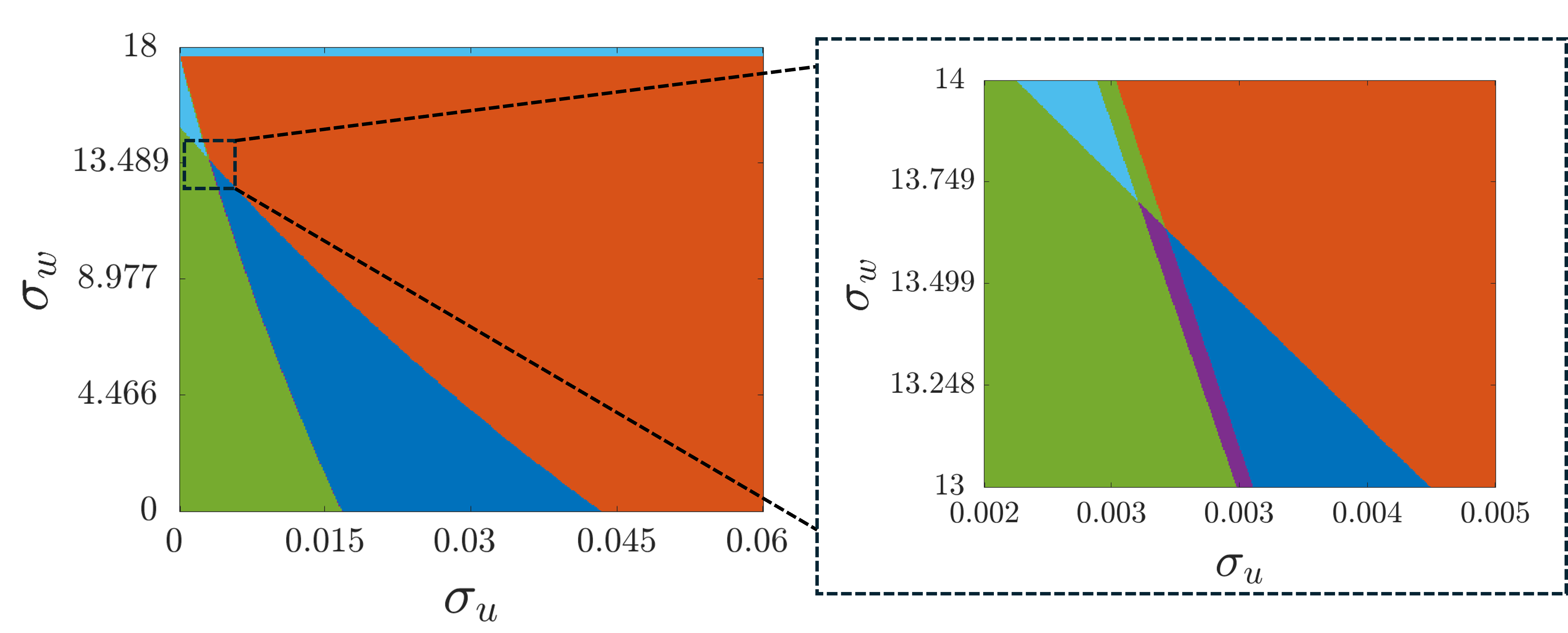}}

    \subfigure[$\alpha=0.07$]{\includegraphics[width=0.8\linewidth]{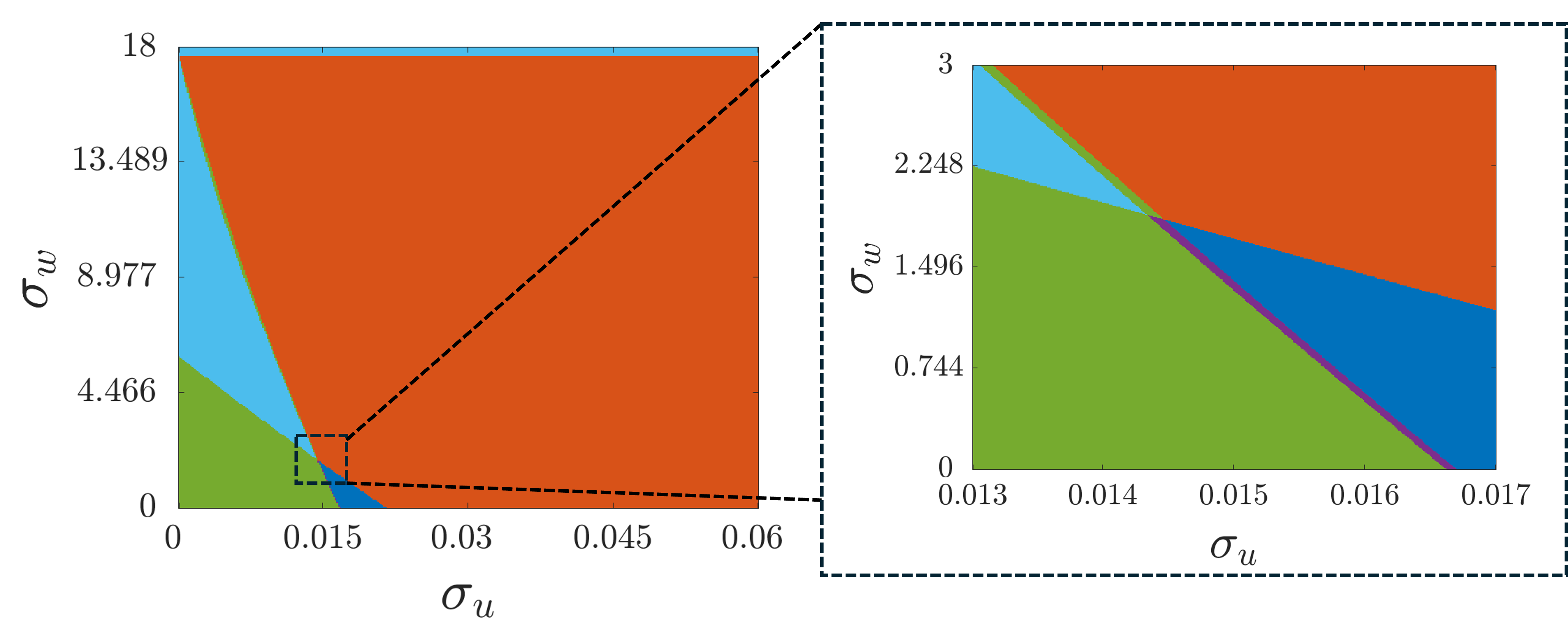}}
    
    \subfigure[$\alpha=0.1$]{\includegraphics[width=0.4\linewidth]{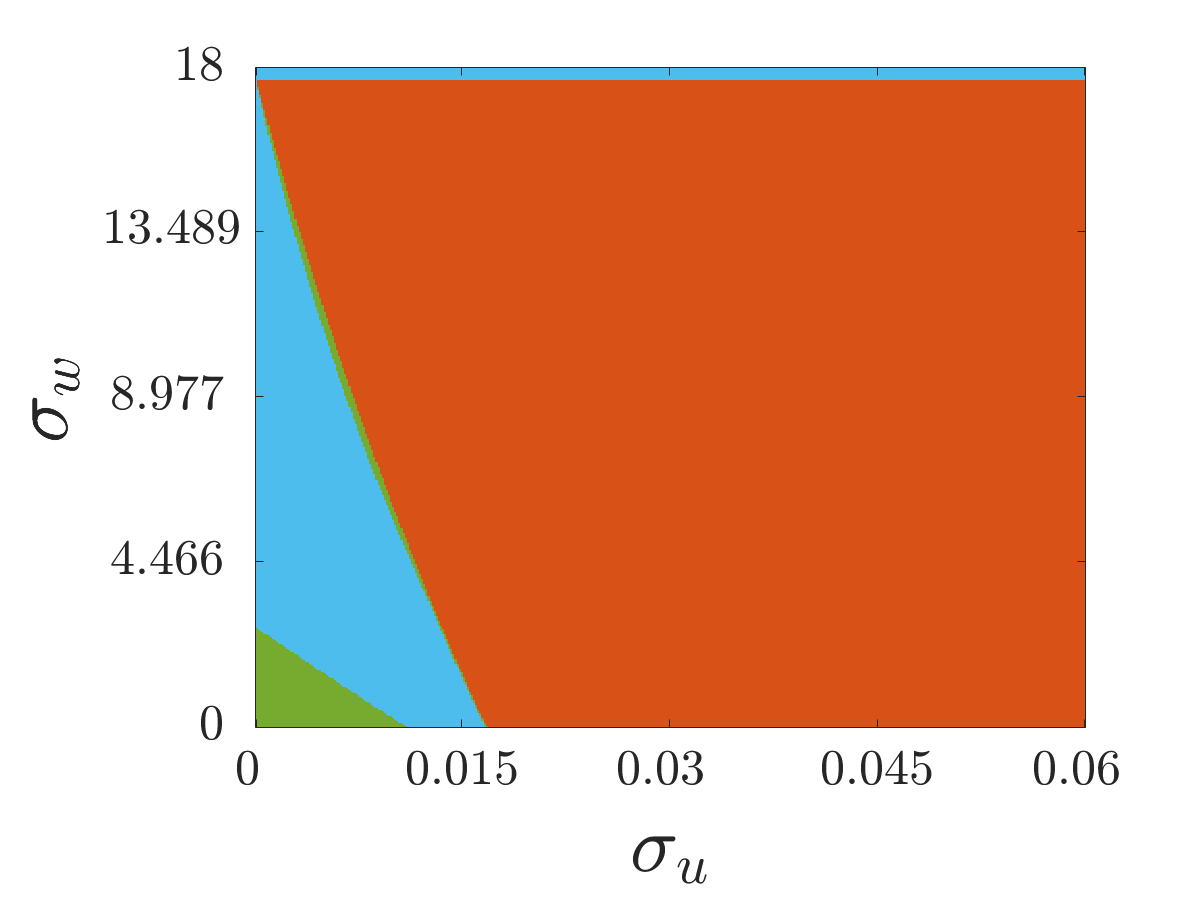}}
    \subfigure[$\alpha=0.2$]{\includegraphics[width=0.4\linewidth]{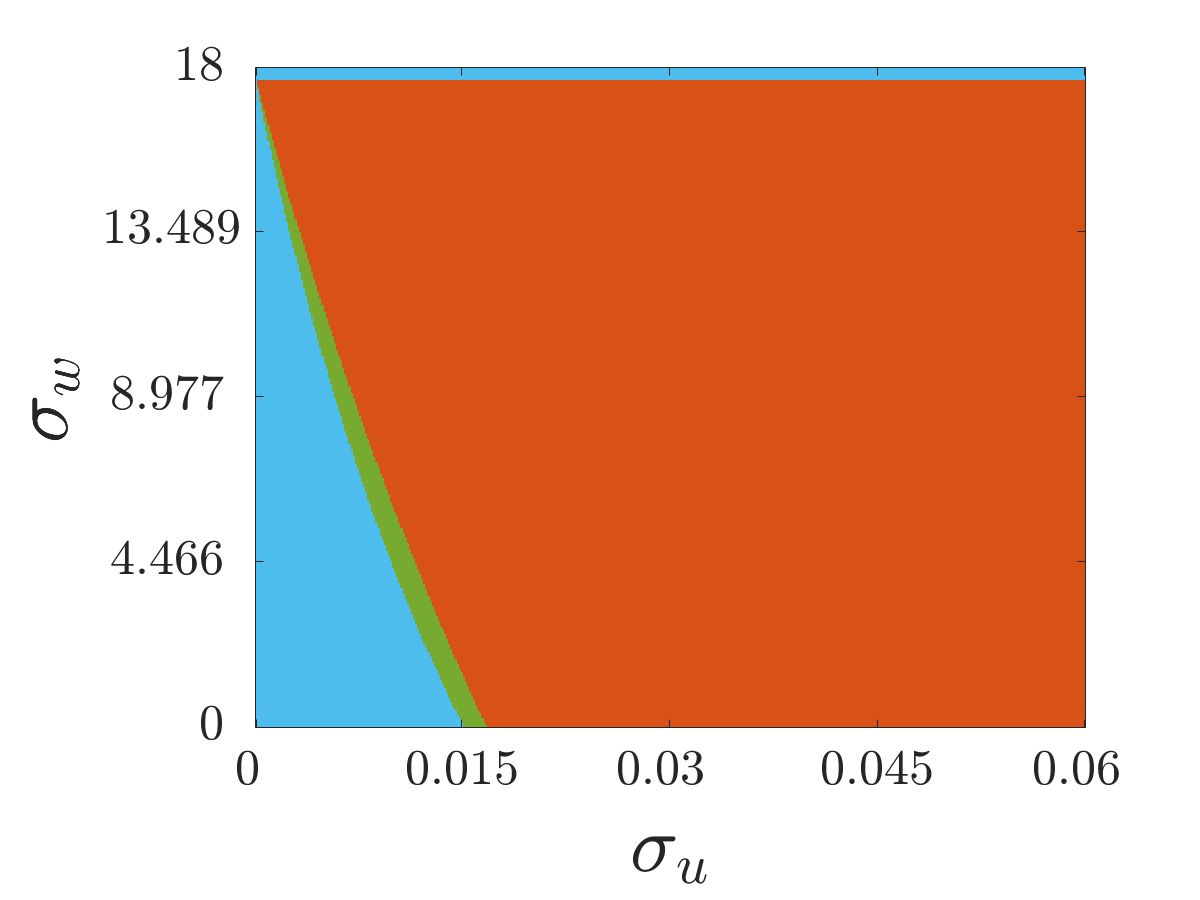}}

    \vspace{0.2cm}
    \includegraphics[width = 0.45\linewidth]{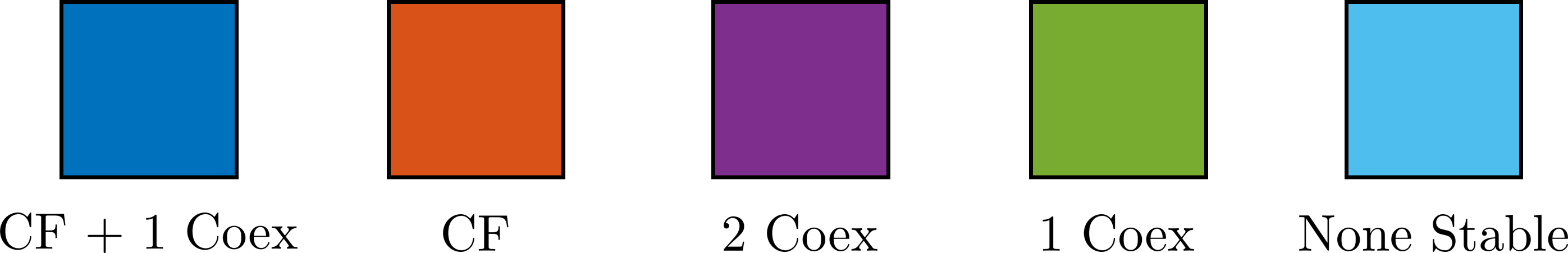}
    \vspace{0.1cm}
    \caption{Classification of the number and type of linearly stable steady states of Equations \eqref{u_hom_eq}-\eqref{w_hom_eq} for varying $\sigma_u$, $\sigma_v$ and $\alpha$. Here, `CF' stands for cancer-free equilibrium, and `coex' for coexistence equilibria.}
    \label{fig:lin_stab_plots}
\end{figure}

We vary the nondimensional treatment parameters $\sigma_u$ and $\sigma_w$ for different levels of $\alpha$ and summarize the results of this stability analysis in Figure \ref{fig:lin_stab_plots} (see the code on GitHub \citep{Molly_Github} for the implementation). We find regions of cancer-free and coexistence equilibria, including bistability regions of two distinct coexistence states as well as bistability regions of a coexistence equilibrium with the cancer-free state. Both bistability regions exist for approximately $0 < \alpha < 0.083$. \temp{Some of} these bounds are consistent with those reported in \citep{ko_stationary_2011} under certain restrictions of the parameters, and a different nondimensionalisation. \temp{Importantly, \cite{ko_stationary_2011} choose parameters that preclude the existence of multiple coexistence states, but also incorporate methods to assess global asymptotic stability of different equilibria in some cases.}

We next analyze the eigenvalues across the boundaries between stability regions using the code in \citep{villar2023computation} to classify the bifurcations in each region, and in particular to determine the criticality of Hopf bifurcations. We find that all boundaries crossing into the `None Stable' (light blue) region are subcritical Hopf bifurcations. The boundaries between the coexistence and cancer-free stability regions are a mix of transcritical bifurcations (with one coexistence equilibrium decreasing in density as $\sigma_u$ is increased until it reaches $v_0=0$), as well as subcritical Hopf bifurcations. The Hopf bifurcations occur for larger values of $\sigma_w$, with the transition in these plots occurring in the cancer-free and coexistence bistability (dark blue) region when it exists. 

We see that increasing values of $\alpha$ increases both the size of the cancer-free stability region, as well as the size of the oscillatory Hopf regime, while decreasing the overall size of the coexistence stability regions. Large values of $\sigma_u$ will always lead to only the cancer-free equilibrium being stable (which can be shown asymptotically), whereas large values of $\sigma_w$ will instead lead to oscillatory states (as all equilibria are no longer feasible). However, such large dosing regimes are unlikely to be sensible in real applications due to dose-dependent side effects, and so the dynamics for small and intermediate parameters are more relevant. We also remark that while larger values of $\alpha$ decrease the overall size of the coexistence stability region, the size of the coexistence stability region (green) bordering the cancer-free stability region (orange) changes somewhat non-monotonically but with a substantially larger extent for $\alpha=0.2$ compared to $\alpha=0.1$. This particular coexistence stability region corresponds to a rather small value of $v_0$ which borders the oscillatory `None Stable' (light blue) region.

\subsection{Spatially Inhomogeneous Stability Analysis}
We now consider a linear stability analysis of the PDE system given by Equations \eqref{u_nondim_eq}-\eqref{w_nondim_eq}. For simplicity of the presentation, we will assume $x \in \mathbb{R}$, so that we can make use of eigenfunctions of the form $e^{ikx}$ with $k^2$ an eigenvalue of the negative Laplacian (see \citep{krause2021modern} for generalizations to complex geometries and an argument why such analysis will give equivalent results in the limit of sufficiently large domains). We then linearize Equations \eqref{u_nondim_eq}-\eqref{w_nondim_eq} using $(u,v,w) = (u_0,v_0,w_0) + \varepsilon (u_1,v_1,w_1)e^{\lambda_k t + ikx}$ with $|\varepsilon| \ll 1$ to find that $\lambda_k$ is an eigenvalue of
\begin{equation}
\bm{M_k} = \begin{pmatrix} J_{11} - \delta_u k^2 & J_{12} & J_{13} \\ J_{21} & J_{22} - k^2  & J_{23} \\ J_{31} & J_{32} & J_{33} - \delta_wk^2 
    \end{pmatrix},
\end{equation}
with the entries given by Equation \eqref{homog_jacobian}. Hence we have that the growth rates $\lambda_k$ satisfy the characteristic equation,
\begin{equation}
    \lambda_k^3 + a_2 \lambda_k^2 + a_1\lambda_k + a_0 = 0,
\end{equation}
where,
\begin{equation}
    a_2 = (\delta_u++\delta_w)k^2-(J_{11}+J_{22}+J_{33}),
\end{equation}
\begin{multline}
    a_1 = (\delta_u+\delta_u\delta_w+\delta_w)k^4-(\delta_u(J_{22}+J_{33})+(J_{11}+J_{33})+\delta_w(J_{11}+J_{22}))k^2\\+J_{11}J_{22}+J_{11}J_{33}+J_{22}J_{33}-J_{12}J_{21}-J_{13}J_{31}-J_{23}J_{32},
\end{multline}
\begin{equation}
    a_0 = Ak^6+Bk^4+Ck^2+D,
\end{equation}
with
\begin{align}
    A &= \delta_u\delta_w, \\ B&=-(\delta_uJ_{33}+\delta_u\delta_wJ_{22}+\delta_wJ_{11})\\
    C &= (\delta_uJ_{22}J_{33}+J_{11}J_{33}+\delta_wJ_{11}J_{22}-\delta_uJ_{23}J_{32}-J_{13}J_{31}-\delta_wJ_{12}J_{21})\\
    D &= -J_{11}J_{22}J_{33}+J_{11}J_{23}J_{32}+J_{12}J_{21}J_{33}-J_{12}J_{23}J_{31}-J_{13}J_{21}J_{32}+J_{13}J_{22}J_{31}.
\end{align}

While this system can in general admit both spatial and spatiotemporal pattern-forming instabilities, we will primarily focus on Turing instabilities where the positive growth rates $\lambda_k$ remain purely real. The Routh-Hurwitz criterion for this system states that we have stability if $a_2>0$, $a_1>0$, $a_0>0$, and $a_2a_1>a_0$. The first of these is always satisfied as long as the homogeneous equilibrium is stable (as that has a necessary condition that $J_{11}+J_{22}+J_{33}<0$). We can discount two of the others as the roots $\lambda_k$ become purely complex when $a_2a_1=a_0$, and hence bifurcations of $a_1$ changing sign, or this inequality being violated, will be of the form of wave instabilities. Looking then for bifurcations where $\lambda_k$ is purely real, we therefore need $a_0<0$ for some range of wavenumbers $k^2>0$. Note that, when $k = 0$, $a_0 = D > 0$ by assumption, and as $A>0$ we have that $a_0 \to \infty$ as $k^2 \to \infty$, so to get a negative range of $a_0$ we require this cubic to have a negative minimum for positive $k^2$. To guarantee this, we can compute the extremal value of $k^2$ for $a_0$, and check if this is negative. This gives us the Turing conditions for instability:
\begin{equation}\label{turing_con}
    a_0 < 0 \quad \text{at} \quad k^2 = k_c^2 = \frac{-2B + \sqrt{4B^2-12AC}}{6A}.
\end{equation}
Hence, assuming stability in the absence of diffusion (i.e~the eigenvalues of $\bm{J}$ given by \eqref{homog_jacobian} all have negative real part), then the conditions \eqref{turing_con} ensure a Turing-type instability for sufficiently large domains \citep{murray2004mathematical}.


For the cancer-free state, any nontrivial spatial perturbations of $v$ will lead to negative cancer densities. Hence, we can consider just the $(u,w)$ subsystem for Turing instability, which has the reduced Jacobian,
\begin{equation}
    \bm{J_{\textrm{CF}}} = 
    \begin{pmatrix}
 \frac{\rho_{u}w}{w+1}-\mu_{u}& \frac{\rho_{u}u}{w+1}-\frac{\rho_{u}uw}{{\left(w+1\right)}^2}\\  0 & -\mu_{w} 
\end{pmatrix},
\end{equation}
which we see cannot admit Turing instabilities due to the 0 element \citep{murray2004mathematical}. Alternatively, the eigenvalues of the full Jacobian given by \eqref{homog_jacobian} are given by the diagonal elements which must all be negative if the cancer-free state is stable, and hence there can be no principal unstable submatrix of the Jacobian. This then precludes Turing instabilities \citep{satnoianu2000turing, villar2023general}, which can then only occur for a coexistence equilibrium.

\begin{figure}
    \centering
    \subfigure[$\alpha=0.01$]{\includegraphics[width=0.85\linewidth]{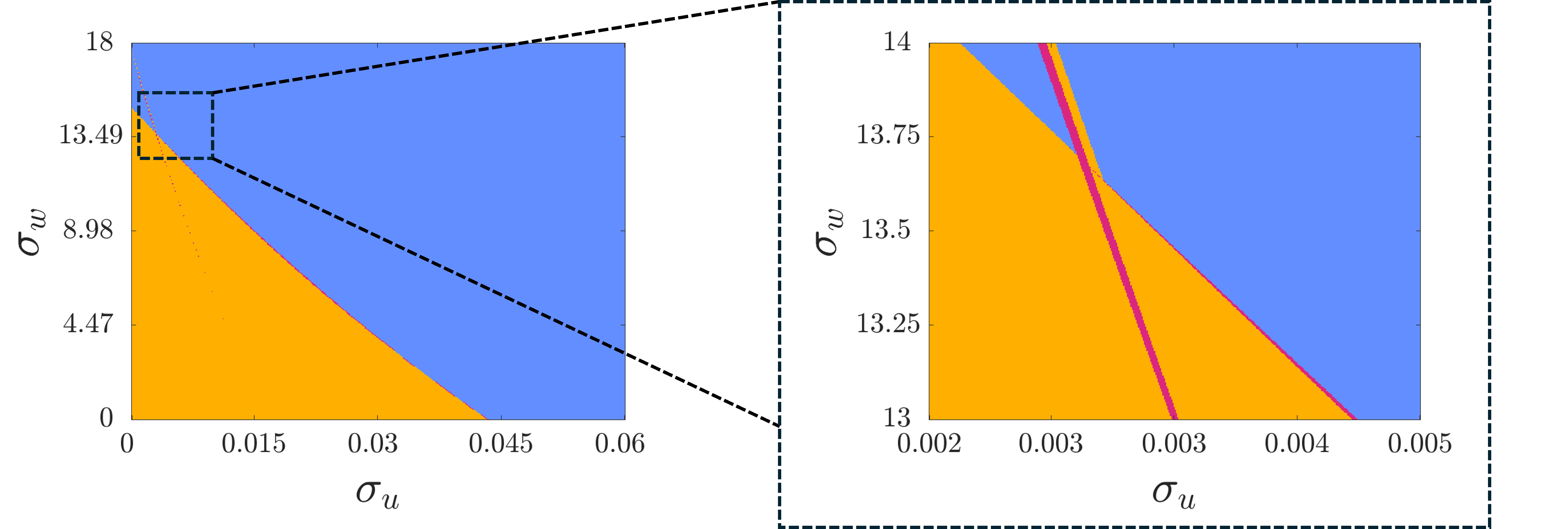}}
    \subfigure[$\alpha=0.07$]{\includegraphics[width=0.85\linewidth]{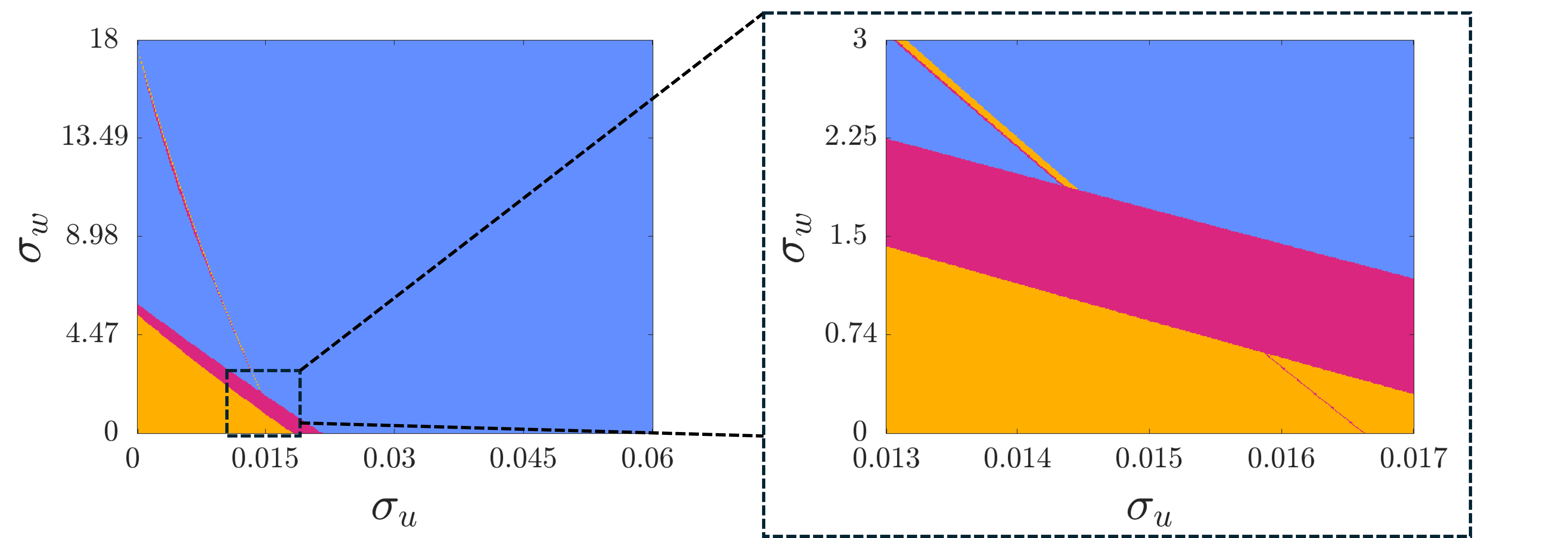}} 
    \subfigure[$\alpha=0.1$]{\includegraphics[width=0.45\linewidth]{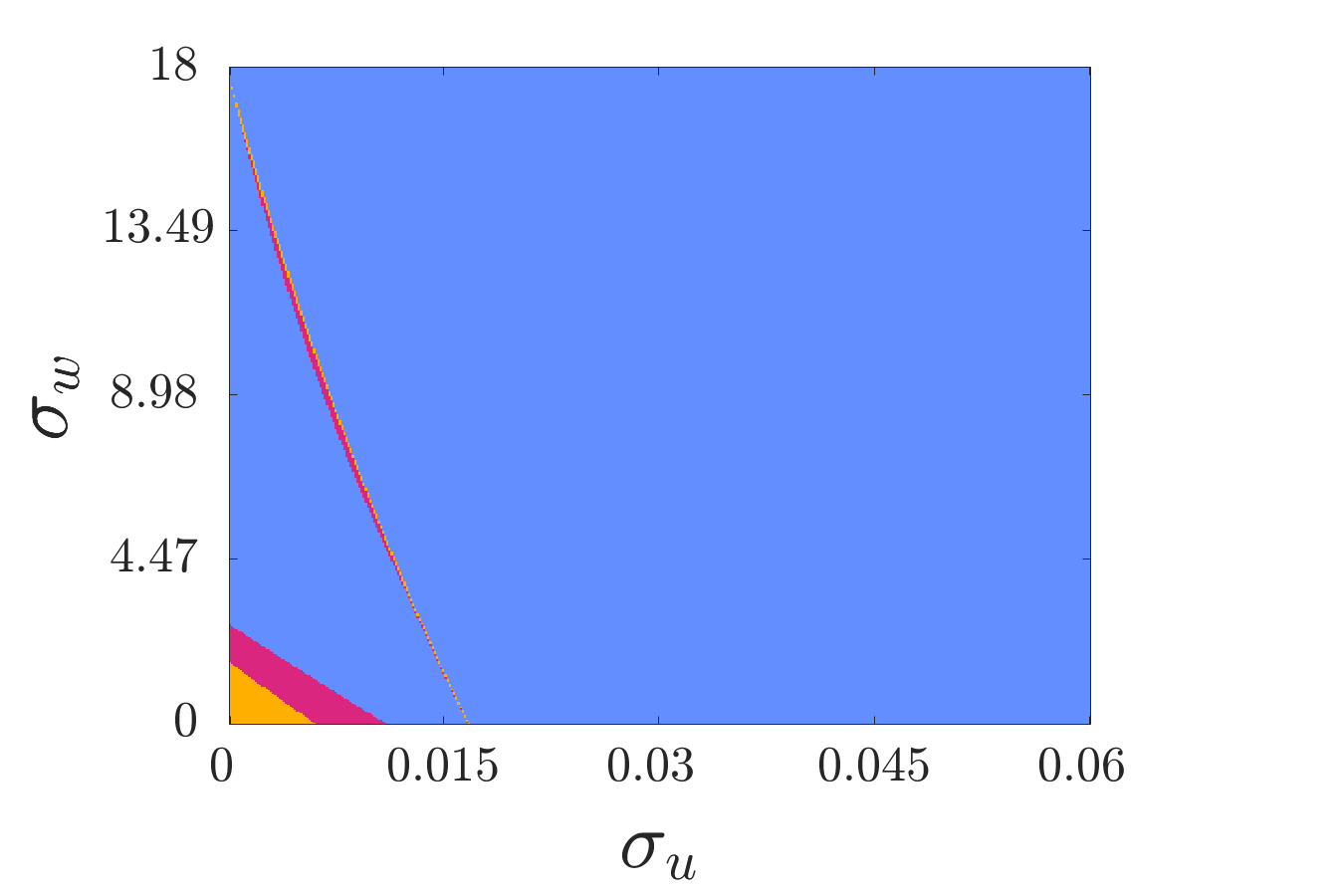}}
    \subfigure[$\alpha=0.2$]{\includegraphics[width=0.45\linewidth]{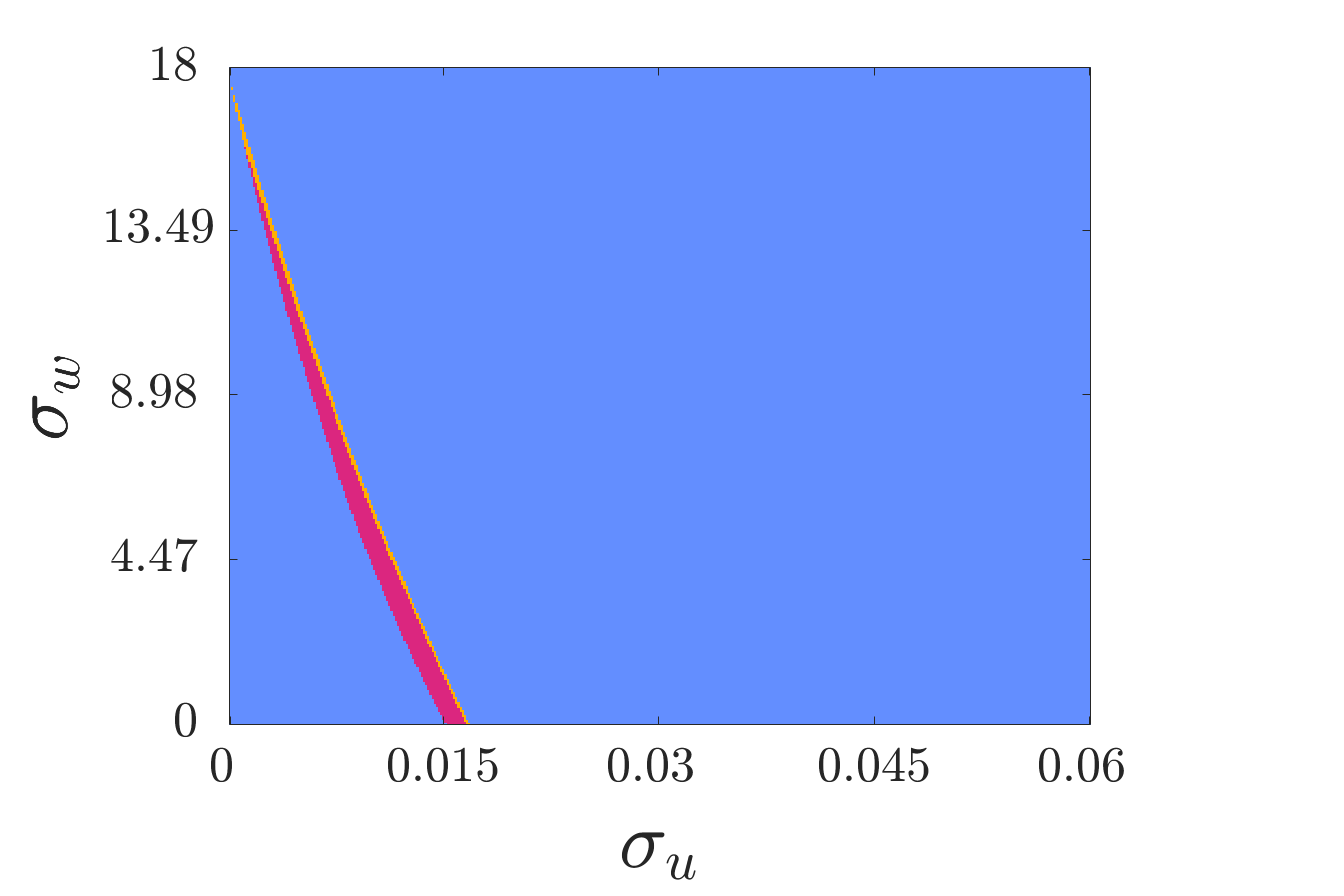}}
    \vspace{0.2cm}
    \includegraphics[width = 0.3\linewidth]{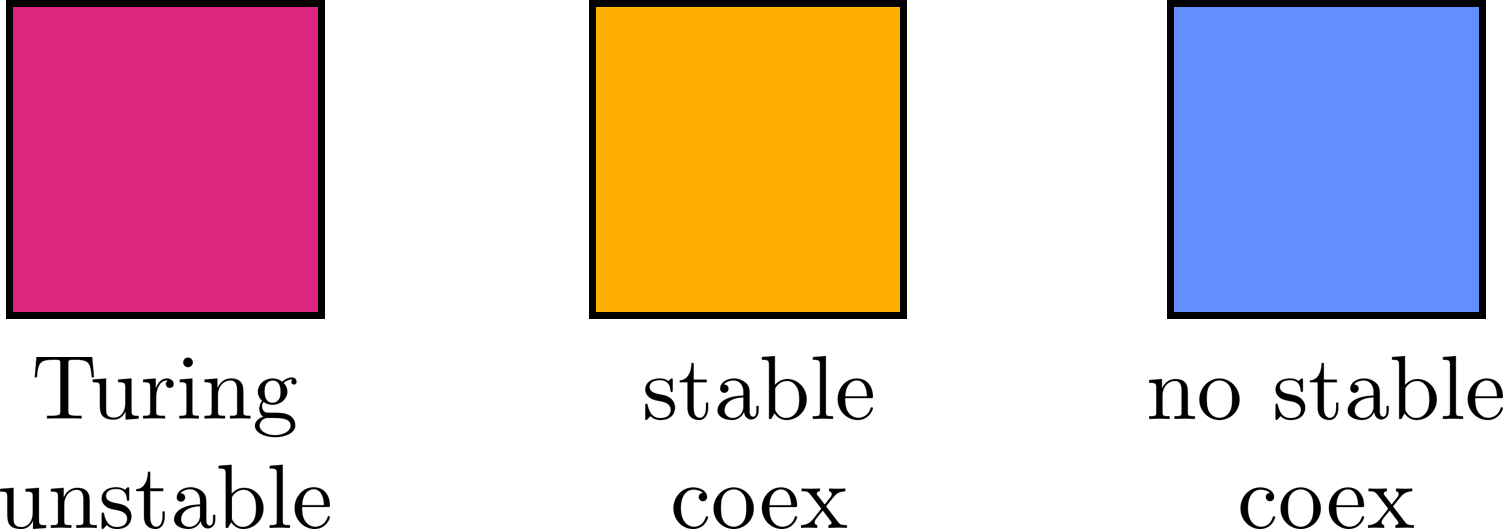}
    \vspace{0.1cm}
    \caption{Classification of parameters where Turing instabilities are possible computed via \eqref{turing_con} for Equations \eqref{u_nondim_eq}-\eqref{w_nondim_eq} across varying $\sigma_u$, $\sigma_v$ and $\alpha$. All parameters are given in Table \ref{parameter_table} with $\delta_u =100$ and $\delta_w =100$.}
    \label{fig:turing_plots}
\end{figure}

We use the conditions given in \eqref{turing_con} to determine parameters that exhibit Turing instabilities (the `Turing space') for coexistence states, and show these in Figure \ref{fig:turing_plots}. These plots show varying levels of $\alpha$, exactly as in Figure \ref{fig:lin_stab_plots}, but where we have simplified the regions to ignore multistability and cancer-free regions. That is, we only plot parameters where a coexistence equilibrium is either Turing unstable, stable to all perturbations, or does not exist at all. Broadly we find that Turing instabilities occur along the boundary of regions where coexistence equilibria are stable, and that the size of Turing unstable regions increases with $\alpha$ for small values, and then varies in a more complicated way with moderate values of $\alpha$, as we saw with the change in the size of coexistence regions in Figure \ref{fig:lin_stab_plots}. We also produced diagrams for varying levels of $\delta_u$ and $\delta_w$, generally finding that $\delta_u$ was key in determining the size of the Turing space, whereas varying $\delta_w$ had no visible effect for all values of $\delta_u$ we considered.

The \emph{criticality} of a Turing or Hopf bifurcation indicates if the emergent limit cycle or periodic pattern, respectively, is stable (supercritical) or unstable (subcritical) after the bifurcation curve is crossed. Often, subcritical Turing instabilities are associated with bistable regions in parameter space with stable homogeneous and patterned states \citep{champneys2021bistability}, though see \citep{krause2024turing} for a collection of counter-examples. We used the code given in \citep{villar2023computation} to show that all Turing bifurcations (the boundaries between the stable coexistence and Turing unstable regions in Figure \ref{fig:turing_plots}) were subcritical in one spatial dimension, as we saw for all Hopf bifurcations in the system. Similarly, we used the code in \citep{villar2024amplitude} to show that there are no wave instabilities for this system in these parameter regimes. While these results do not preclude wave, supercritical Hopf, or supercritical Turing bifurcations for other parameter regimes (or in higher dimensions, where the criticality of Turing bifurcations can depend on geometric details), they give some insight on the structure of bifurcations in the region of the parameter space explored here. 

\subsection{Numerical Continuation}\label{sec_continuation}

\begin{figure}
     \centering
     \subfigure[]{\includegraphics[width=0.49\linewidth]{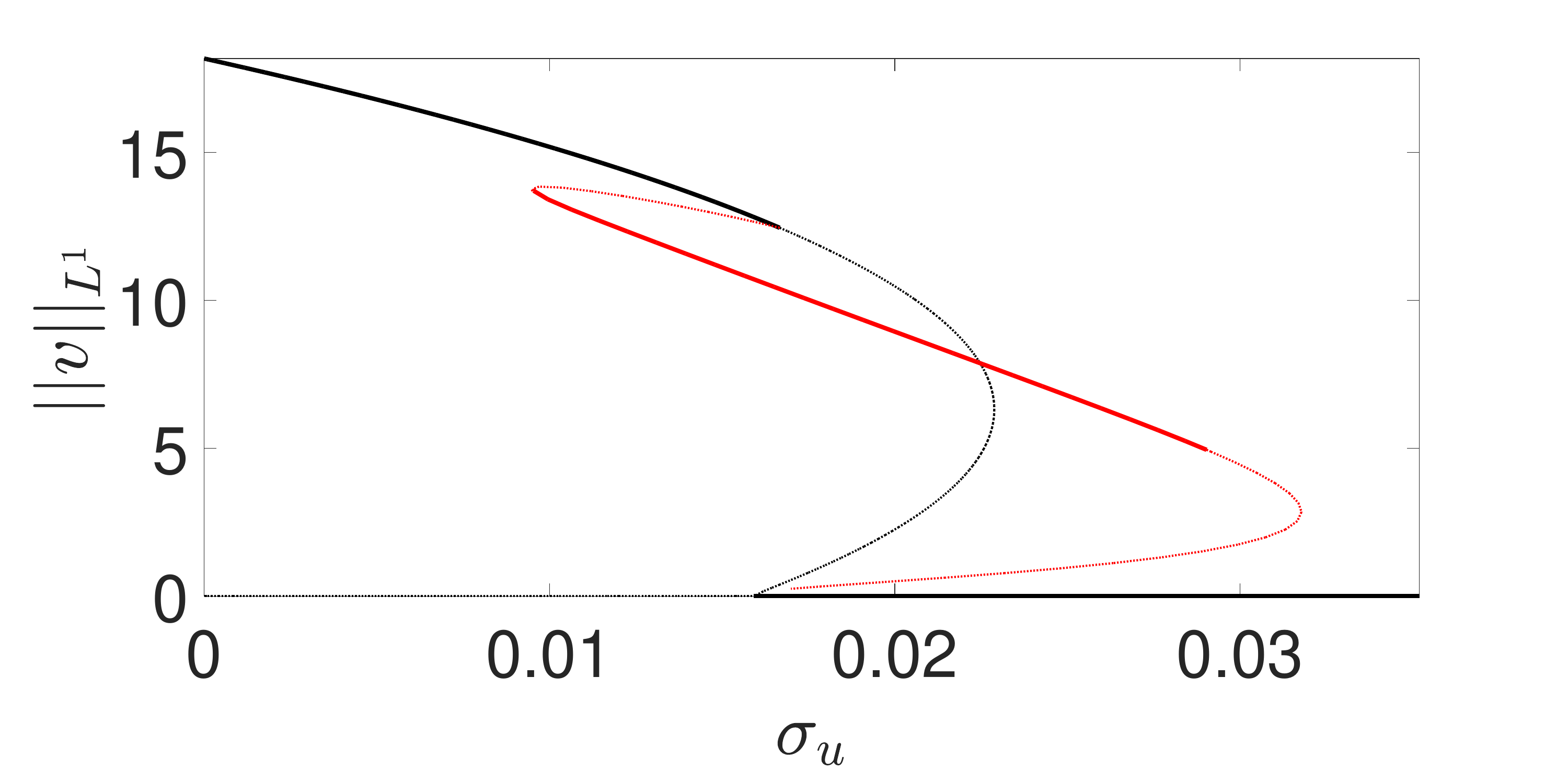}}
     \hfill
     \subfigure[]{\includegraphics[width=0.49\linewidth]{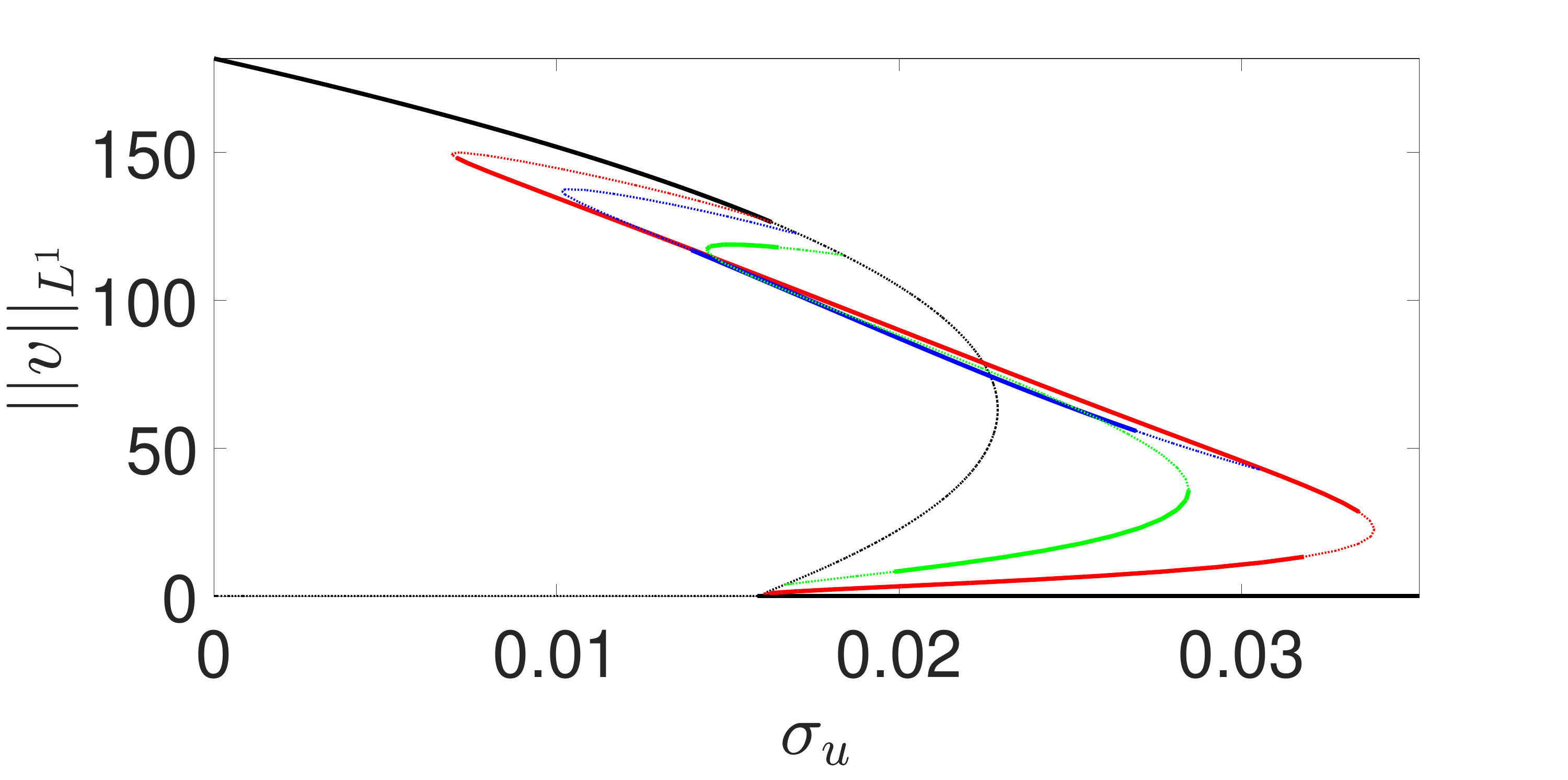}}
    \caption{Bifurcation diagram for model \eqref{u_nondim_eq}--\eqref{w_nondim_eq} showing steady-state values of $\lVert v\rVert_{L^1}$ for the same parameter values as in the simulation shown in Fig \ref{fig:overview_figure}(b)-(c) except treating $\sigma_u$ as a constant bifurcation parameter. Panel (a) is on a domain of size $L = 30$, and panel (b) is on a domain of size $L=300$. Thick solid curves are linearly stable steady states, and dotted curves are unstable. The black curve corresponds to a homogeneous equilibrium (i.e.~a steady state of \eqref{u_hom_eq}--\eqref{w_hom_eq}), whereas the coloured curves represent patterned solution branches.}
    \label{fig:bifurcation_sigma_u}
\end{figure}

Lastly, in order to connect the stability analysis of equilibria with the time-dependent simulation shown in Figure \ref{fig:overview_figure} and in the next section, we plot a bifurcation diagram focusing on the stable steady states found for Equations \eqref{u_nondim_eq}--\eqref{w_nondim_eq} using the numerical continuation software \texttt{pde2path} \citep{uecker2014pde2path, uecker2019pattern}, focusing for simplicity on a one-dimensional spatial domain. We plot the equilibria and their stability in Figure \ref{fig:bifurcation_sigma_u} corresponding to Figure \ref{fig:overview_figure}(b)-(c). We use two domain lengths to demonstrate how the number of inhomogeneous branches (and hence the complexity of such diagrams) grows rapidly as the domain size increases. The black curves in these plots correspond to spatially homogeneous equilibria, with solid curves indicating stable branches and dotted curves indicating unstable solutions. The coloured branches correspond to spatially inhomogeneous patterned branches. Panel (a) shows a small domain of size $L=30$, where only a single patterned branch bifurcates from the homogeneous coexistence state in a subcritical Turing instability, as predicted. There are 8 other (unstable) patterned branches not shown in this panel, and likely further branches of unknown stability after secondary bifurcations. For the larger domain in panel (b), with a domain size of $L=300$ corresponding to panels (b) and (c) of Figure \ref{fig:overview_figure}, we instead see four patterned branches with some stable region in the parameter space. In this larger domain, there are an additional 22 unstable patterned branches that we do not plot, as they are all unstable (again neglecting subsequent secondary bifurcations).

We carried out a numerical continuation of the spatially homogeneous model in this parameter regime using AUTO \citep{doedel1998auto97}, finding that it exhibits a subcritical Hopf bifurcation at $\sigma_u \approx 0.019608174019$, and this leads to unstable transient oscillations, eventually falling into the cancer-free state as shown in Figure \ref{fig:overview_figure}. Importantly, the Turing bifurcation occurs before this Hopf point, leading to the resilience of the system in the presence of spatial instabilities for a much larger range of treatments than the spatially homogeneous model would predict.

\begin{figure}
     \centering
     \subfigure[]{\includegraphics[width=0.49\linewidth]{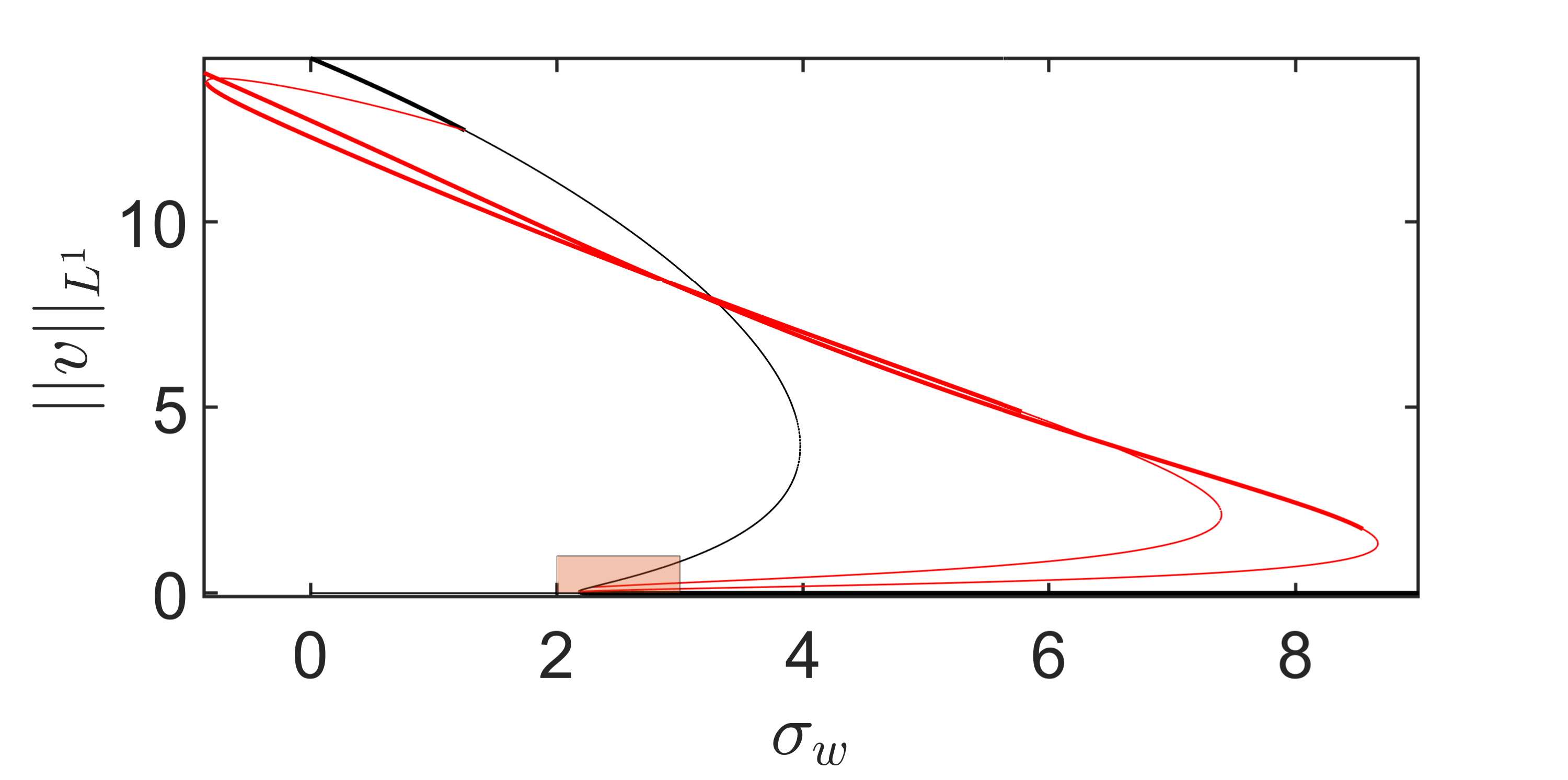}}
     \hfill
     \subfigure[]{\includegraphics[width=0.49\linewidth]{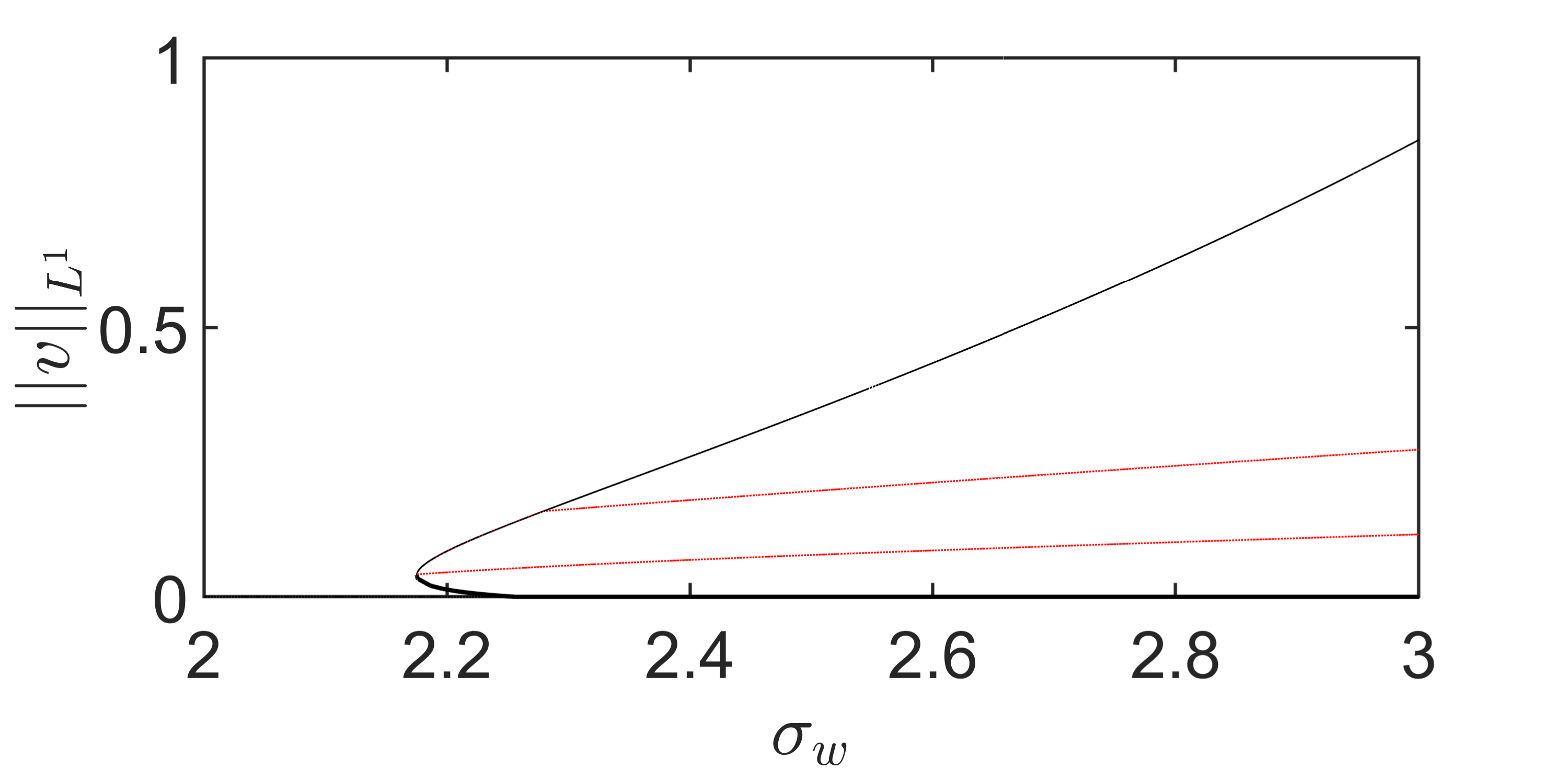}}
    \caption{Bifurcation diagram for model \eqref{u_nondim_eq}--\eqref{w_nondim_eq} showing steady state values of $\lVert v\rVert_{L^1}$ for the same parameter values as in the simulation shown in Fig \ref{fig:overview_figure}(e)-(f) except treating $\sigma_w$ as a constant bifurcation parameter and using a domain of size $L=30$. Panel (b) is a zoomed version showing the complexity of branches near the cancer-free equilibrium for $\sigma_w \in [2, 3]$ (the red boxed region in panel (a)). Thick solid curves are linearly stable steady states, and dotted curves are unstable. The black curves correspond to homogeneous equilibria (i.e.~steady states of \eqref{u_hom_eq}--\eqref{w_hom_eq}), whereas the red curves represent patterned solution branches.}
    \label{fig:bifurcation_sigma_w}
\end{figure}

We also carried out numerical continuation by varying $\sigma_w$ in Figure \ref{fig:bifurcation_sigma_w}, finding a similar mechanism but with additional complexity. Here we fix $L=30$ due to the number of branches, and we remark that we have to continue $\sigma_w$ to unphysical negative values in order to resolve the fold point which occurs for $\sigma_w = - 0.84536$. We remark that there are at least two stable patterned solutions which exist for $\sigma_w=0$, alongside the stable homogeneous solution. The second of these is not due to the primary Turing bifurcation of the first homogeneous equilibrium, but actually arises from the fold of a subcritical Turing bifurcation of another homogeneous equilibrium, which is shown in more detail in panel (b). This second homogeneous equilibrium can also be seen in panel (b) of Figure \ref{fig:turing_plots}, where a stable cancer equilibrium is stable for only a small range of $\sigma_w$, and has a small region of Turing instability for smaller values of $\sigma_w$, whereas it undergoes a transcritical bifurcation with the cancer-free state for larger $\sigma_w$. Overall, we observe an intricate interplay between the two Turing instabilities of these two homogeneous equilibria, some of which connect and others which apparently do not (at least at the level of the primary bifurcation curves). Importantly, here the range of stable patterned branches is vastly larger than the range of stable homogeneous cancer equilibria, suggesting a strong role of this resilience mechanism in the presence of treatment via IL-2 compounds, which we explore numerically in the next section.

\section{Simulations} \label{sec_sim}

We simulated the ODE model \eqref{u_hom_eq}-\eqref{w_hom_eq} and the PDE model \eqref{u_nondim_eq}-\eqref{w_nondim_eq} in a number of different scenarios using a set of MATLAB codes that can be found at \citep{Molly_Github}, as well as the interactive web-based tool VisualPDE \citep{walker2023visualpde}. Both implementations used a standard method-of-lines discretization with the centred three-point (1D) or nine-point (2D) stencil for the Laplacian. The MATLAB code used the implicit integrator, \textsc{ode15s}, which implements a BDF15 method with variable time-stepping. Default absolute and relative tolerances were set to $10^{-11}$, with $m=10^4$ discrete nodes in 1D, and $m=100$ nodes in each direction in 2D square domains. Additional convergence checks were carried out in space, though broadly even rather crude numerical methods (e.g.~simulations in VisualPDE) gave qualitatively similar behaviour. Typical simulations started on one of the homogeneous cancer states which was stable in the absence of diffusion. We took random perturbations of this steady state for the initial data. For example, for $u$ we set an initial condition of the form $u_0(1+\eta\xi(\bm{x}))$, where $\xi$ was an independently and identically distributed Normal random variable with zero mean and unit variance, and $\eta$ then scaled the standard deviation (typically with the value $\eta=0.1$). Several interactive and editable simulations of the model can be found at \url{https://visualpde.com/mathematical-biology/immunotherapy-model}, which loosely correspond to the simulations shown below. We strongly encourage a reader to view and interact with these in order to rapidly explore the behaviours illustrated below in a variety of settings.

\subsection{One-Dimensional Simulations}\label{sec_1d_sims}

\begin{figure}
    \centering
    \subfigure[]{\includegraphics[width=0.3\linewidth]{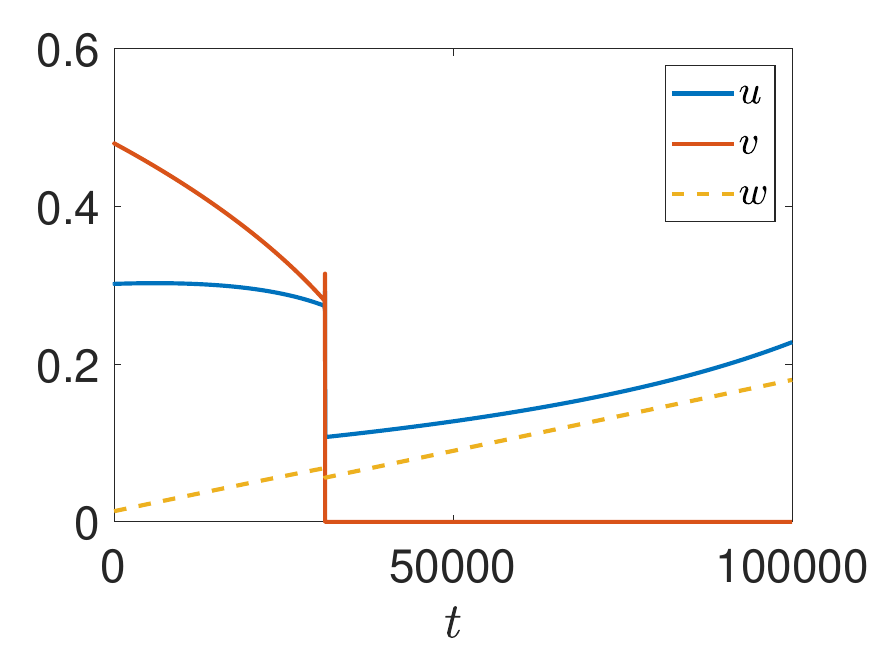}}
    \subfigure[]{\includegraphics[width =0.3\linewidth]{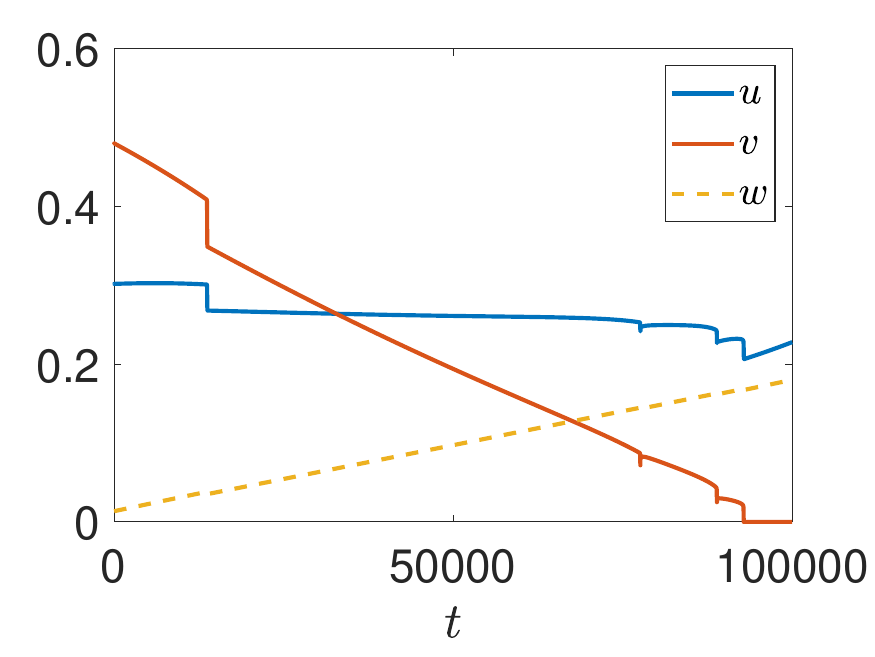}}
    \subfigure[]{\includegraphics[width = 0.3\linewidth]{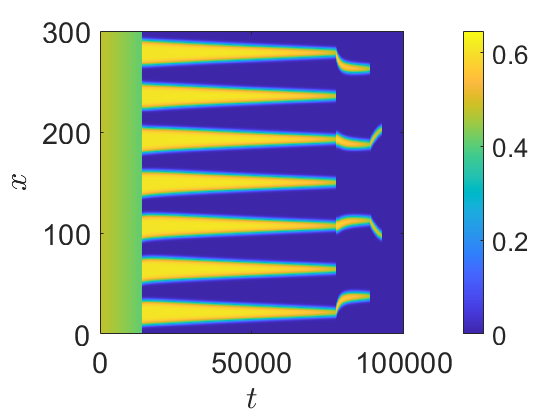}}
    \caption{{Simulations of the ODE (panel (a)) and PDE (panels (b)-(c)) models as in Figure \ref{fig:overview_figure} with  $\sigma_u=0.014$, and $\sigma_w = 10\frac{t}{T}$, with the final simulation time given by $T = 10^5$ in all cases.}}
    \label{fig:overview_figure_sigma_w}
\end{figure}

{We first demonstrate that the qualitative impact of increasing treatment shown in Figure \ref{fig:overview_figure} can also occur due to increasing treatment via IL-2. We use a quasi-static increasing function given by $\sigma_w = 10\frac{t}{T}$ for a large time $T=10^5$. We show an example of a comparison between ODE and PDE simulations in Figure \ref{fig:overview_figure_sigma_w}, finding qualitatively comparable behaviour in terms of tumour resilience due to spatial restructuring of the tumour cell density to the case of effector-cell treatment shown in Figure \ref{fig:overview_figure}.}

\begin{figure}
    \centering
    \subfigure[]{\includegraphics[width=0.3\linewidth]{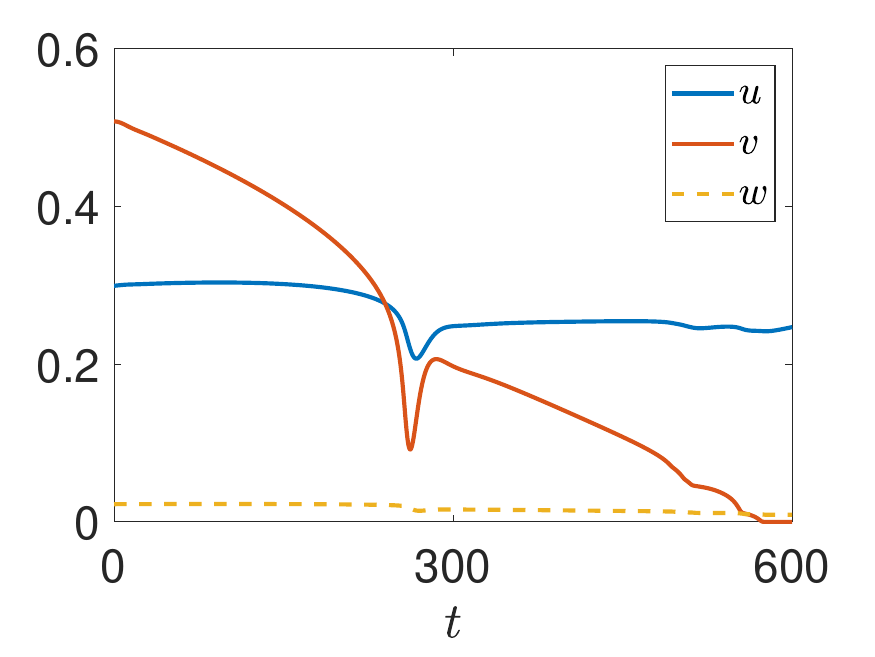}}
    \subfigure[]{\includegraphics[width=0.3\linewidth]{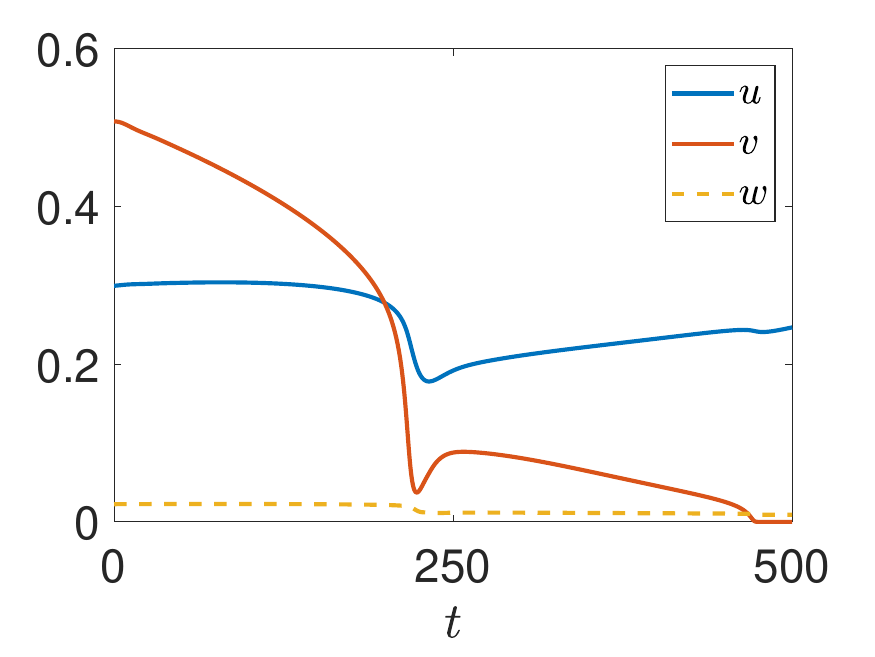}}
    \subfigure[]{\includegraphics[width=0.3\linewidth]{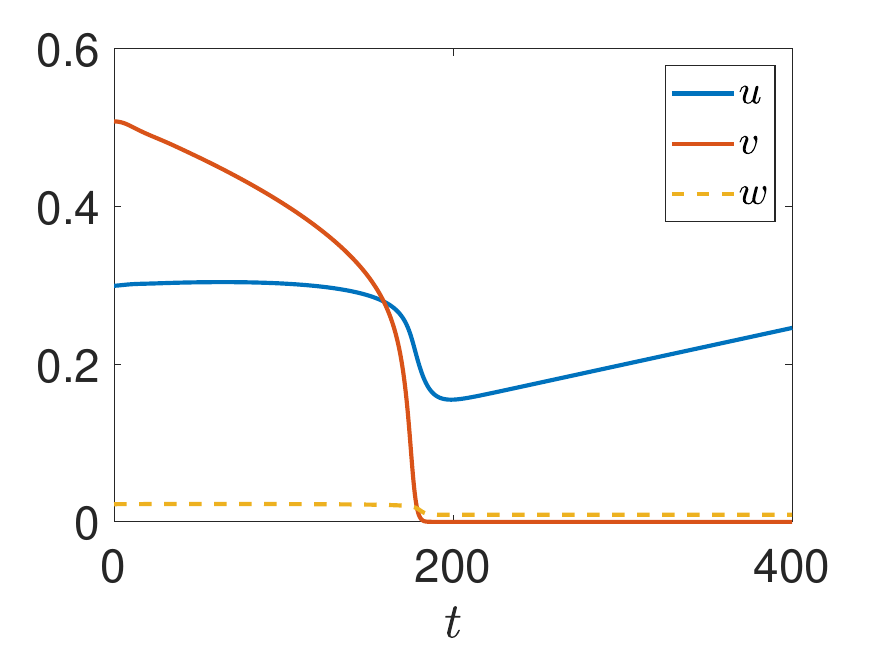}}
    
    \subfigure[]{\includegraphics[width=0.3\linewidth]{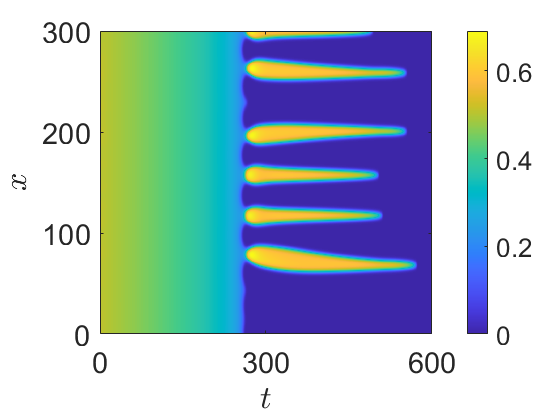}}
    \subfigure[]{\includegraphics[width=0.3\linewidth]{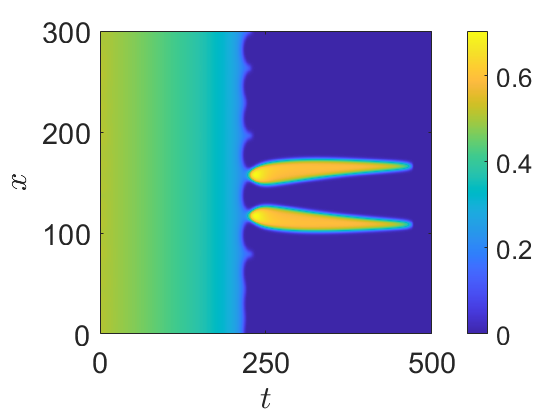}}
    \subfigure[]{\includegraphics[width=0.3\linewidth]{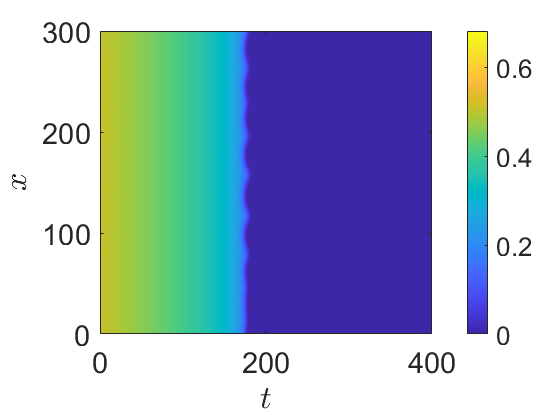}}
    \caption{Simulations of the PDE model \eqref{u_nondim_eq}-\eqref{w_nondim_eq} as in Figure \ref{fig:overview_figure} with a different final simulation time of $T=600$ in (a), (d),  $T=500$ in (b), (e), and $T=400$ in (c), (f). Spatial averages are given in (a)-(c), whereas kymographs of $v$ are shown in (d)-(f). Recall that the varying treatment parameter is given as $\sigma_u = 0.01 + 0.03t/T$, so that the treatment rate increases along each row.}
    \label{fig:treatment_rate_figure}
\end{figure}

Realistic treatment scenarios will be time-dependent, as in Figure \ref{fig:overview_figure}, albeit typically with more complexity than linear ramping. We first investigate the impact of the rate at which the bifurcation is passed in these somewhat simplified time-dependent simulations. In Figure \ref{fig:treatment_rate_figure} we simulate the same scenario of varying $\sigma_u$ linearly in time {(i.e.~$\sigma_u = 0.01 + 0.03t/T$)} as in Figure \ref{fig:overview_figure}(b)-(c), but vary the total simulation time{, $T$,} so that the treatment is applied more quickly as one moves across each row. We see that the nature of the Turing-induced resilience remains until the treatment parameter is varied sufficiently quickly, at which point the dynamics of the total cancer density behaves similarly between the ODE and PDE models (cf. Figures \ref{fig:overview_figure}(a) and \ref{fig:treatment_rate_figure}(b)). This could in principle be studied using weakly nonlinear approaches, as in the work by \cite{dalwadi2023universal}, albeit due to the interplay of multiple equilibria this goes well beyond the scope of investigation here. Nevertheless, these rate-induced bifurcations are an important caveat to the static analysis shown in Section \ref{sec_continuation}, and the quasi-static `slow' simulations given in Figure \ref{fig:overview_figure}.

\begin{figure}
    \centering
    \subfigure[]{\includegraphics[width=0.3\linewidth]{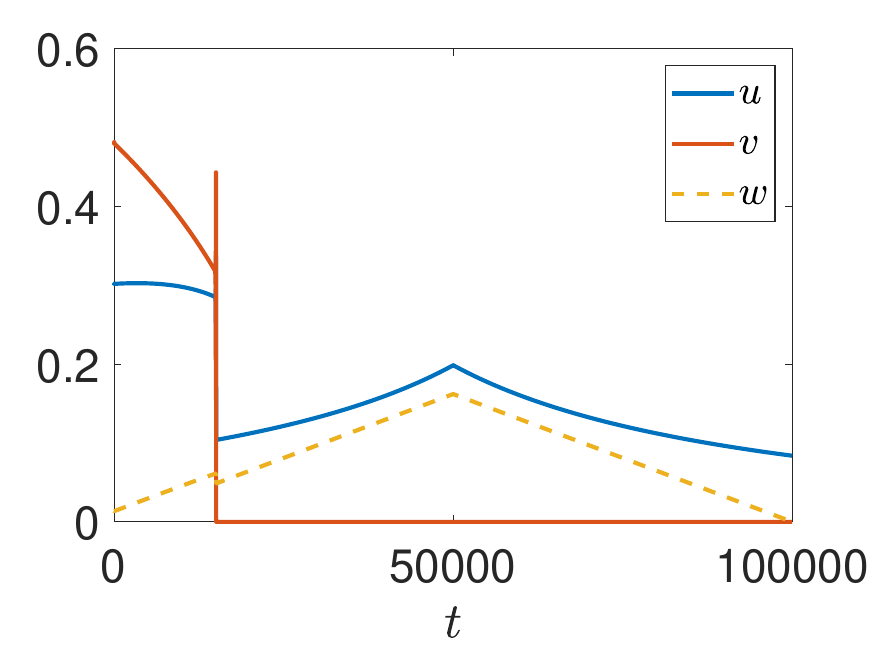}}
    \subfigure[]{\includegraphics[width=0.3\linewidth]{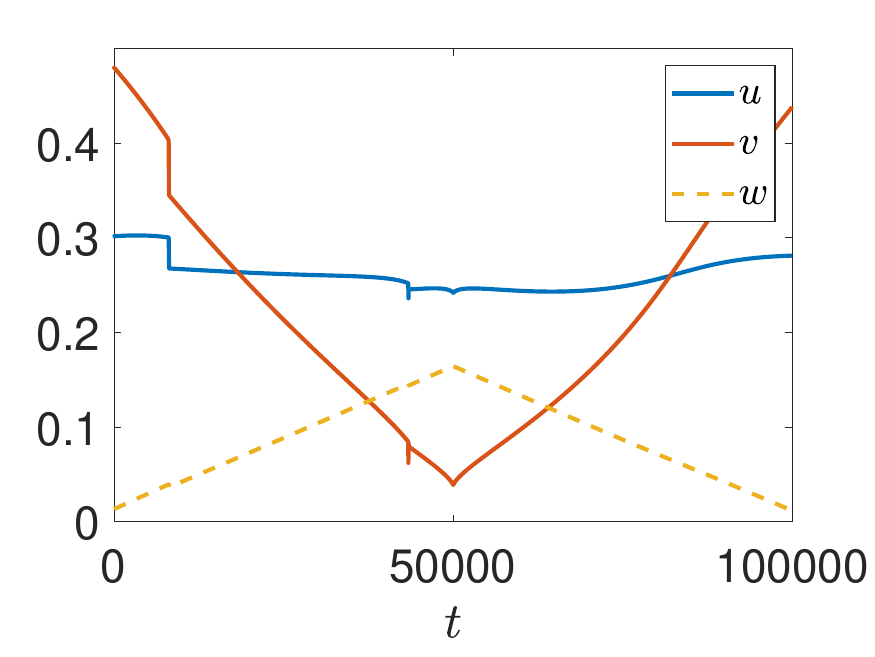}}
    \subfigure[]{\includegraphics[width=0.3\linewidth]{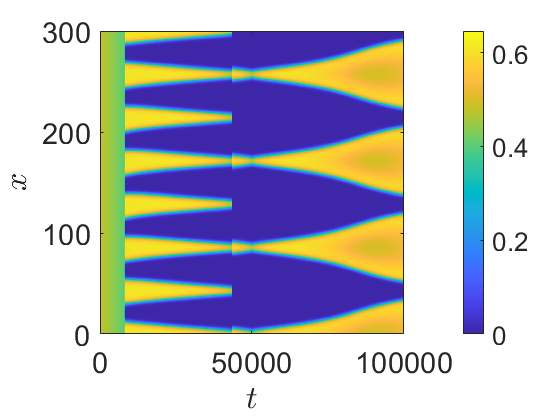}}
    \caption{Simulations of the ODE (panel (a)) and PDE (panels (b)-(c)) models as in Figure \ref{fig:overview_figure} with $\sigma_u=0.014$ and $\sigma_w = 18(\frac{t}{T}H\left(0.5-\frac{t}{T}\right)+(1-(\frac{t}{T}))H\left(\frac{t}{T}-0.5\right))$, where $H$ denotes the heaviside step-function. }
    \label{fig:reversing_treatment}
\end{figure}

The bifurcation diagram shown in Figure \ref{fig:bifurcation_sigma_w} has an emergent hysteresis loop whereby multiple branches are stable for $\sigma_w=0$. We now visualize how this hysteresis loop can emerge as a factor in time-dependent treatment, again using a slow variation in $\sigma_w$ as before. Namely, in Figure \ref{fig:reversing_treatment}, we show ODE and PDE simulations where we ramp up $\sigma_w$ linearly in time to a maximal value of $9$, and then decrease it back to $0$ at the reversed linear rate. As anticipated, the existence of stable patterned branches for a wide range of parameters leads to different patterned states emerging after the primary Turing bifurcation, without returning to a spatially homogeneous state even as $\sigma_w=0$ at the end of the simulation. 

We remark that in the ODE simulation shown in Figure \ref{fig:reversing_treatment}(a), the system can exhibit numerical instabilities due to the quasi-static change of stability of the cancer-free equilibrium. This equilibrium is unstable for $t=0$, becomes stable for intermediate times as the treatment is increased, and then becomes unstable again as $t$ approaches the final simulation time.  The cancer cell density, $v$, reaches extremely small but nonzero densities ($v<10^{-40}$) at the maximal treatment time $t=T/2$, where these low populations no longer are meaningfully represented by ODEs, and the behaviour of the system becomes sensitive to the numerical integration methods used. This is an example of the `Atto-fox problem' \citep{fowler2021atto}. Adding a term $-H(10^{-15}-v)v$ to the right-hand side of Equation \eqref{v_nondim_eq} prevents this behaviour, and the presence or absence of this term has no qualitative or even quantitative effect on the results in Figure \ref{fig:reversing_treatment}(b) or (c), and in general had no impact on other simulation results presented throughout the paper. Importantly, this gives a difference in the kind of spatial resilience shown in Figure \ref{fig:reversing_treatment} from the discussion of emergent limit cycles as a resilience mechanism discussed by \cite{kirschner_mathematical_2003,matzavinos2004mathematical} and others, which may be subject to these kinds of `Atto-fox' problems of the modelling framework. That is, oscillations in the cancer cell population which undergo periods of very low cell density may no longer be well-represented by continuous state models, and hence such oscillations may not be relevant for understanding cancer recurrence.

\subsection{Two-Dimensional Simulations}

We carried out simulations in circular and square geometries using both VisualPDE and MATLAB. We record a few snapshots here to give an idea of the behaviours observed, but again encourage the reader to explore these in more detail at the page \url{https://visualpde.com/mathematical-biology/immunotherapy-model}. 

\begin{figure}
    \centering
    \subfigure[$t=50$]{\includegraphics[width=0.27\linewidth]{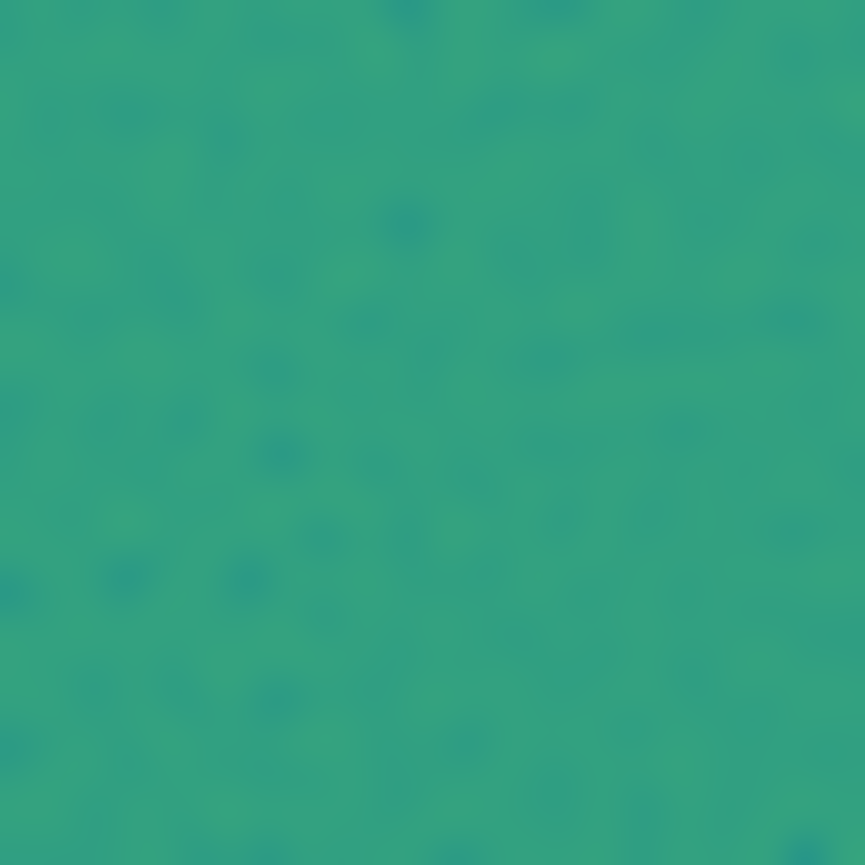}}
    \hspace{0.2cm}
    \subfigure[$t=100$]{\includegraphics[width=0.27\linewidth]{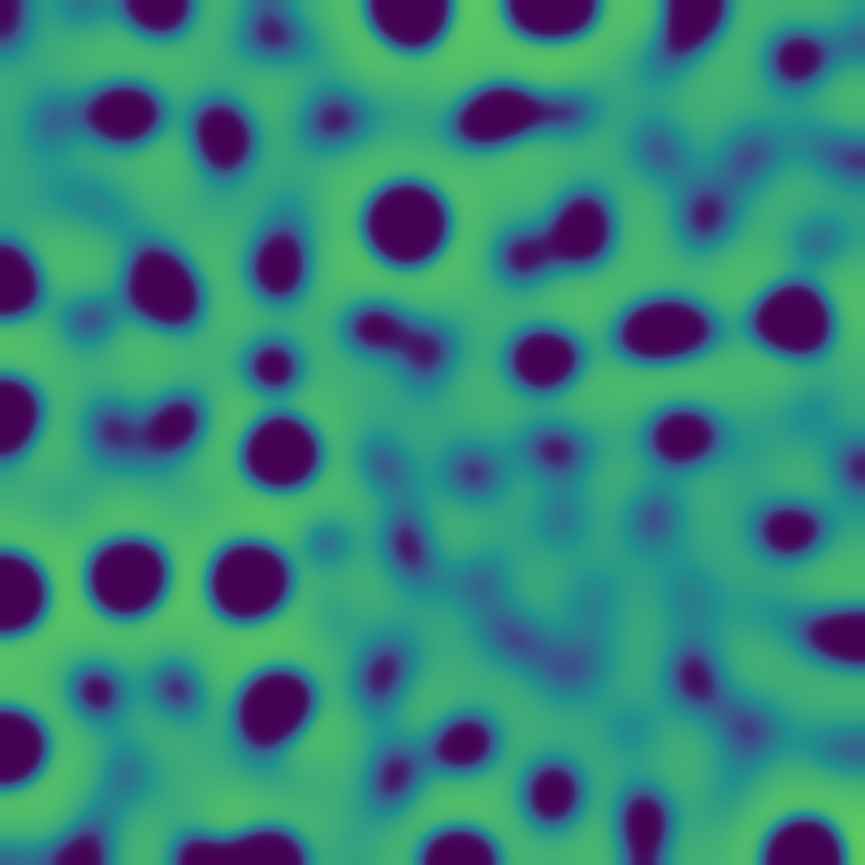}}
    \hspace{0.2cm}
    \subfigure[$t=150$]{\includegraphics[width=0.27\linewidth]{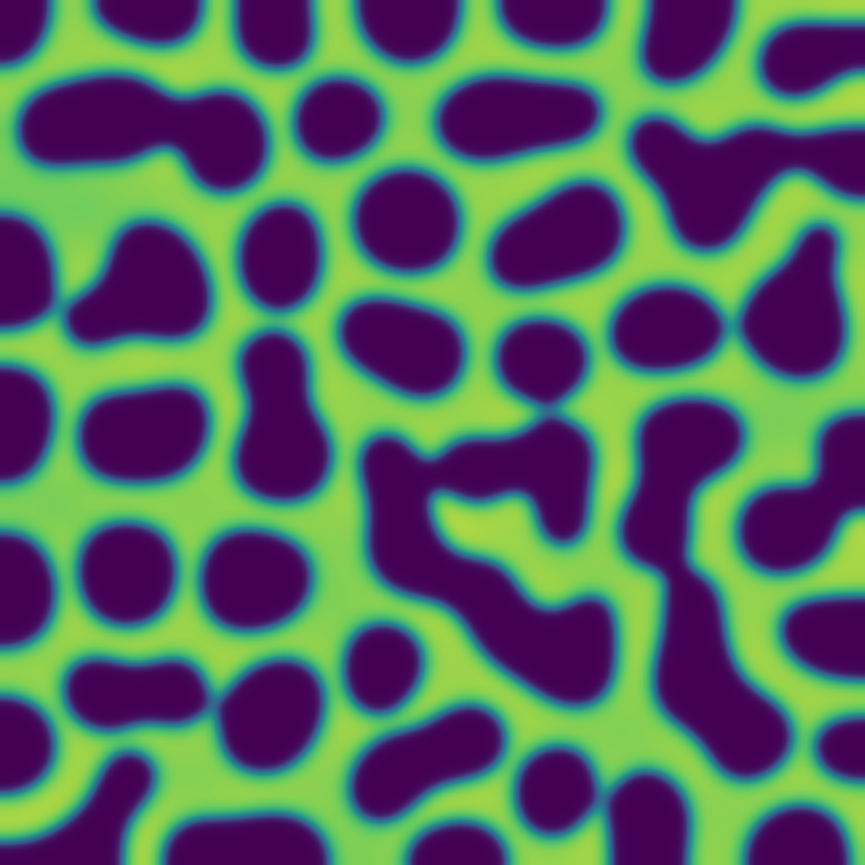}}
    \hspace{0.2cm}
    \raisebox{-0.1cm}{\subfigure{\includegraphics[width=0.05\linewidth]{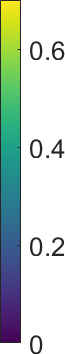}}}

    \subfigure[$t=300$]{\includegraphics[width=0.27\linewidth]{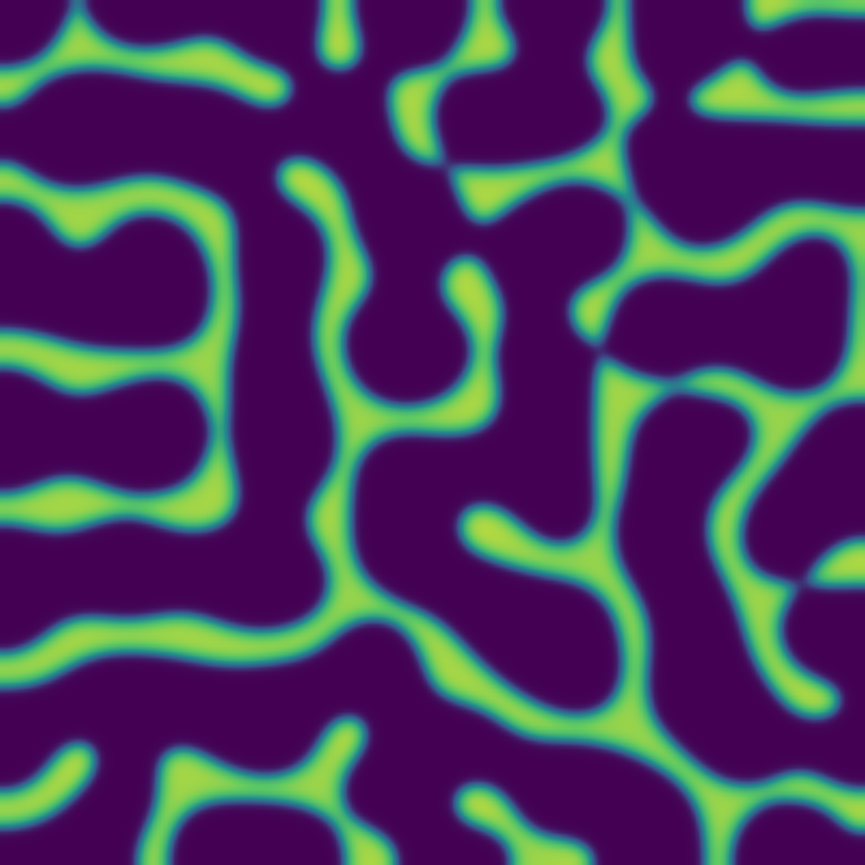}}
    \hspace{0.2cm}
    \subfigure[$t=400$]{\includegraphics[width=0.27\linewidth]{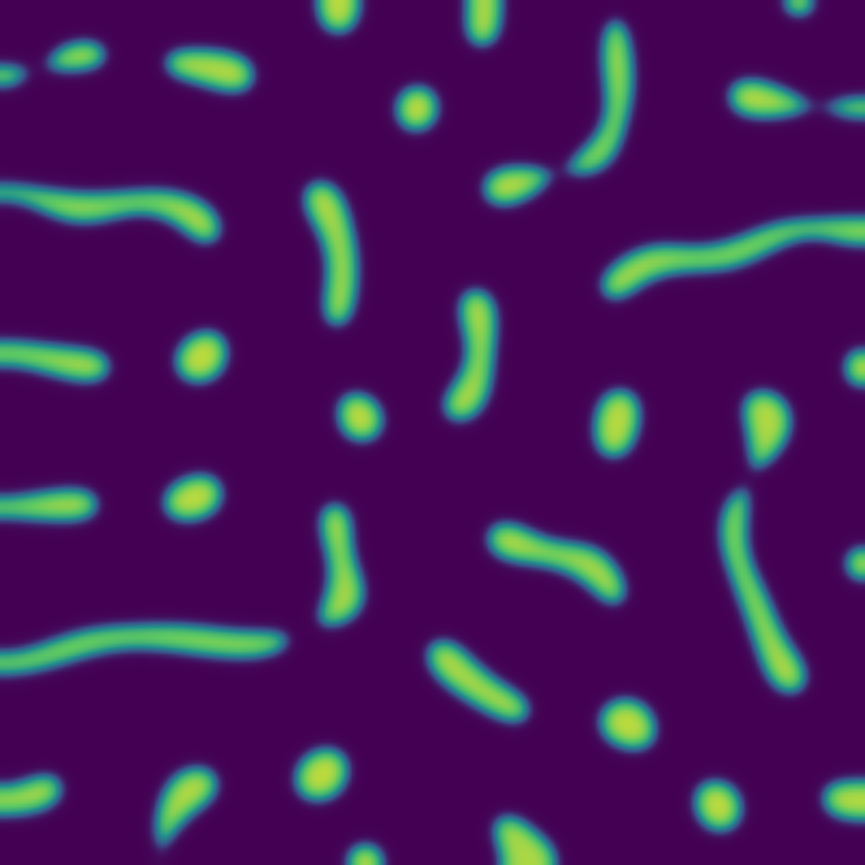}}
    \hspace{0.2cm}
    \subfigure[$t=500$]{\includegraphics[width=0.27\linewidth]{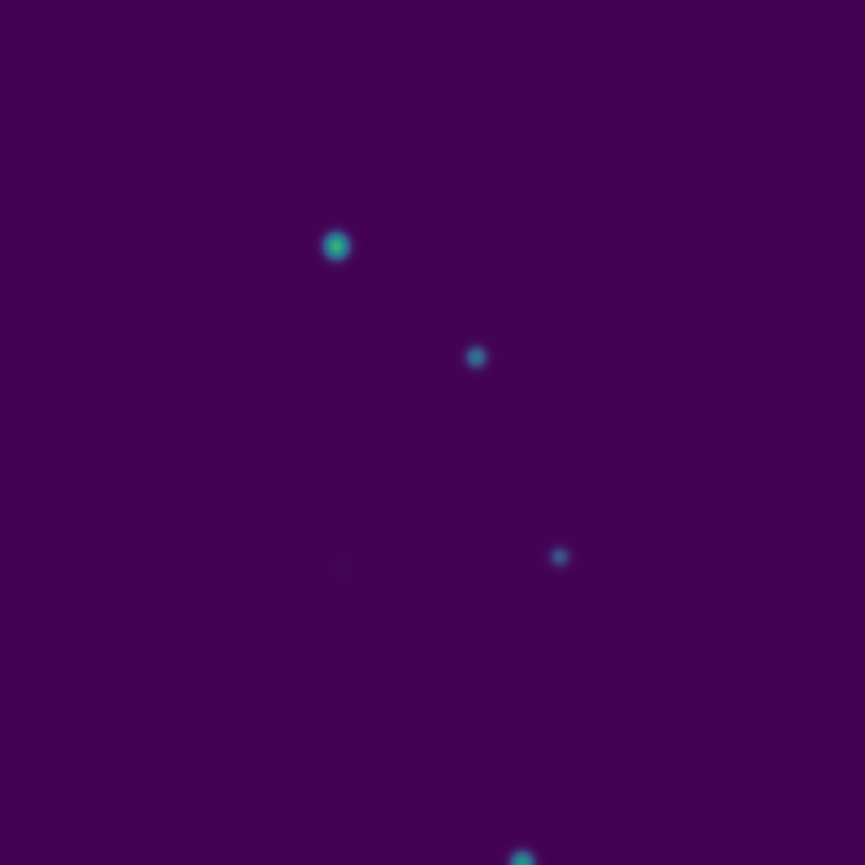}}
    \hspace{0.2cm}
    \raisebox{-0.1cm}{\subfigure{\includegraphics[width=0.05\linewidth]{figures/2D_Treatment_sigma_u_colorbar.png}}}
    \caption{Snapshots from simulations of Equations \eqref{u_nondim_eq}-\eqref{w_nondim_eq} in a square domain of side length $L=300$ {with homogeneous Neumann boundary conditions,} $\sigma_u=0.015 + 5\times 10^{-5}t$, $\sigma_w = 0.5$, $\alpha = 0.07$, $\delta_u=\delta_w=100$, and all other parameters as in Table \ref{parameter_table}. The cancer cell density $v$ is plotted, with $u$ and $w$ exhibiting in-phase patterned solutions in each panel.}
    \label{fig:2D_sigma_u_treatment}
\end{figure}

We start with an analogous simulation to Figure \ref{fig:overview_figure} in a square domain, shown in Figure \ref{fig:2D_sigma_u_treatment} where effector cell treatment ($\sigma_u$) is increased {linearly} over time. As in the one-dimensional case, the system undergoes a spatial patterning process as the treatment parameter $\sigma_u$ is increased. Simulations with no initial spatial perturbations (i.e.~$\eta = 0$) instead show $v$ uniformly in space approaching the cancer-free equilibrium approximately by $t=160$. We remark that while this simulation is on a shorter timescale than would merit a quasi-static understanding, if the treatment parameter is fixed at $t=400$ (at the level $\sigma_u=0.035$), then there is a small rearrangement of the patterns, but they persist for over $10^3$ units of time, suggesting that there are asymptotically stable patterned states that persist into high treatment regimes.

\begin{figure}
    \centering
    \subfigure[$t=50$]{\includegraphics[width=0.27\linewidth]{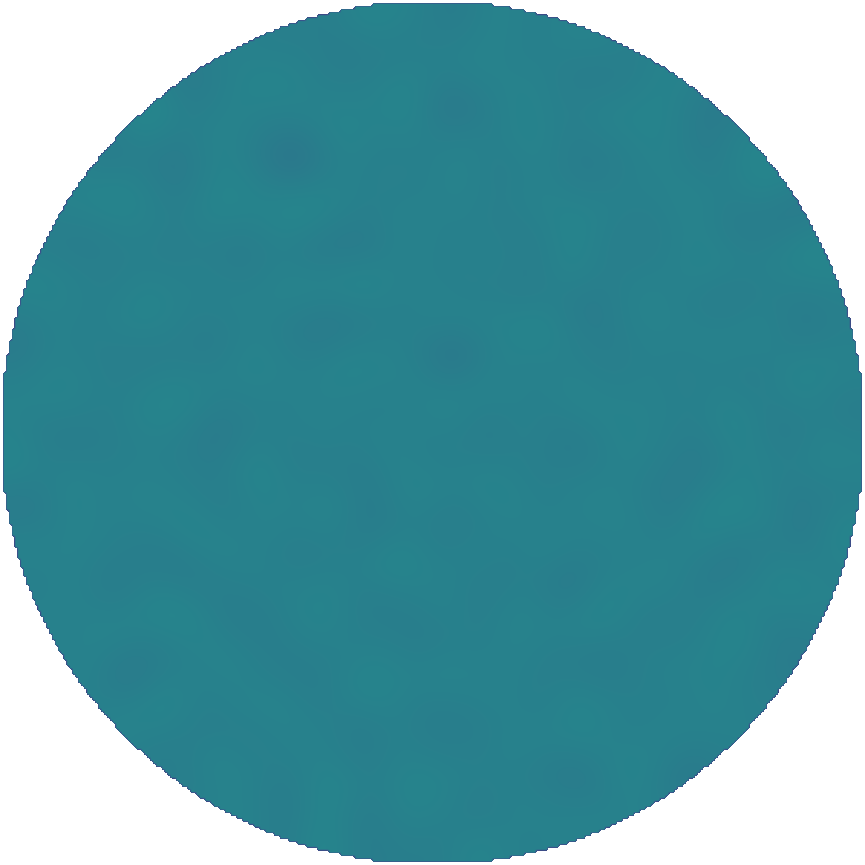}}
    \hspace{0.2cm}
    \subfigure[$t=75$]{\includegraphics[width=0.27\linewidth]{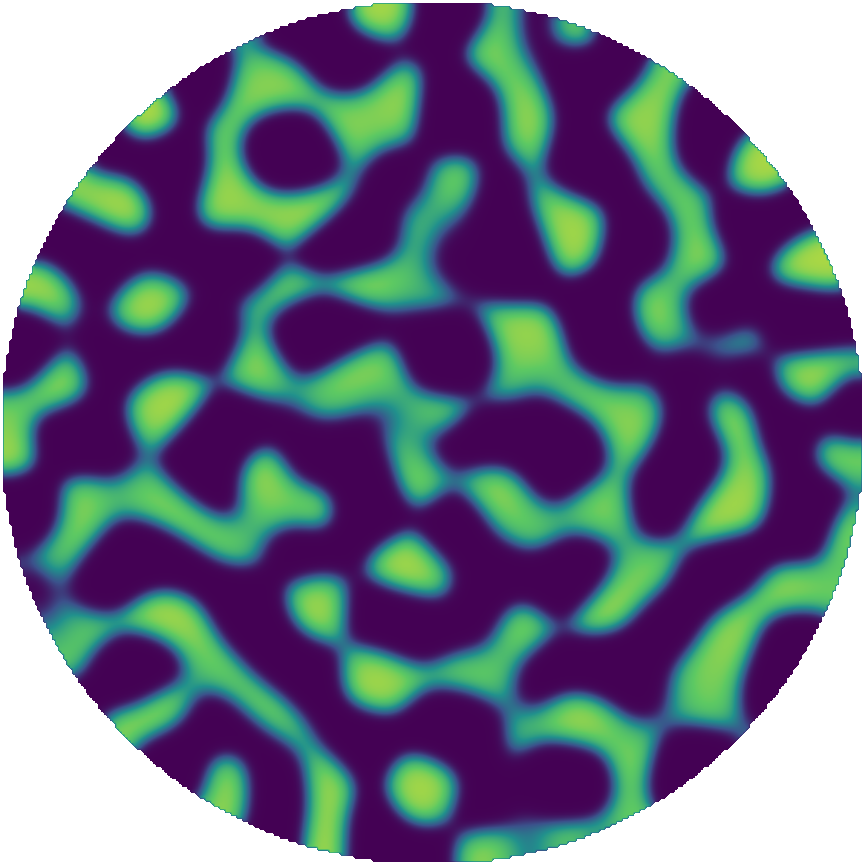}}
    \hspace{0.2cm}
    \subfigure[$t=200$]{\includegraphics[width=0.27\linewidth]{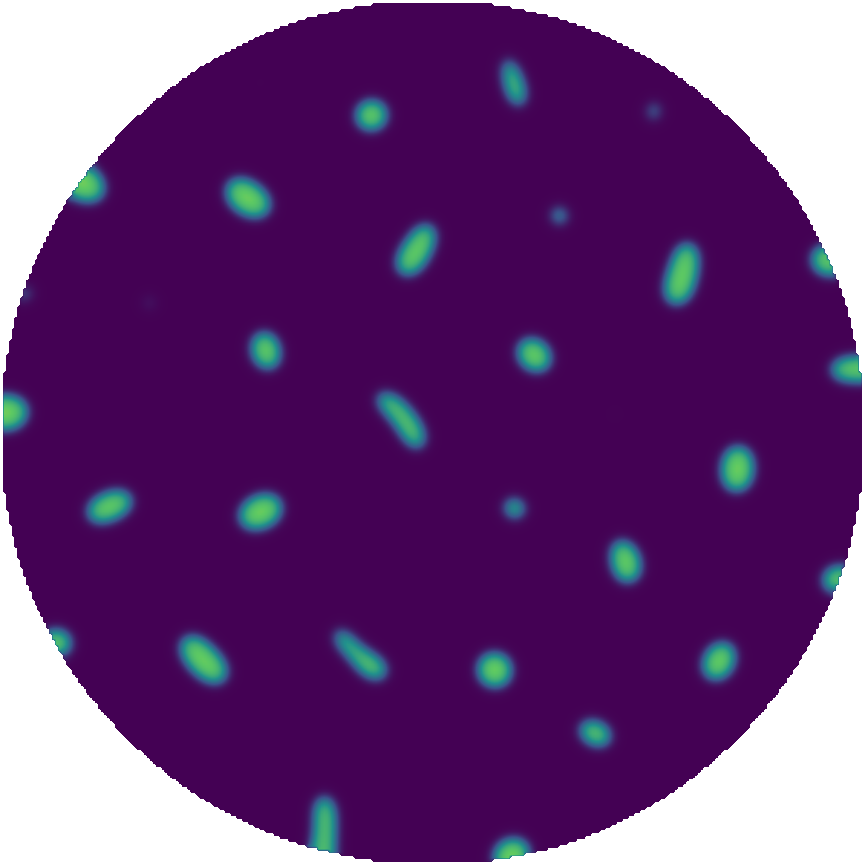}}
    \hspace{0.2cm}
    \raisebox{-0.1cm}{\subfigure{\includegraphics[width=0.05\linewidth]{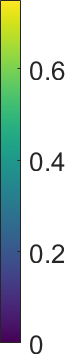}}}

    \subfigure[$t=250$]{\includegraphics[width=0.27\linewidth]{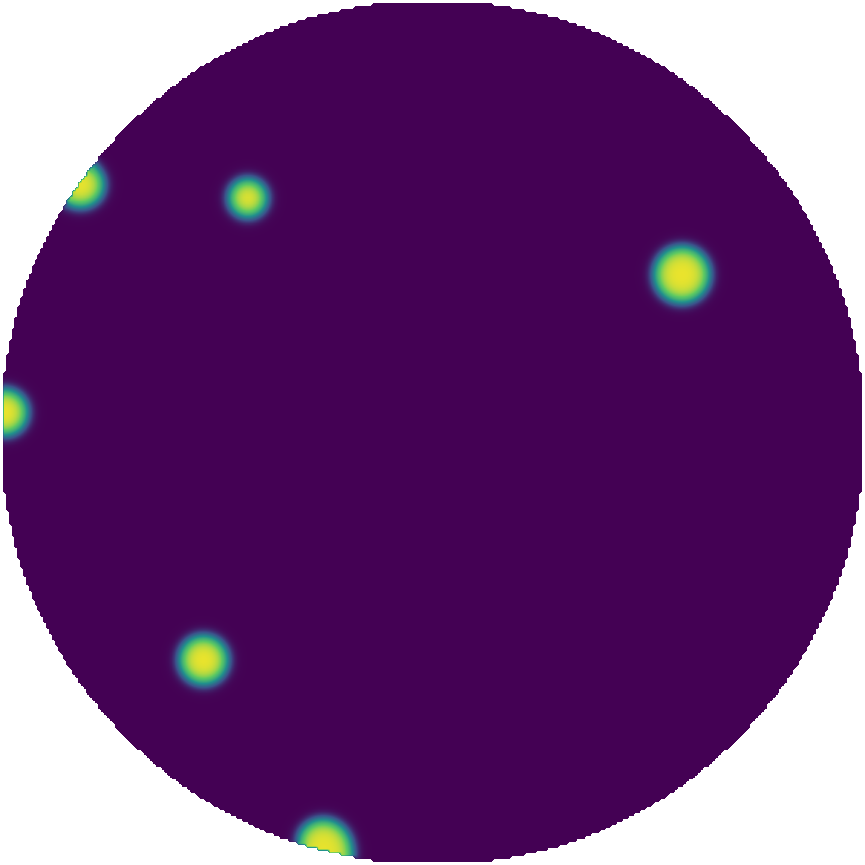}}
    \hspace{0.2cm}
    \subfigure[$t=350$]{\includegraphics[width=0.27\linewidth]{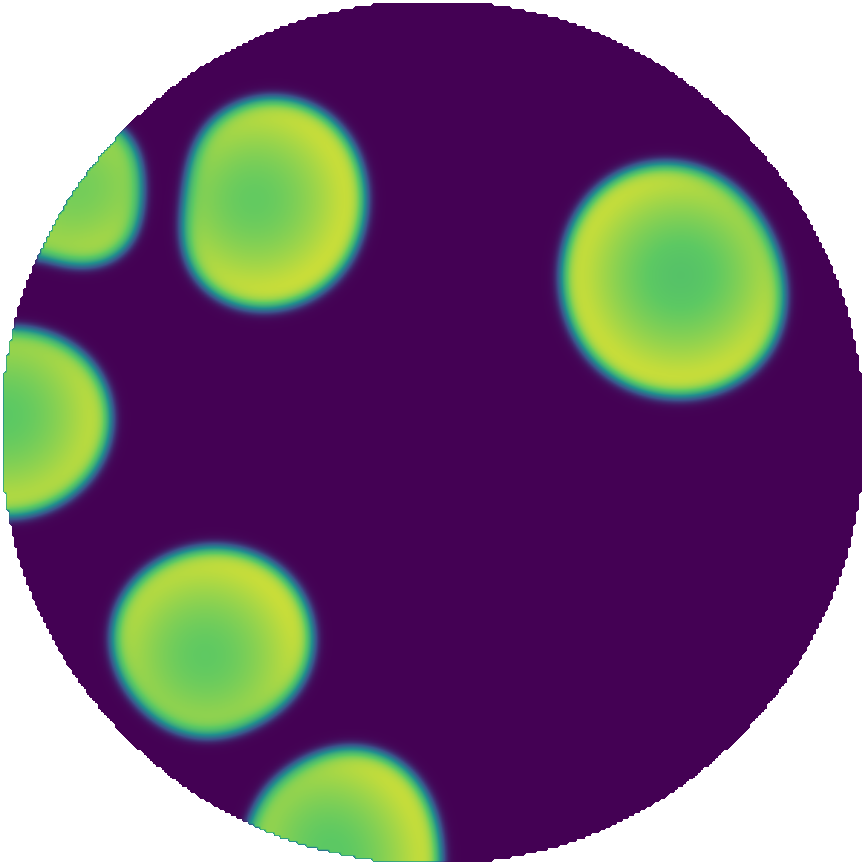}}
    \hspace{0.2cm}
    \subfigure[$t=500$]{\includegraphics[width=0.27\linewidth]{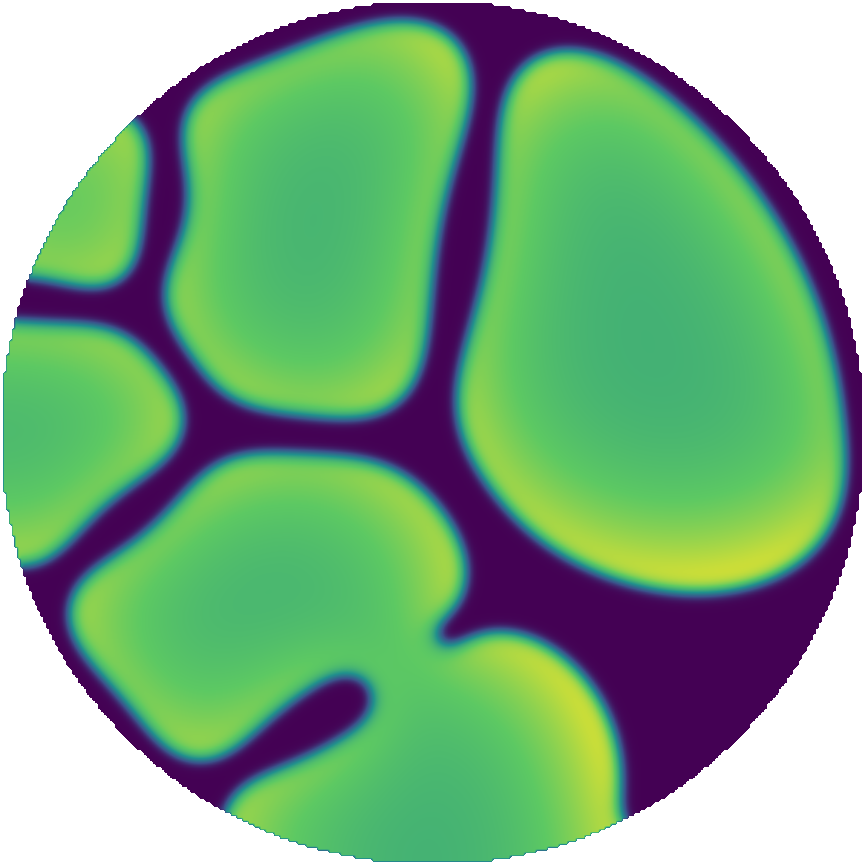}}
    \hspace{0.2cm}
    \raisebox{-0.1cm}{\subfigure{\includegraphics[width=0.05\linewidth]{figures/2D_Treatment_sigma_w_colorbar.png}}}
    \caption{Snapshots from simulations of Equations \eqref{u_nondim_eq}-\eqref{w_nondim_eq} in a circular domain of diameter $L=300$ {with homogeneous Neumann boundary conditions,} $\sigma_u=0.014$, $\sigma_w = 20\left(\frac{t}{400}H\left(0.5-\frac{t}{400}\right)+\left(1-\frac{t}{400}\right)H\left(\frac{t}{400}-0.5\right)\right)H\left(1-\frac{t}{400}\right)$, $\alpha = 0.07$, $\delta_u=\delta_w=100$, and all other parameters as in Table \ref{parameter_table}. The cancer cell density $v$ is plotted, with $u$ and $w$ exhibiting in-phase patterned solutions in each panel.}
    \label{fig:2D_sigma_w_treatment}
\end{figure}

Next we consider the hysteresis effect shown in Figure \ref{fig:reversing_treatment} on a circular domain. We use a similar linear ramping of the IL-2 treatment parameter, $\sigma_w$, from $\sigma_w=0$ to a maximum value of $\sigma_w=20$ at $t=200$, and then back to $\sigma_w=0$ at $t=400$, after which we turn off the treatment entirely. We show snapshots from these simulations in Figure \ref{fig:2D_sigma_w_treatment}, where we see a progression of increasingly sparse patterns as the treatment parameter is increased, and then a subsequent growth of these sparse remnants into larger regions of high tumour density. Simulating further in time in panel (g), the system appears to approach a steady inhomogeneous distribution, as one might expect from the one-dimensional numerical continuation shown in Figure \ref{fig:bifurcation_sigma_w}. In two spatial dimensions however, we expect there to be many more stable inhomogeneous equilibria, and this can be seen by running this same simulation again using a different seed of the initial random perturbation of the homogeneous equilibrium.

We now consider a more realistic model of treatment which accounts for local transport of the IL-2-compound into the tumour via tissue boundaries (e.g.~through blood vessels and capillaries on the boundary of the tissue region being modelled). We propose a simple model of this using inhomogeneous Dirichlet conditions for $w$ given by
\begin{equation}\label{eq_Dirichlet_BC}
     w = B_w(t),
\end{equation}
where $B_w(t)$ is a given time-dependent function modelling absorption of the IL-2-compound at the boundary of the tissue. We also considered inhomogeneous Neumann conditions specifying fluxes on the boundaries, with qualitatively comparable results that we omit for brevity. We refer to \citep{dillon1994pattern, krause2021isolating} for further discussion on the biological interpretation of such mixed boundary conditions, and their impacts on pattern formation.

\begin{figure}
    \centering
    \subfigure[$t=100$]{\includegraphics[width=0.27\linewidth]{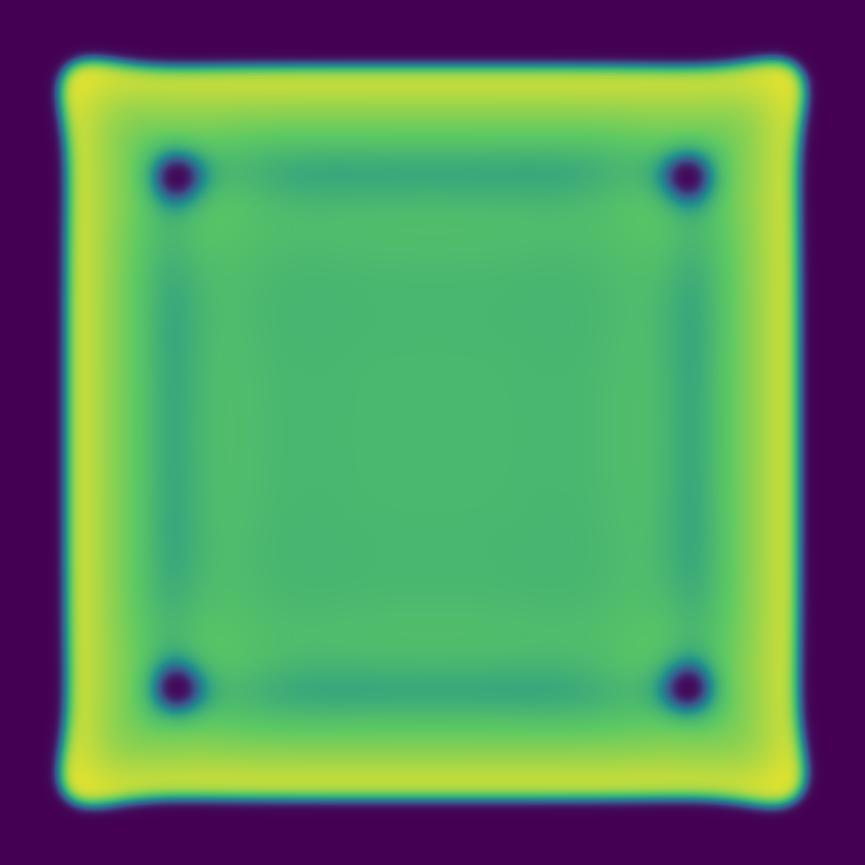}}
    \hspace{0.2cm}
    \subfigure[$t=150$]{\includegraphics[width=0.27\linewidth]{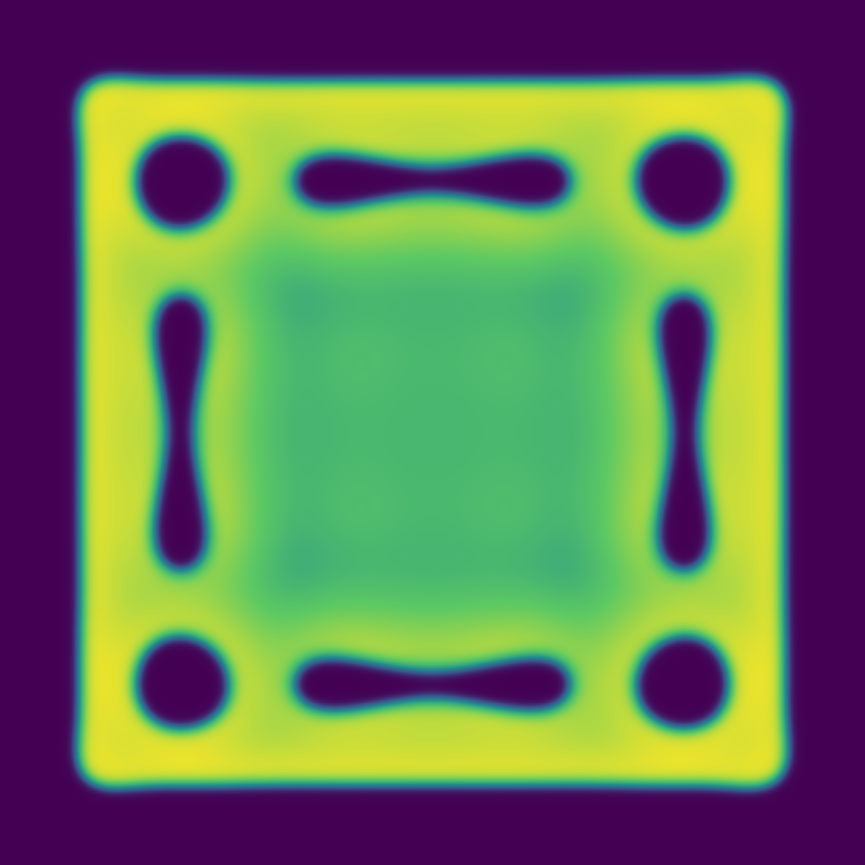}}
    \hspace{0.2cm}
    \subfigure[$t=250$]{\includegraphics[width=0.27\linewidth]{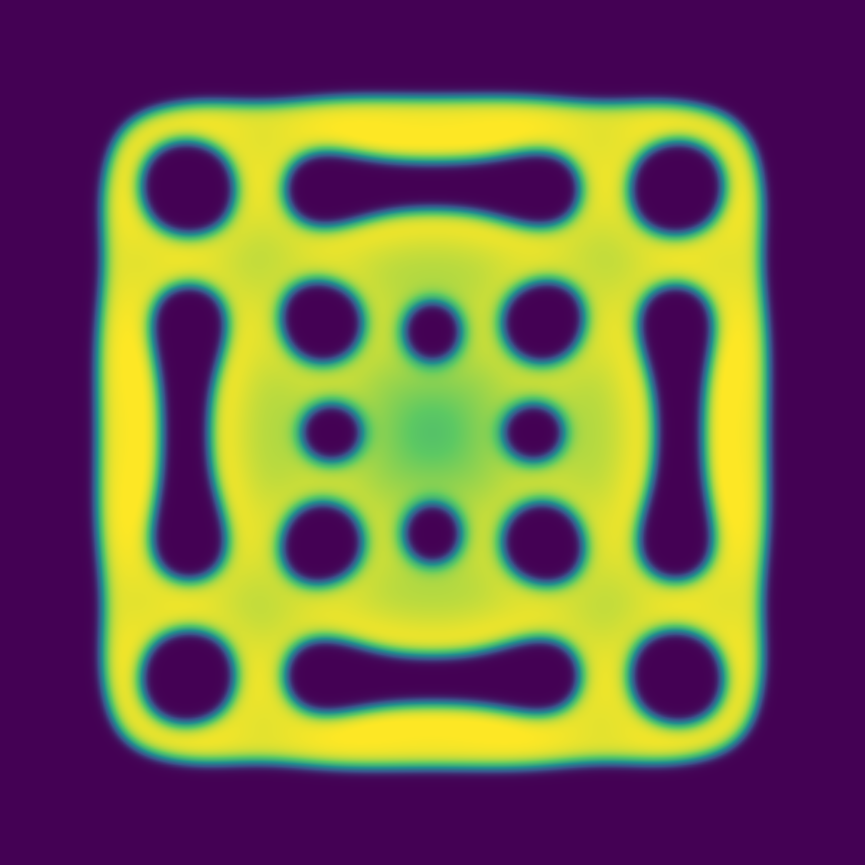}}
    \hspace{0.2cm}
    \raisebox{-0.1cm}{\subfigure{\includegraphics[width=0.05\linewidth]{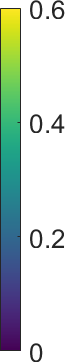}}}

    \subfigure[$t=350$]{\includegraphics[width=0.27\linewidth]{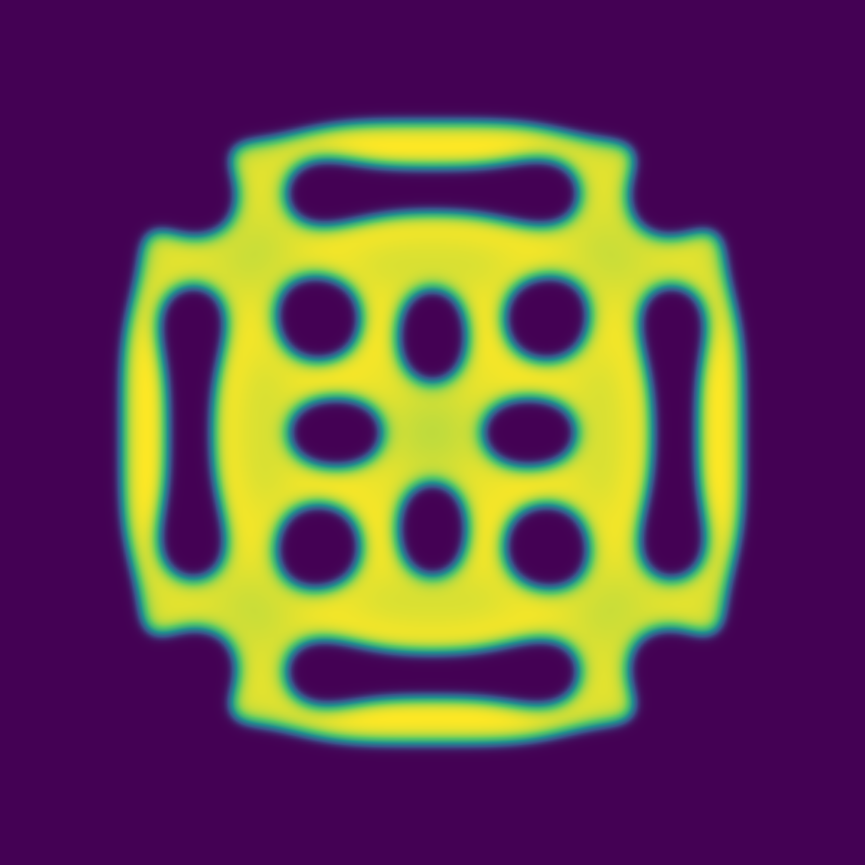}}
    \hspace{0.2cm}
    \subfigure[$t=450$]{\includegraphics[width=0.27\linewidth]{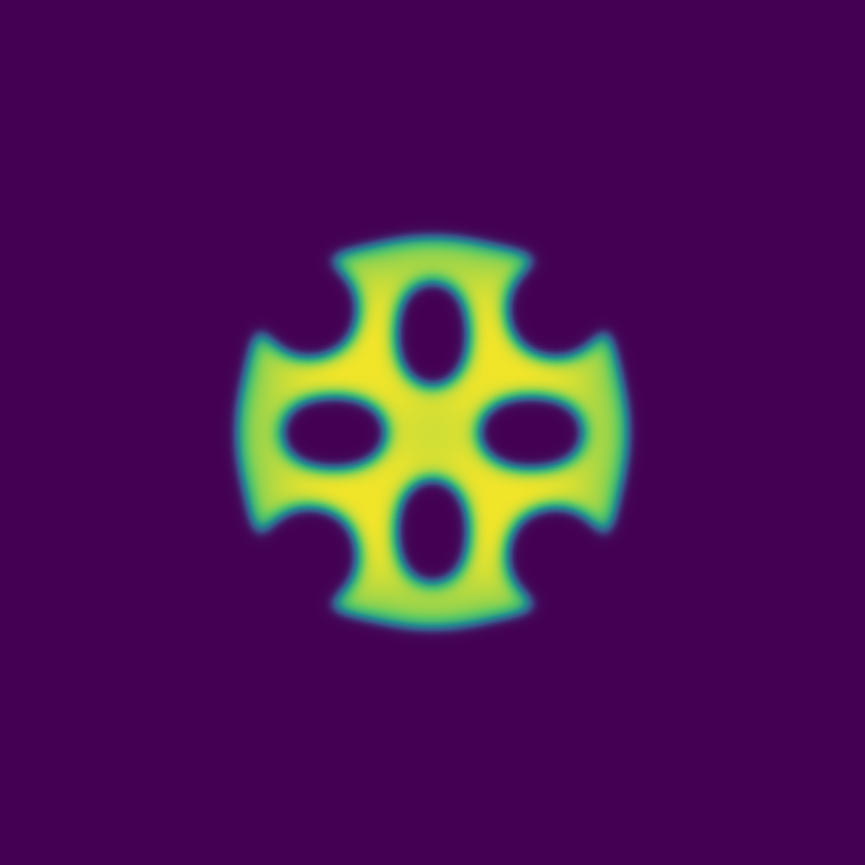}}
    \hspace{0.2cm}
    \subfigure[$t=500$]{\includegraphics[width=0.27\linewidth]{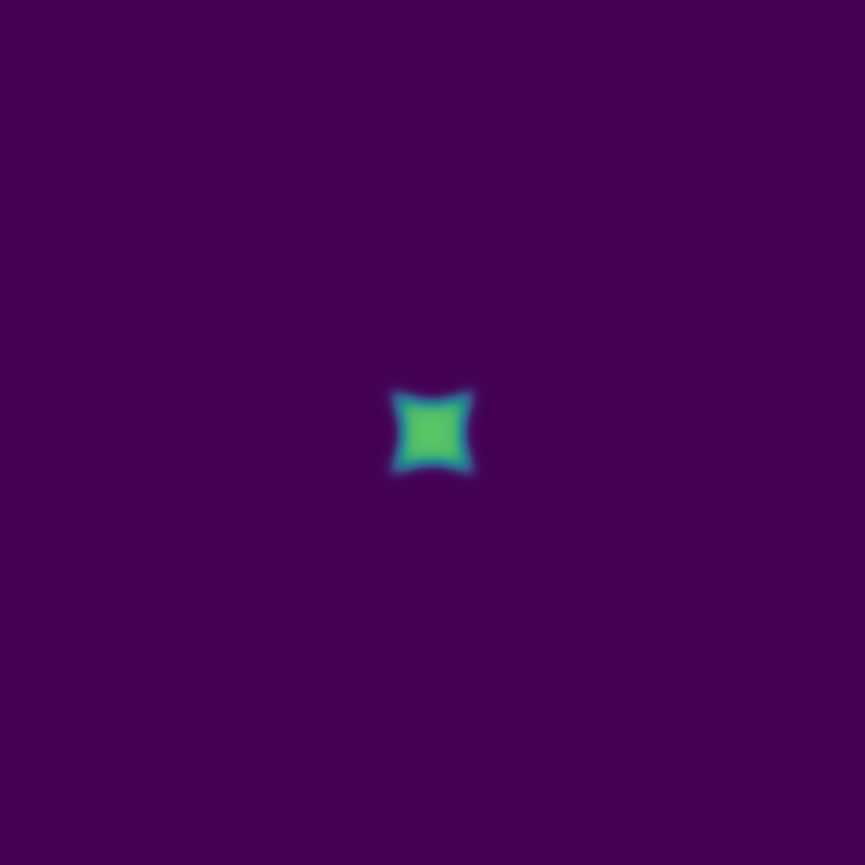}}
    \hspace{0.2cm}
    \raisebox{-0.1cm}{\subfigure{\includegraphics[width=0.05\linewidth]{figures/2D_Boundary_colorbar.png}}}
    \caption{Snapshots from simulations of Equations \eqref{u_nondim_eq}-\eqref{w_nondim_eq} in a square domain of side length $L=300$, $\sigma_u=0.007$, $\sigma_w = 0$, $\alpha = 0.1$, $\delta_u=100$, $\delta_w=1000$, and all other parameters as in Table \ref{parameter_table}. The boundary conditions for $u$ and $v$ are homogeneous Neumann, whereas we implement Equation \eqref{eq_Dirichlet_BC} for $w$ on the four boundaries with $B_w(t) = 0.01t$. The cancer cell density $v$ is plotted, with $u$ and $w$ exhibiting in-phase patterned solutions in each panel.}
    \label{fig:2D_boundary_treatment}
\end{figure}

We show an example of this situation in Figure \ref{fig:2D_boundary_treatment}, with a linearly-in-time increasing treatment specified at the domain boundaries. We see that the tissue recedes from the boundary, as one might expect, but also that it forms an internal periodic pattern of holes. As the boundary treatment is increased, the region where the cancer cell density is high recedes towards the center of the domain, with a slow restructuring of the patterned regions. 

\begin{figure}
    \centering
    \subfigure[$t=100$]{\includegraphics[width=0.27\linewidth]{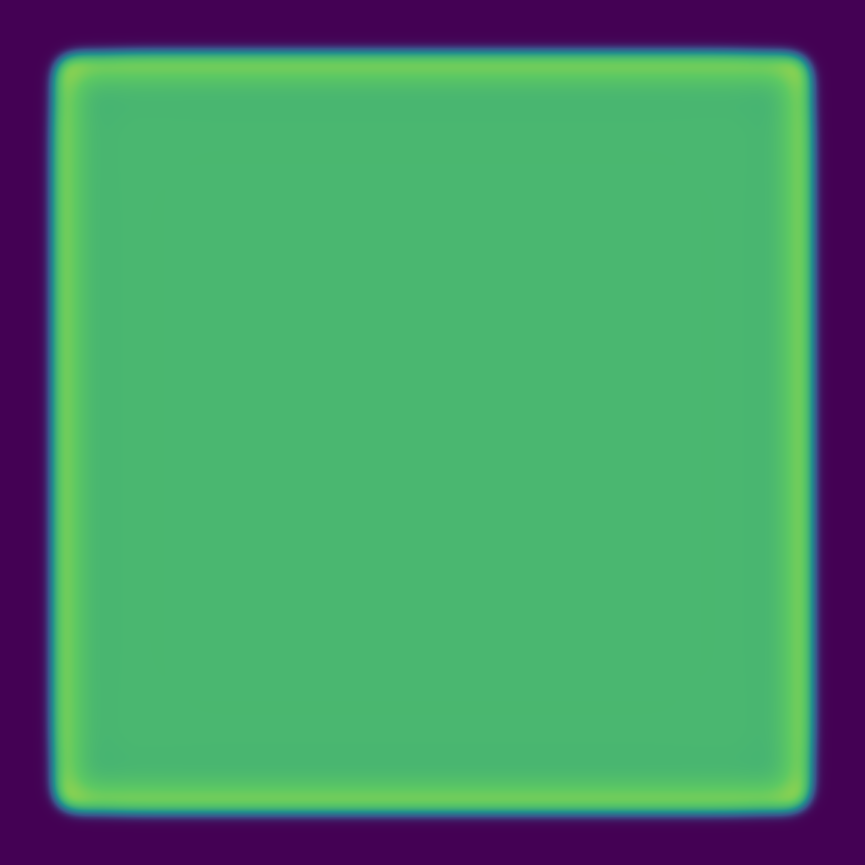}}
    \hspace{0.2cm}
    \subfigure[$t=250$]{\includegraphics[width=0.27\linewidth]{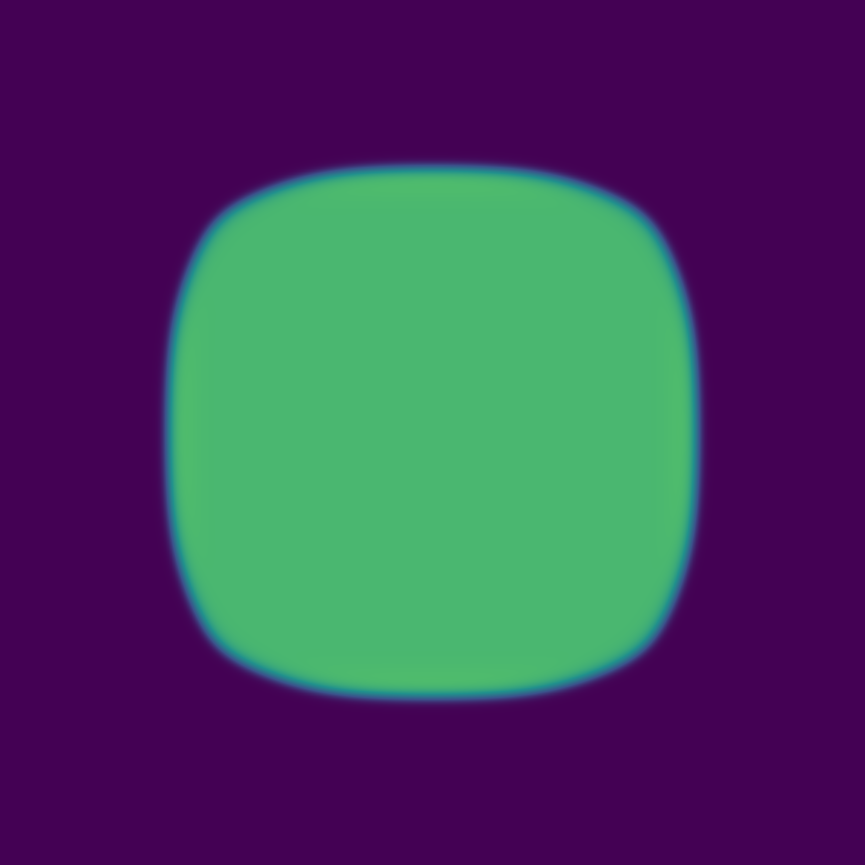}}
    \hspace{0.2cm}
    \subfigure[$t=350$]{\includegraphics[width=0.27\linewidth]{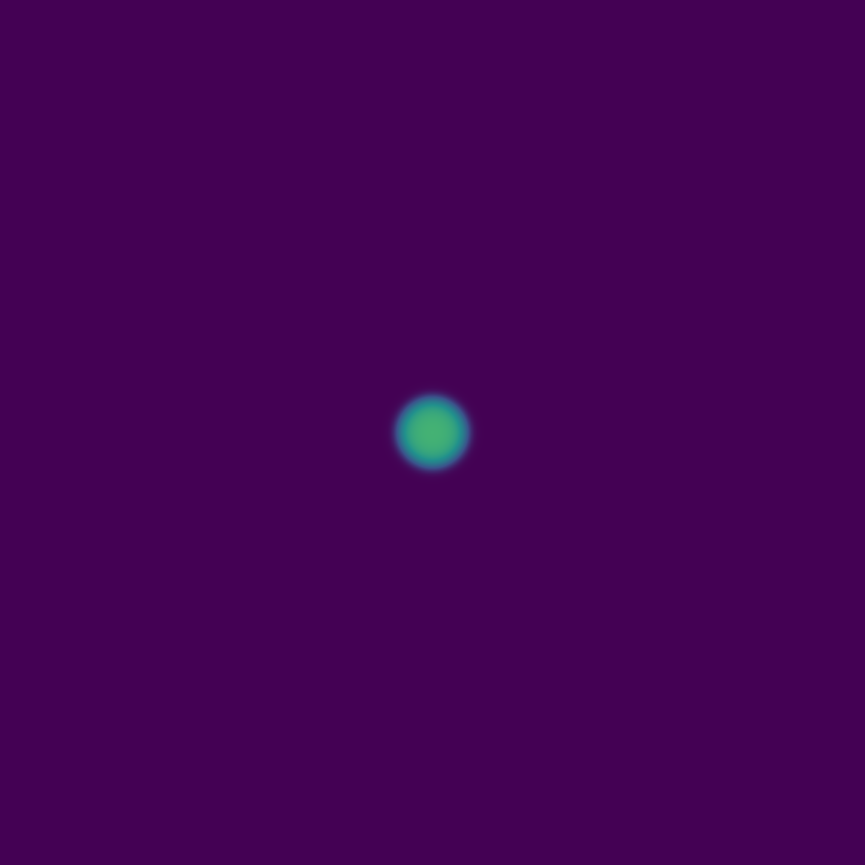}}
    \hspace{0.2cm}
    \raisebox{-0.1cm}{\subfigure{\includegraphics[width=0.05\linewidth]{figures/2D_Boundary_colorbar.png}}}
    \caption{Snapshots from simulations exactly as in Figure \ref{fig:2D_boundary_treatment} except with $\delta_u=10$.}
    \label{fig:2D_boundary_treatment_NoTuring}
\end{figure}

We contrast this with a simulation in a parameter regime that is not Turing unstable (i.e.~by decreasing the value of $\delta_u$) in Figure \ref{fig:2D_boundary_treatment_NoTuring}. We see comparable regions of where the tumour is up to time $t=250$ in this simulation compared to those in Figure \ref{fig:2D_boundary_treatment}, though without the internal patterning as before. Importantly, without the pattern formation, the tumour is eradicated earlier (just after $t=350$ here, compared to after $t=500$ in the previous case). 

\begin{figure}
    \centering
    \subfigure[$t=1000$]{\includegraphics[width=0.27\linewidth]{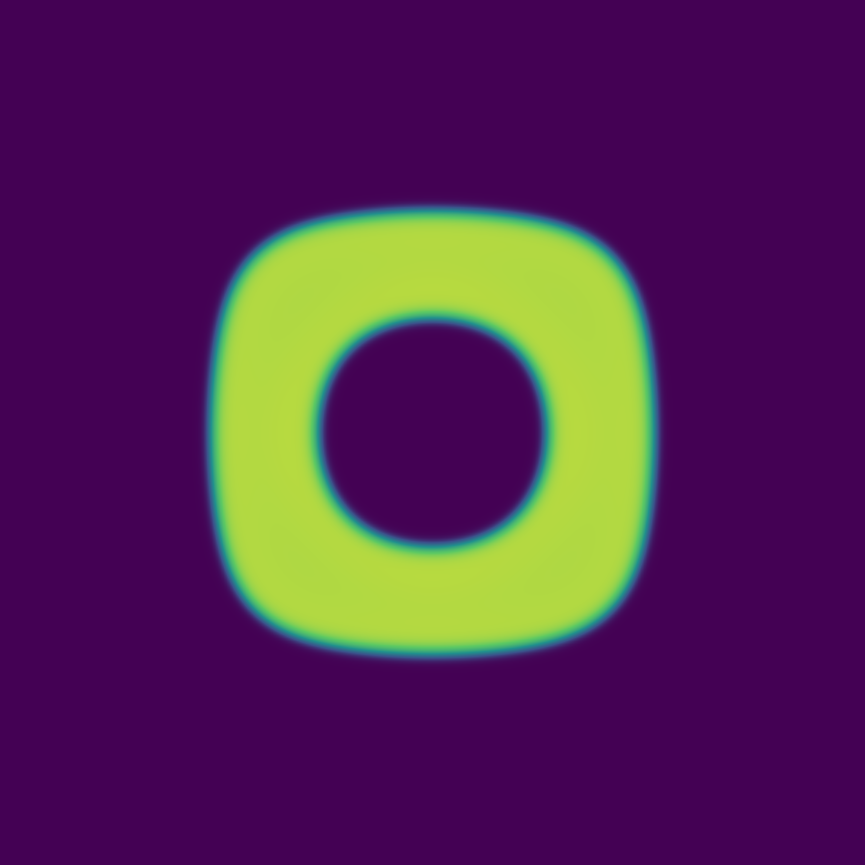}}
    \hspace{0.2cm}
    \subfigure[$t=1500$]{\includegraphics[width=0.27\linewidth]{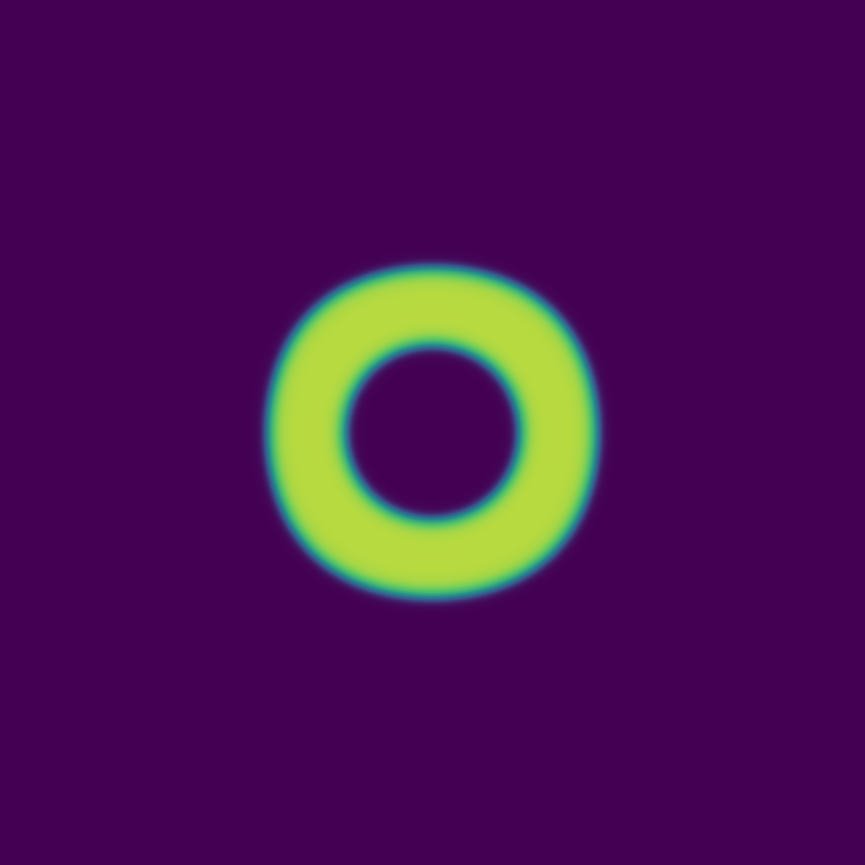}}
    \hspace{0.2cm}
    \subfigure[$t=1900$]{\includegraphics[width=0.27\linewidth]{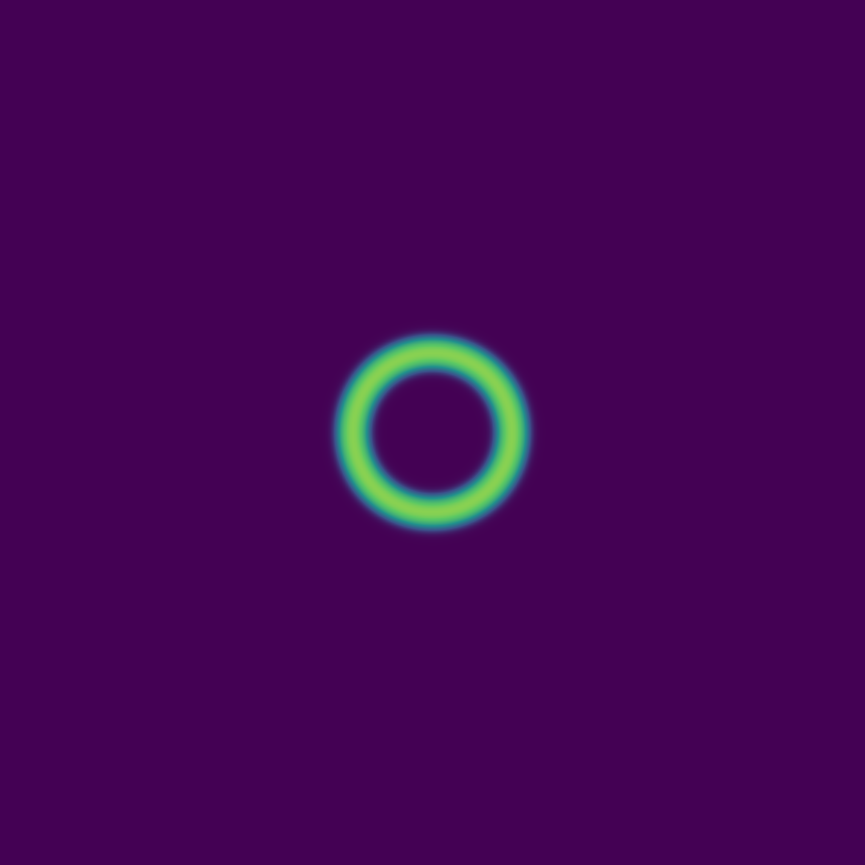}}
    \hspace{0.2cm}
    \raisebox{-0.1cm}{\subfigure{\includegraphics[width=0.05\linewidth]{figures/2D_Treatment_sigma_u_colorbar.png}}}

    \caption{Snapshots from simulations exactly as in Figure \ref{fig:2D_boundary_treatment} except with $\delta_u = 1000$}
    \label{fig:2D_boundary_treatment_MoreTuring}
\end{figure}

We also consider moving further into the Turing space by increasing $\delta_u$ to see how this changes the timescale of extinction. We set $\delta_u=1000$ and run the same simulation again, which we plot in Figure \ref{fig:2D_boundary_treatment_MoreTuring}. Here, while the outer boundary of the cancerous tissue is similar to the previous two Figures for short times (not shown), the only internal structure which forms is a circular region at the center of the domain. This structure persists for very long timescales, eventually thinning into a circular region of cancer cells which are eliminated just after $t=1950$. This is almost four times longer than in the case of $\delta_u=100$ and over five times longer than $\delta_u=10$, clearly implicating a counter-intuitive role of effector cell mobility on the resilience of tumours to boundary-driven treatment by IL-2 compounds. Recall that these timescales loosely correspond to the doses being administered at the boundaries, indicating substantially more treatment being required to successfully eliminate the cancer. 

\subsection{Nonlinear tumour cell diffusion}

\begin{figure}
    \centering
    \subfigure[$t=125$]{\includegraphics[width=0.27\linewidth]{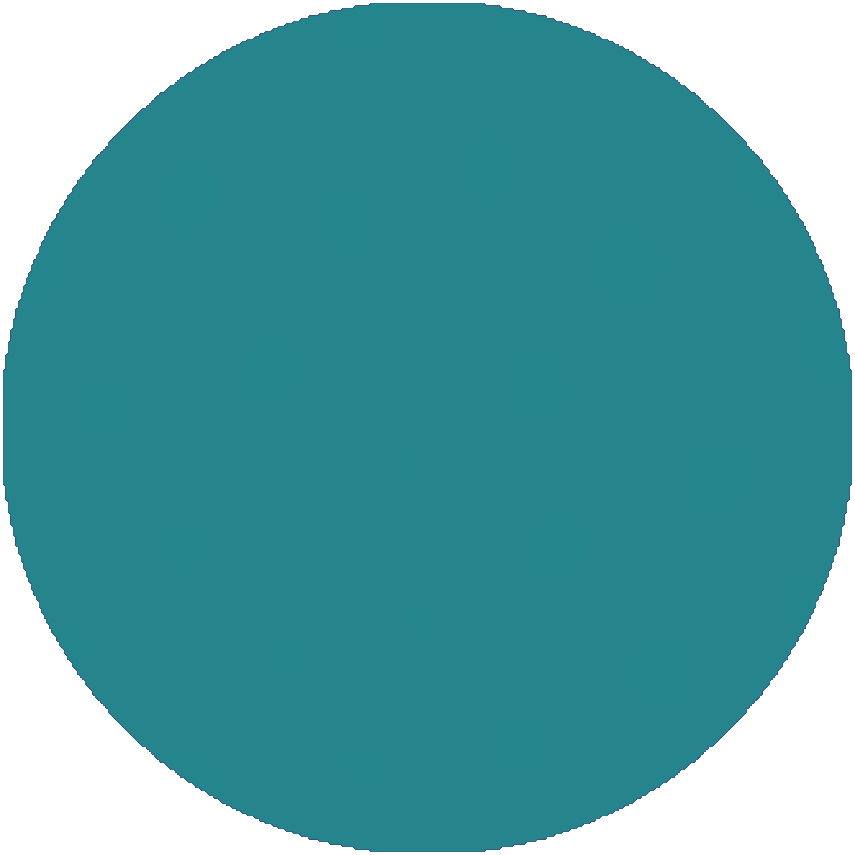}}
    \hspace{0.2cm}
    \subfigure[$t=150$]{\includegraphics[width=0.27\linewidth]{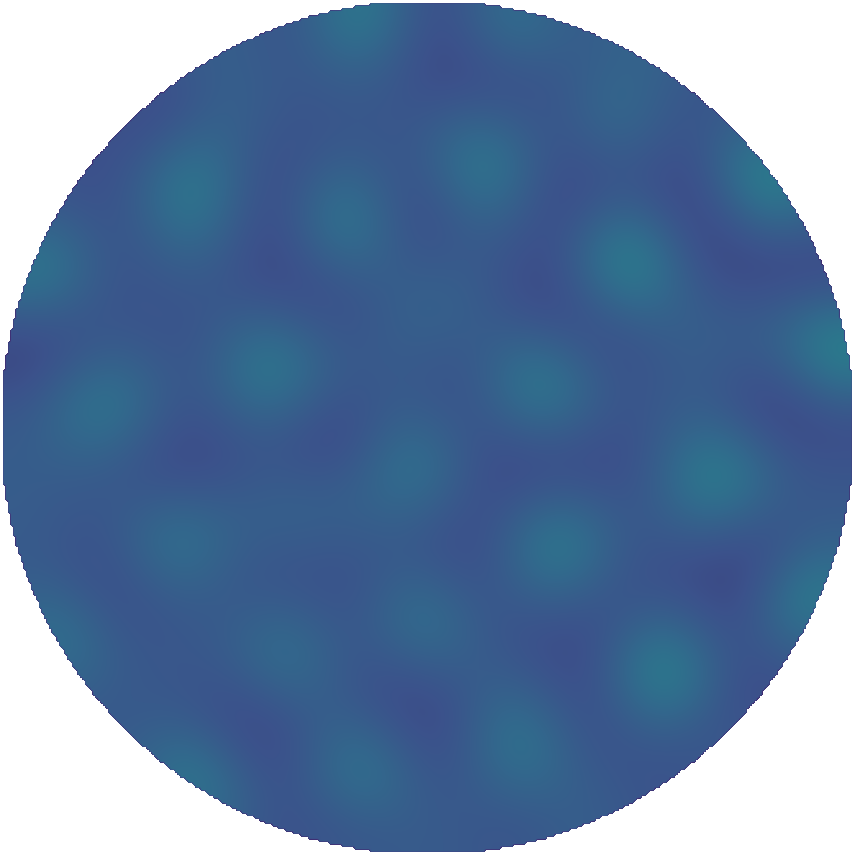}}
    \hspace{0.2cm}
    \subfigure[$t=175$]{\includegraphics[width=0.27\linewidth]{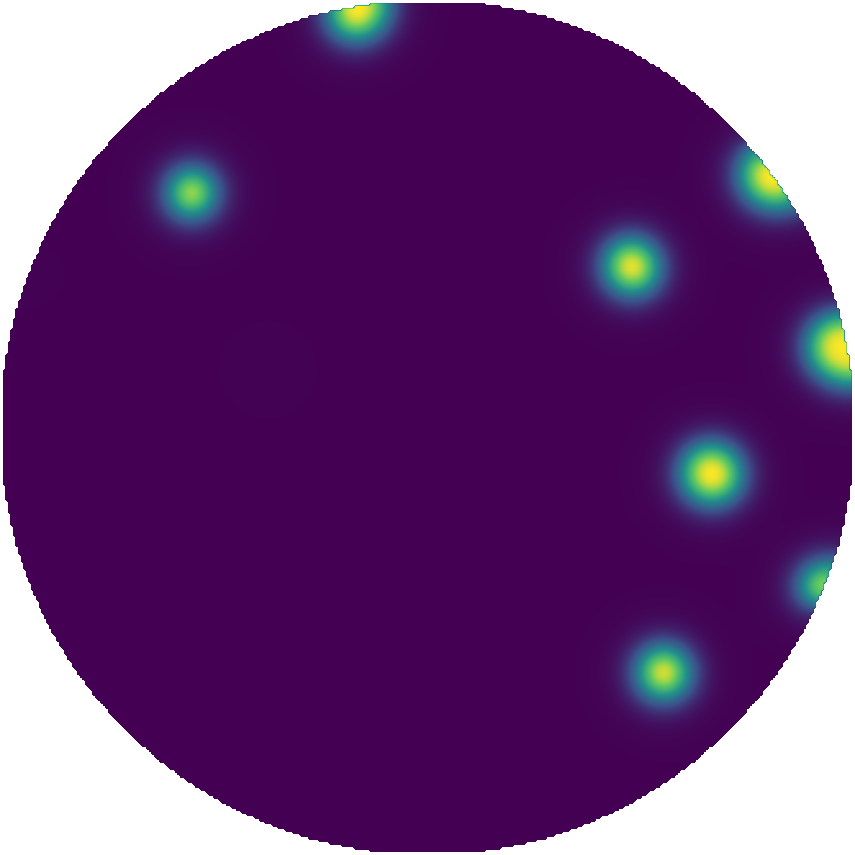}}
    \hspace{0.2cm}
    \raisebox{-0.1cm}{\subfigure{\includegraphics[width=0.05\linewidth]{figures/2D_Treatment_sigma_u_colorbar.png}}}
    
    \subfigure[$t=200$]{\includegraphics[width=0.27\linewidth]{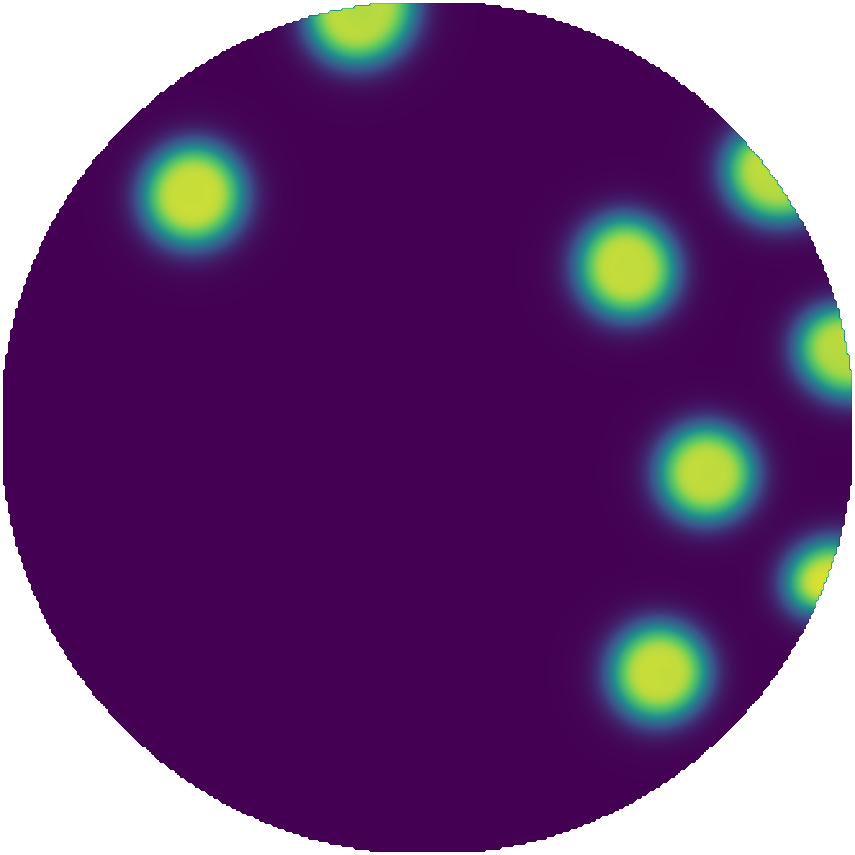}}
    \hspace{0.2cm}
    \subfigure[$t=250$]{\includegraphics[width=0.27\linewidth]{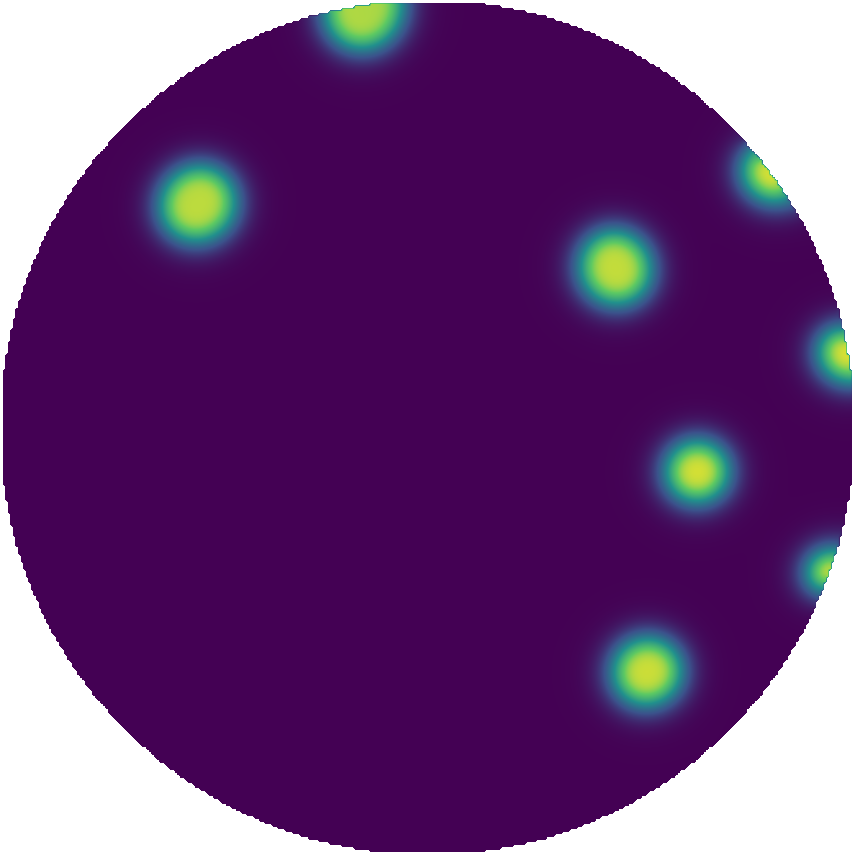}}
    \hspace{0.2cm}
    \subfigure[$t=300$]{\includegraphics[width=0.27\linewidth]{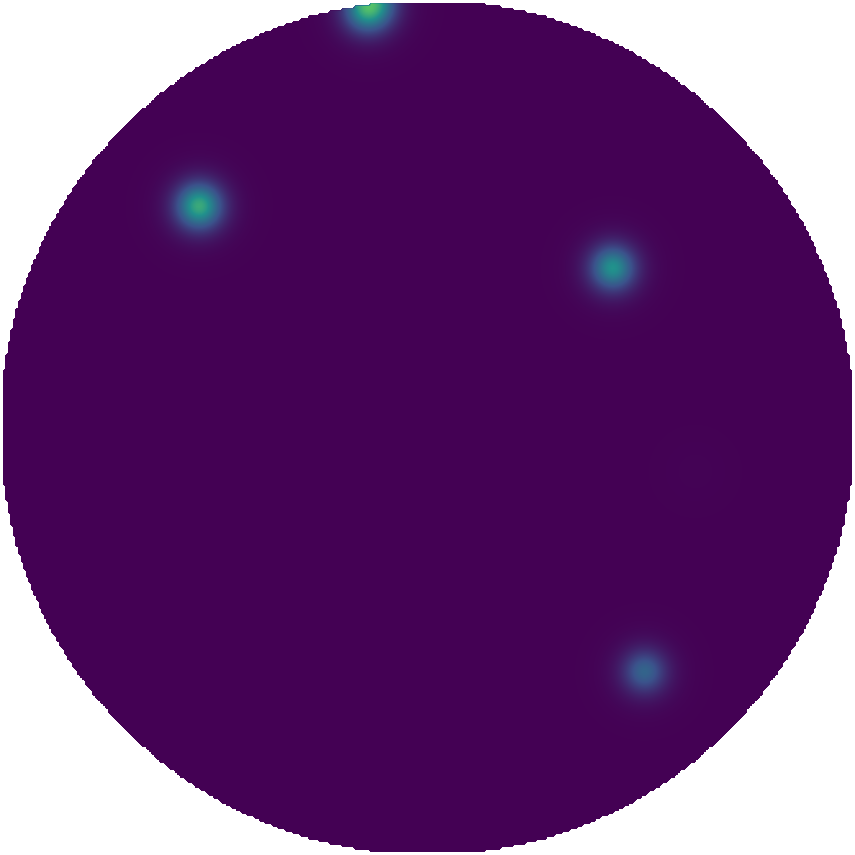}}
    \hspace{0.2cm}
    \raisebox{-0.1cm}{\subfigure{\includegraphics[width=0.05\linewidth]{figures/2D_Treatment_sigma_u_colorbar.png}}}
    \caption{{Snapshots from simulations of Equations \eqref{u_nondim_eq}, \eqref{w_nondim_eq}, and \eqref{v_nonlinear_diffusion_eq} in a circular domain of diameter $L=100$ with homogeneous Neumann boundary conditions, $\sigma_u=0.015 + 5\times 10^{-5}t$, $\sigma_w = 0.5$, $\alpha = 0.07$, $\delta_u=\delta_w=5$, and all other parameters as in Table \ref{parameter_table}. The cancer cell density $v$ is plotted, with $u$ and $w$ exhibiting in-phase patterned solutions in each panel.}}
    \label{fig:nonlinear_diffusion} 
\end{figure}

{As discussed earlier, linear diffusion is a rather crude model of cell motility, especially for cancer cell populations. We show a simple example of the impact of nonlinear cell diffusion by incorporating a volume-filling effect in the diffusion of the cancer cell population (see, e.g., \citep{strobl2020mix, andasari2011mathematical} for examples of such terms in the literature, which model a competition for space to grow). We replace Equation \eqref{v_nondim_eq} with the following equation,
\begin{equation}
    \frac{\partial v}{\partial {t}} = \nabla\cdot \left((1-v)^2 \nabla v \right) +  v(1-v) -\frac{uv}{\gamma_v + v},
    \label{v_nonlinear_diffusion_eq}
\end{equation}
so that the effective random motility of the cancer cell population decreases as it approaches its treatment-free carrying capacity of $v=1$. 

This modification of the diffusion term gave rise to comparable simulations in several of the parameter regimes studied. Importantly, it also had the impact of widening the region of Turing instabilities as we now demonstrate. Using parameters comparable to Figure \ref{fig:2D_sigma_u_treatment}, except with reduced effector and IL-2 diffusivities of $\delta_u=\delta_w=5$, we show simulations of this nonlinear-diffusion effect in Figure \ref{fig:nonlinear_diffusion}. For these parameters, the original model with linear diffusion will not undergo Turing patterning, and will instead lead to extinction of the cancer cell population around $t=160$. In this Figure, we instead see the formation of sporadic blobs of cancer cells that decay but regrow after spatially restructuring themselves. Eventually the immunotherapy is strong enough to eliminate them, but this occurs around twice the dose needed to eliminate the cancer without the effect of the nonlinear diffusion. We note that the lengthscale of Figure \ref{fig:nonlinear_diffusion} is smaller than many other simulations, as the smallest effective diffusion (and hence the lengthscale of the patterned states) is smaller than in previous examples.

Due to the complexity involved in modelling other forms of transport, we do not pursue further examples of this here. Rather, we emphasise that the resilience mechanisms described in the case of linear diffusion can persist, and in fact be enhanced, with the inclusion of more realistic transport mechanisms.} 

\section{Discussion}\label{sec_discussion}

In summary, we found that the mechanism of resilience to immunotherapy via Turing-like patterning persists across a range of geometries and approaches to treatment. The ability of a cancerous cell population to form inhomogeneous spatial structures allowed for persistence well beyond cases where models neglecting this spatial structure would predict elimination of the cancer. Perhaps counter-intuitively, this implicated a role of increased effector cell mobility on resilience to immunotherapy via IL-2 compounds, though this may be an artifact of the particular choices of {transport} made here, or the simplification of the full interactions involved in such immunotherapies. While the {use of linear diffusion} is likely overly simplistic given what we know about effector and tumour cell motility, we conjecture that this mechanism of spatial structuring persists in more complicated models, and may be an under-explored mechanism of treatment resistance in cancer therapies more generally. {There is also an important need to consider more realistic treatment regimes, such as more complicated dosing schedules which balance efficacy with side-effects. The brief discussion of the Atto-fox problem at the end of Section \ref{sec_1d_sims} suggests some of the complexities involved in interpreting outputs of time-dependent continuous-state models biologically.}

Besides exploring this mechanism in more detailed models, there is likely an interplay between the spontaneous spatial restructuring observed here, and the emergence of phenotypic heterogeneity in tumours undergoing treatment with different therapies. The spontaneous evolution of drug resistance among cancerous tissues is an increasingly well-studied area \citep{chisholm2016cell, bull2023quantification}. In particular, there is a growing literature on spatially resolving phenotypic heterogeneity in order to better understand adaptive treatment strategies \citep{strobl2022spatial, west2023survey}. While the complexity of incorporating patterning-induced resilience in such scenarios may make analysis more challenging, understanding the interplay between these distinct mechanisms of treatment resilience is undoubtedly important for mathematical modelling. This is especially important due to the increasing collection of complicated spatial omics datasets \citep{ahmed2022spatial}. The emergence of novel oncolytic therapies have recently been studied in the context of reaction-diffusion models \citep{baabdulla2024oscillations}, and exploring the interplay of spatial resilience and spatiotemporal pattern formation would be another important avenue for future work.

A major caveat of the work presented here is on the comparison between spatial and non-spatial models. In particular, Figure \ref{fig:overview_figure} and many other examples compare ODE and PDE simulations with the same kinetic parameters, but in actuality this is a crude comparison. Given a microscale model of cellular responses to individual chemical stimuli (e.g.~an agent-based model, or a stochastic reaction scheme), one could derive mean-field continuum models which aim to average out spatial variation (see, e.g., \citep{bull2022hallmarks, erban2020stochastic}). In some cases, these will have the same kinetic parameter values, but that is not always the case, depending on the details of the microscale modelling and the limiting procedures used. Nevertheless, our main point here is to demonstrate a nuanced phenomenon which exists in spatial models that does not occur for their non-spatial counterparts. Capturing the impact of this mechanism in non-spatial mean-field models would be an interesting, albeit difficult, goal of future work. \temp{\cite{schonfeld2022environmental} used a different, data-driven approach to capture spatial variations via an auxiliary ``environmental stress" variable, which may yield more quantitatively comparable outcomes. Importantly, this variable (or any analogous approaches) would need to successfully account for the spatial resilience phenomenon demonstrated here to replace the use of explicit spatial modelling. Given the role of multistability on the emergence of this effect, one would need data from varying initial conditions to correctly capture this phenomenon.}

Beyond these practically focused questions, there is a need for better theoretical frameworks to explain and explore these kinds of mechanisms. Our stability analyses and numerical continuation results satisfactorily explain the behaviours explored here in this simple model, though even still we have omitted many details by, e.g., only showing a handful of continuation branches in Section \ref{sec_continuation}, and entirely neglecting the complexity of 2D continuation. We have also evidenced impacts of rate-induced bifurcations, but have not explored these systematically. As discussed by \cite{krakovska_resilience_2024}, there are different formalisms of resilience that go beyond these local kinds of analysis, though it is not apparent which of these might be appropriate for more realistic models in modelling cancer. In population ecology, where these notions of resilience are more well-studied, authors such as \cite{scheffer2015generic} argue that new quantitative tools are needed to operationalize qualitative notions of resilience, and that the `correct' formalism to use will be application-dependent. The work in this paper, then, is ideally a good representation of the need to explore such questions within the context of mathematical oncology more generally.

\begin{acknowledgements}
E.~V-S. has received PhD funding from ANID, Beca Chile Doctorado en el extranjero, number 72210071.
\end{acknowledgements}

\noindent\textbf{\small Data Availability Statement:} {\small All data in this manuscript come from simulations described in the GitHub repository, \citep{Molly_Github}.}

\bibliographystyle{abbrvnat}
\bibliography{refs}
\appendix
\section{Calculating coexistence equilibria}\label{appendix_equilibria}
When solving for the coexistence equilibria we assume $v_0 \neq 0$ and rearrange \eqref{v_hom_eq} and \eqref{w_hom_eq} to obtain equations for $u_0$ and $w_0$ in terms of $v_0$, given by \eqref{uss} and \eqref{wss}. We can then substitute these back into \eqref{u_hom_eq} in order to solve for $v_0$. In doing this we obtain a quintic equation for $v_0$,
\begin{equation}
    a_5 v_0^5 + a_4 v_0^4 + a_3 v_0^3 + a_2 v_0^2 + a_1 v_0 + a_0 = 0, \label{quintic}
\end{equation}
with coefficients depending on our system parameters as follows:
\begin{align}
  a_5 =& \rho_w(\rho_u - \mu_u),\\
  a_4 =& - \rho_w(\alpha + 2(\gamma_v -1)(\mu_u - \rho_u)),\\
  a_3 =&  \rho_w (\alpha(1- \gamma_v) + \rho_u(1 + \gamma_v(\gamma_v -4)) - \sigma_u) - \rho_u \sigma_w + \mu_u(\mu_w - \rho_w(1 + \gamma_v(\gamma_v - 4)) + \sigma_w),\\
  a_2 =& -(\rho_w(\gamma_v -1)(2 \gamma_v \rho_u + \sigma_u) + (\gamma_v + \gamma_w - 1)(\mu_u \mu_w + \sigma_w(1 - \rho_u)) + \alpha(\mu_w + \gamma_v \rho_w + \sigma_w) + 2 \gamma_v \rho_w(\gamma_v - 1)),\\
  a_1 =& \rho_w(\sigma_u \gamma_v + \gamma_v^2(\rho_u - \mu_u)) + \sigma_u(\mu_w + \sigma_w) + \mu_w \gamma_w(\alpha - \mu_u) + \gamma_w \sigma_w(\alpha - \mu_u + \rho_u) + \gamma_v(\gamma_w -1)(\mu_u(\mu_w + \sigma_w) - \rho_u \sigma_w)\\
  a_0 =&\gamma_w(\gamma_v\rho_u\sigma_w +(\mu_w + \sigma_w)(\sigma_u - \gamma_v \mu_u)).
\end{align}

 \end{document}